\documentclass[fleqn,usenatbib]{mnras}

\usepackage{newtxtext,newtxmath}

\usepackage[T1]{fontenc}

\DeclareRobustCommand{\VAN}[3]{#2}
\let\VANthebibliography\thebibliography
\def\thebibliography{\DeclareRobustCommand{\VAN}[3]{##3}\VANthebibliography}

\usepackage{graphicx}	
\usepackage{amsmath}	

\title[Cluster dynamical states and mass accretion]{Tracing the dynamical states and mass accretion histories of galaxy clusters in IllustrisTNG}

\author[R. Reid et al.]{
Rashaad Reid,$^{1,2}$\thanks{E-mail: yreid@uwaterloo.ca}
Syeda Lammim Ahad,$^{1,2,3}$
Roan Haggar,$^{1,2}$
Charlie T. Mpetha$^{4}$
and James E. Taylor$^{1,2}$
\\
$^{1}$Waterloo Centre for Astrophysics, University of Waterloo, Waterloo, Ontario N2L 3G1, Canada\\
$^{2}$Department of Physics and Astronomy, University of Waterloo, Waterloo, Ontario N2L 3G1, Canada\\
$^{3}$Center for Astronomy, Space Science and Astrophysics, Independent University, Bangladesh, Dhaka 1229, Bangladesh\\
$^{4}$NASA Goddard Space Flight Center, 8800 Greenbelt Rd, Greenbelt, MD 20771, USA
}

\date{Accepted XXX. Received YYY; in original form ZZZ}

\pubyear{2026}

\begin{document}
\label{firstpage}
\pagerange{\pageref{firstpage}--\pageref{lastpage}}
\maketitle

\begin{abstract}
As the largest and most recently formed stage of hierarchical structure, present-day galaxy clusters are predicted to have a broad range of late-time assembly histories. This diversity may explain much of the scatter in scaling relations and other cluster properties. Observationally, systems with more or less recent accretion should appear as unrelaxed and relaxed clusters, respectively. However, it is unclear which of the many possible structural measures best correlate with assembly history. Using the IllustrisTNG simulations, we explore the correlation between structural parameters and assembly history. To assess the effectiveness of different structural selection criteria, we define subsamples of the most and least relaxed clusters based on the values of various intrinsic, projected, and stellar structural parameters, and then compare the median assembly history of the subsamples in each case. We find that several observable quantities, including the magnitude gap between the brightest galaxies and the asymmetry of the stellar mass distribution, are very effective in selecting cluster samples with more or less recent accretion, even when applied in projection. Given the strong correlations between assembly history and present-day cluster structure, we suggest that structural classification be included explicitly in any analysis of catalogue completeness, scaling relations, or mean density profiles.
\end{abstract}




\section{Introduction} 

Galaxy clusters are the most massive gravitationally bound structures in the universe. As such, they constitute the most recent phase of hierarchical structure formation and continue to grow significantly up to the present day \citep{Kravtsov2012}. Thus, clusters provide an important probe of the growth of structure in the late-time universe, most often through studies of cluster abundance \citep[e.g.][] {Allen2011,Lesci2022,Ghirardini2024,Aymerich2024}. These studies estimate the abundance of dark matter halos by measuring the cluster mass function and assuming an approximate correspondence between observed clusters and halos. Individual cluster mass determinations require considerable lensing or dynamical data, and thus they can only be made accurately for small numbers of systems. A more common approach is to estimate mass from simpler observable quantities, typically through scaling relations \cite[][]{Kaiser1986,Giodini2013,Lovisari2022}. The mean formation history or instantaneous growth rate of clusters also depends on cosmology, and estimates of these quantities may provide complementary cosmological tests \citep{Richstone1992,Evrard1993, Mohr1995,Amoura2021,Amoura2023}. However, practical methods applicable to large samples have only recently been developed \citep{Haggar2024b,Mpetha2024}. These cosmological tests based on galaxy clusters are of particular interest given recent tensions between low-redshift measures of growth and measures based on the cosmic microwave background \citep[][and references therein]{Abdalla2022,Karim2025}.

Beyond their cosmological applications, galaxy cluster catalogues play a number of other important roles. Given their contrast relative to the field, they provide easily identified samples of high-z galaxies. Due to their high density, clusters are the site of significant morphological change for galaxies \citep[e.g.][ and references therein]{Jeon2022,Oxland2024}, and they also allow many other tests of physics at high densities, temperatures, and/or gravitational curvature \citep{Eckert2022,Vogt2024,Roche2024,Butt2025,Chen2025,Supanitsky2026}. These tests will become increasingly powerful as cluster samples grow.

All current methods of cluster detection depend strongly on mass and/or redshift. Given the strong observational selection effects, cluster samples are naturally grouped using these two observed properties. The implicit assumption is that clusters in a small range of mass and redshift form a homogeneous sample appropriate for, e.g., calibrating scaling relations. However, clusters are continuously evolving systems; those with recent accretion experience significant short-term changes in dark matter structure \citep[][]{Smith2008, Wang2020, Drakos2019a, Drakos2019b,Richardson2022,Wang2025,Valles2025} and detailed gas properties \citep[][]{Chen2019,Zhang2022,Kelkar2023, Valles2026}. Extensive theoretical work has studied the intrinsic variation in the properties of dark matter halos and galaxy clusters \citep{Jeeson2011, Skibba2011, Wong2012, Chen2020, Haggar2024a}. These studies typically find one or more important forms of variation correlated with the dynamical state and growth history of the system. 

Observationally, part of this structural variation has long been recognised as the distinction between ``relaxed'' and ``unrelaxed'' clusters. Originally, the distinction between the two categories was clearest in X-ray studies and corresponded to the presence or absence of a cool core of enhanced emission. Other X-ray features also trace the degree of dynamical relaxation, including centre of mass offsets \cite[e.g.][]{Zenteno2020,Seppi2023}, or asymmetries and substructure in X-ray maps (\citealt{Jeltema2005}; see \citealt{Yuan2020,Yuan2022} for recent references). In the optical, a rich literature has developed studying and classifying tracers of dynamical activity or substructure, including the relative dominance of the brightest cluster galaxy (BCG), the concentration, and the projected shape of the halo \citep[e.g.][and references therein]{Pinkey1996,WenHan2013,Zarattini2016, Contreras2022}, or the phase-space distribution of the member galaxies \citep{Dressler1988, Benavides2023}. These dynamical features are known to correlate with scatter and offsets in scaling relations \citep[][]{Mulroy2019,Cerini2025} and can thus introduce bias in estimates of, e.g., mean cluster mass. They may also affect sample selection more broadly, introducing differences in purity and completeness correlated with dynamical state (e.g.~\citealt{Lovisari2017}; Reid et al.~in preparation).

Ongoing or forthcoming large optical surveys such as Euclid \citep{Euclid2019} and Vera C. Rubin Observatory’s Legacy Survey of Space and Time (LSST) \citep{Ivezic2019} will expand the available samples of galaxy clusters by several orders of magnitude. On the other hand, most of the objects in these samples will be low signal-to-noise ratio (SNR) detections, with membership information for only a few of the brightest galaxies. X-ray, Sunyaev-Zel'dovich (SZ) or weak lensing information will be available only after stacking large numbers of systems and will thus require pre-selection using optical information alone. Given the intrinsic variation in cluster properties at fixed mass and redshift, a practical question arises: can we structurally classify clusters based on their optical properties alone to form more homogeneous subsamples for stacking?

In this paper, we take the following ``split sample'' approach to evaluating possible structural classification methods. We consider a number of different potential indicators of the dynamical state, including centre of mass offset, projected shape, concentration, asymmetry, and mass or luminosity ratios of some of the brightest members. We sort a sample of simulated clusters according to the strength of the indicator and consider the extremes (roughly quintiles, i.e.,~the bottom and top 20\%) of the distribution. We then determine the median mass accretion history (MAH) for each subset and compare how different these are for splits based on different indicators. We do this first for 3D indicators based on the full mass (dark matter + baryonic) distribution to understand how much intrinsic correlation exists between a structural feature and the growth history. Next, we consider projected versions of these measures; these are quantities that could be estimated for small samples of well-studied, massive, low-redshift systems for which it is possible to make accurate lensing maps of the mass distribution. Finally, we consider measures based on the stellar mass or galaxy distribution that should be obtainable for much larger samples of optically detected clusters. (We note that we have already tested the split sample approach in practice on one large catalogue --- see \cite{ahad2025arXiv}.)

The outline of the paper is as follows. In Section~\ref{sec:2simulations}, we review the IllustrisTNG simulations and describe our structural parameter measurements. In Section~\ref{sec:3correlations} we examine how structural properties are correlated with median growth history, and we develop the split sample approach on this basis. In Section~\ref{sec:4structure-selection}, we compare samples split by different structural parameters. In Section~\ref{sec:5discussion}, we compare to previous results from the literature and demonstrate a simple application of split samples to the study of scaling relations. In Section~\ref{sec:6conclusions}, we review our results and conclude on the potential of the split sample approach.

\section{Simulations and structural analysis}
\label{sec:2simulations}

\subsection{IllustrisTNG}

The Next Generation Illustris Simulations (IllustrisTNG) is a suite of cosmological simulations that model the evolution of large sections of the universe from early fluctuations in the matter field to the formation of galaxies and galaxy clusters \citep{Nelson2018,Nelson2019,Springel2018,Pillepich2018b,Marinacci2018,Naiman2018}. The suite features both dark-matter-only and magnetohydrodynamical simulations with various volumes and resolutions. The dark-matter-only simulations are N-body simulations that model the gravitational interactions between representative dark matter particles. The magnetohydrodynamical simulations also include gas, stellar and black hole particles that interact via magnetic fields, radiative cooling, star formation and feedback, metal enrichment, and active galactic nucleus (AGN) feedback. All IllustrisTNG simulations assume a $\Lambda$CDM cosmology with cosmological parameters from the \citet{Planck2015} results: $\Omega_\Lambda=0.6911$, $\Omega_m=0.3089$, $\Omega_b=0.0486$, and $H_0=67.74\;\text{km/s/Mpc}$. This work makes use of the TNG300-1 simulation in particular. This is the highest-resolution realisation of IllustrisTNG's largest volume baryonic simulation. TGN300-1 is run in a cubic box with 205~comoving Mpc/$h$ (cMpc$/h$) side lengths and periodic boundary conditions. It contains $2500^3$ dark matter particles with masses of $5.9\times10^7\,\mathrm{M}_\odot$ and an equal number of baryonic particles.

In the publicly available IllustrisTNG data, subhalos represent galaxies and satellites, while halos represent groups and clusters of galaxies. Halos are identified using a friends-of-friends (FoF) algorithm with a linking length $b=0.2$ times the mean inter-particle distance. The FoF algorithm considers any two particles within the maximum linking length $b$ as ``friends''. It then groups all particles that are ``friends of friends'' into a single halo \citep{Press1982,More2011}. Subhalos are defined using the \textsc{subfind} algorithm, which identifies gravitationally bound overdensities of particles within halos \citep{Springel2001,Onions2012}. The most massive subhalo within each parent halo contains the most massive galaxy in the parent and most of the particles from the diffuse regions of the halo. The rest of the subhalos typically contain only local overdensities within the parent.

The IllustrisTNG simulations also feature full merger trees for all subhalos. These trees track groups of gravitationally bound particles as the simulation evolves. They then record how subhalos merge together between the simulation's 100 available snapshots to form progressively more massive systems. Due to hierarchical structure formation, any given subhalo has many progenitors that gradually merge together. IllustrisTNG provides a ``main progenitor branch'' of the merger tree for each subhalo that tracks the most massive progenitor of the most massive progenitor iteratively back through the simulation snapshots. We use this set of main progenitor branches to define the MAH, $M(z)$ \citep{vandenBosch2002}, that indicates the mass of the main progenitor at each redshift. All of the MAHs and related parameters presented in this work use the $M_{200c}$ masses of halos. $M_{200c}$ is defined as the mass inside a radius $R_{200c}$ around a cluster centre, within which the mean density is 200 times the critical density of the universe. The choice to include only the particles within $R_{200c}$ provides a more stable definition of halo mass over multiple simulation snapshots than the mass of the full FoF halo, which is in general larger. The full halo mass $M_\text{FoF}$ is a less stable mass measurement, experiencing sudden variations during halo mergers when two systems suddenly come within the grouping algorithm's linking length $b$. In contrast, $M_{200c}$ increases more gradually as an infalling group merges into the inner regions of a more massive cluster. Our method is also consistent with previous work by \citet{Amoura2023} that compared the dynamical states of galaxy clusters to their MAHs.

\subsection{Structural parameters}

This section provides definitions of the structural measurements of halos that are used as indicators of dynamical state and MAH throughout this work. We include definitions of intrinsic halo properties based on their 3D mass distributions, as well as projected and optical properties that are more observationally attainable. We also include brief explanations of why each structural parameter is expected to correlate with recent mass accretion and the dynamical state of a halo.

\subsubsection{Centre of mass offset}
\label{subsec:offset-explanation}

\begin{figure} 
    \centering
    \includegraphics[width=\linewidth]{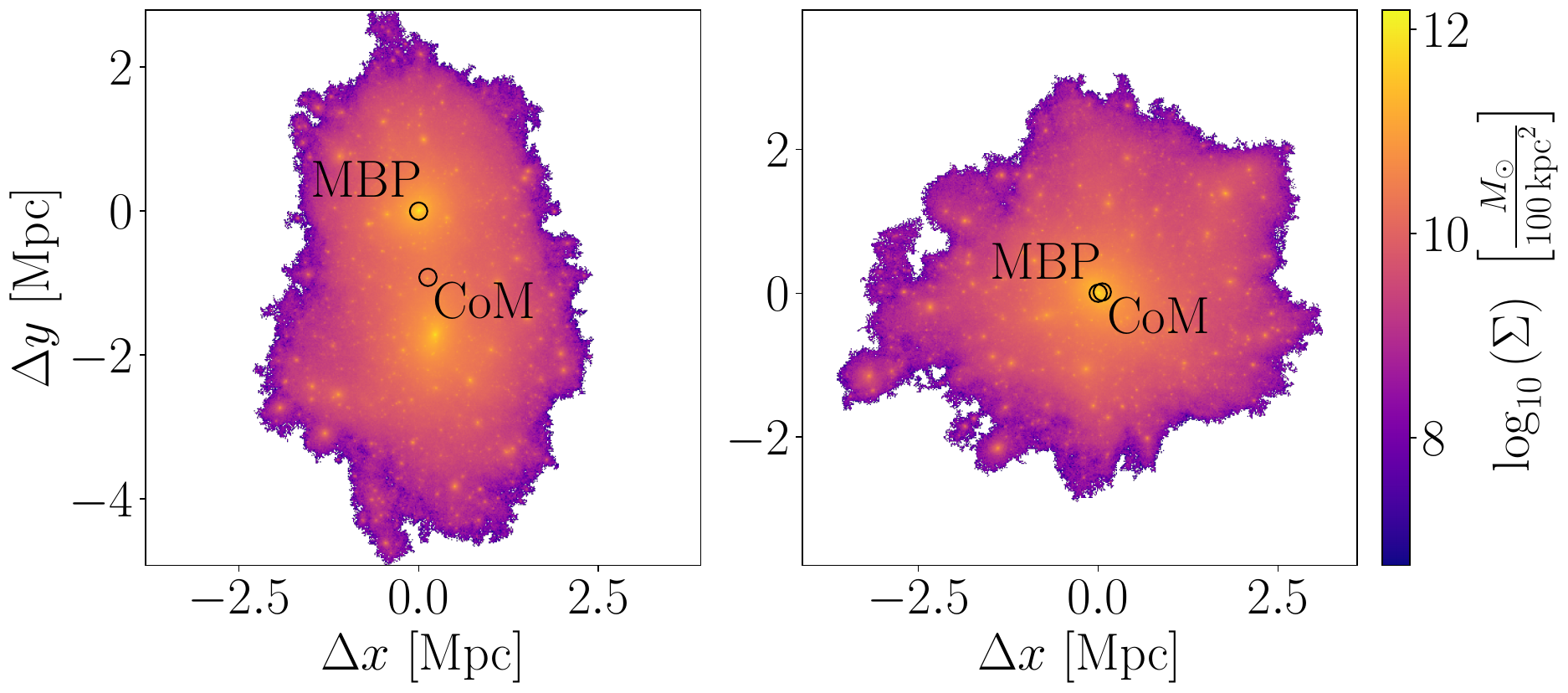}
    \caption{The surface mass densities $\Sigma$ and projected centre of mass offsets of two massive halos at $z=0$ in TNG300-1. The circles labelled ``MBP'' are centred on the most bound particle, while the circles labelled ``CoM'' are at the centre of mass of the entire halo mass distribution. The projected centre of mass offsets computed from the entire mass distributions for the left and right halos are 0.928 Mpc and 0.059 Mpc, respectively. The projected offsets computed from material within $R_{200c}$ for the two halos are 0.402 Mpc and 0.004 Mpc, respectively.}
    \label{fig:offset-visual}
\end{figure}

The centre of mass offset of a halo is defined as the distance between the centre of mass and the peak of the density profile \citep{Maccio2007,Power2012,Amoura2023}. This is similar to the centre of mass shift introduced by \citet{Crone1996} and used by \cite{Thomas1998}, which measures how the centre of mass changes as the density threshold which defines the halo's boundary is varied. In this work, we define the centre of mass offset relative to the location of the most gravitationally bound particle in a halo, as given in the simulation data. We assume this point to be the peak of the density profile, as in \citet{Maccio2007}. In practice, the most bound particle generally lies in the centre of the BCG. The variation of the centre of mass offset is illustrated in Figure~\ref{fig:offset-visual}, where the left panel shows a halo with a large offset, and the right panel shows a halo with a small offset.

We compute centres of mass from only the particles within the $R_{200c}$ of each halo's centre. The centre of mass offset within this radius is more stable during halo mergers than the offset computed from the full FoF halo, and it correlates well with the MAH of the halo. Three-dimensional centres of mass are computed using the full mass distribution of halos with 3D position information. Projected centres of mass are computed from the same mass distribution, except we discard position information along the chosen line of sight axis. Both the 3D and projected centre of mass offsets are normalised to $R_{200c}$ for each halo in order to compare similar relative scales across halos of different sizes.

We also consider projected stellar offsets, for which the centre of mass is calculated from only the star particles in a halo rather than from the full mass distribution. Rather than using the star particles within $R_{200c}$, our projected stellar offsets are computed from the star particles within a cylinder with a $0.5$ Mpc radius tangent to the line of sight and with the full depth of the simulation box along the line of sight. The $0.5$ Mpc radius of this cylinder is approximately the mean $R_{500c}$ of clusters in our sample (where $R_{500c}$ is defined analogously to $R_{200c}$, but with a higher density contrast). We chose this smaller radius since measurements of the dense inner regions of clusters are often more robust to projection effects. We chose to project star particles along the full $\sim\!300$~Mpc depth of the simulation box because this distance is comparable to photometric redshift constraints. The impact of more precise redshift constraints, such as those achievable with spectroscopic redshifts, is discussed briefly in Section~\ref{subsec:structure-offsets}.

Lastly, we consider the projected galaxy offset. Rather than using each particle in a halo, the projected galaxy offset computes the centre of mass using the centres of galaxies as the only tracer particles. Each galaxy within a cylinder, as defined for the stellar offset, is projected into each cluster and is weighted by its stellar mass. A galaxy's stellar mass is taken to be the total mass of the stellar particles within a 30 kpc radius of the galaxy's centre. This aperture radius was found to be effective for comparing simulated galaxies to observations, as it provides mass estimates comparable to observational measurements \citep{Schaye2015,Pillepich2018b,Engler2021}. Given the mass resolution of TNG300-1, we consider only galaxies with stellar masses $M_\star\geq10^9M_\odot$ when calculating the projected galaxy offset to ensure that each object is well defined. Neither the stellar nor the galaxy offsets are normalised to $R_{200c}$ since these observationally accessible measurements are designed not to require any knowledge of the full mass distribution.

We expect the centre of mass offset to be correlated with recent mass accretion through major mergers because material falling into a halo should pull the centre of mass away from the most bound particle and toward the infalling mass. Systems with large centre of mass offsets should thus be young and dynamically unrelaxed. An example of such a cluster is shown in the left panel of Figure~\ref{fig:offset-visual}. Conversely, systems with small centre of mass offsets should have a deficit of infalling material and symmetric mass distributions, which enables the centre of mass to settle toward the most bound particle in the centre of the halo. Systems with small centre of mass offsets are thus expected to be old and dynamically relaxed systems. An example of such a cluster is shown in the right panel of Figure~\ref{fig:offset-visual}.

\subsubsection{Axis ratio}

The axis ratio of a halo describes the relative lengths of its principal axes. The principal axes of a halo are equivalent to the preferred axes of rotation of an ellipsoid with the same moment of inertia \citep{Bett2012}. This structural measurement indicates the elongation of the system and is computed from its total mass distribution.

The primary axes of a halo are defined by \citet{Bett2012,Wong2012} as the eigenvectors of the mass quadrupole moment tensor:
\begin{equation}
    M_{ij}=\sum_{k=1}^{N}m_k\left(x_{k,i}-\langle x_i\rangle\right)\left(x_{k,j}-\langle x_j\rangle\right)\,,
    \label{eq:axis-ratio-tensor}
\end{equation}
where $x_{k,i}$ is the $i^{\textrm{th}}$ Cartesian component of the $k^{\textrm{th}}$ particle within $R_{200c}$ of the halo, and $\langle x_i\rangle$ is the $i^{\textrm{th}}$ Cartesian component of the centre of the halo. In IllustrisTNG, we take the centre of each halo to be its most gravitationally bound particle. The square roots of the eigenvalues of $M$ are then the relative lengths of these principal axes \citep{Bett2012,Wong2012,Anbajagane2022}. More explicitly:
\begin{equation}
    a=\sqrt{\lambda_1} \quad>\quad b=\sqrt{\lambda_2} \quad>\quad c=\sqrt{\lambda_3}\,\,,
    \label{eq:axes-eigenvalues}
\end{equation}
where $\lambda_1>\lambda_2>\lambda_3$ are the eigenvalues of the matrix $M$, and $a>b>c$ are the relative lengths of the axes. Note that the magnitudes of $a$, $b$ and $c$ are not taken here to be physically meaningful in themselves. Rather, their ratios are used as measurements of the elongation of a halo. The axis ratios are thus naturally given by:
\begin{equation}
    s=\frac{c}{a}=\sqrt{\frac{\lambda_3}{\lambda_1}}\quad,\quad q=\frac{b}{a}=\sqrt{\frac{\lambda_2}{\lambda_1}}\quad,\quad p=\frac{c}{b}=\sqrt{\frac{\lambda_3}{\lambda_2}}\,\,.
    \label{eq:ratios}
\end{equation}
We consider only the minor-to-major axis ratio $s=c/a$ in this work, as it was found to best correlate with a halo's MAH.

\begin{figure} 
    \centering
    \includegraphics[width=0.59\linewidth]{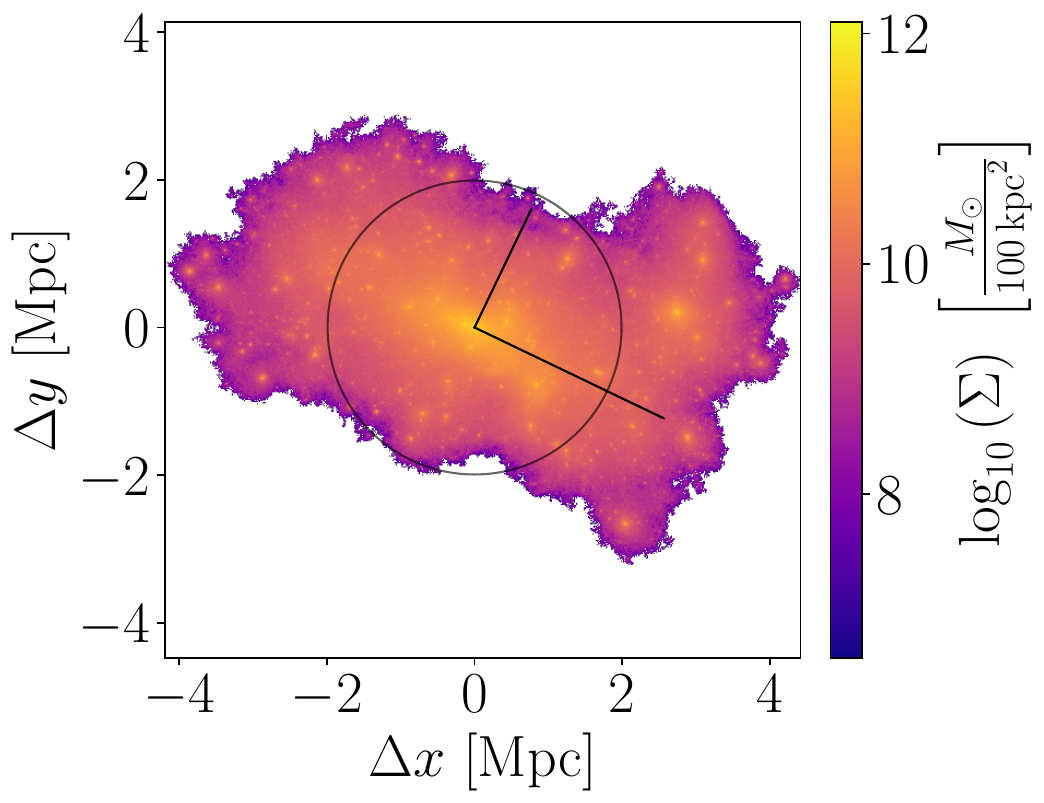}
    \caption{The surface mass density $\Sigma$ and principal axes of a massive projected halo at $z=0$ in TNG300-1. The black circle represents $R_{200c}$. The two black lines extend in the direction of the principal axes of the projected particle distribution. The relative lengths of the lines indicate the elongation of the mass distribution in each direction. The projected axis ratio in this example is 0.627.}
    \label{fig:axis-ratio-visual}
\end{figure}

We compute 3D axis ratios from the total mass distribution within $R_{200c}$ for each halo. To compute a projected axis ratio, we use the same mass distribution and construct the quadrupole moment tensor from Equation~\ref{eq:axis-ratio-tensor} while omitting the Cartesian coordinate along the line of sight axis. Due to the reduced dimension of the tensor, the projected distribution only produces two eigenvalues, which result in a single axis ratio for the projected halo. A visual example of the principal axes of a projected halo mass distribution and their relative lengths is shown in Figure~\ref{fig:axis-ratio-visual}. We note that the axes identified from the projected mass distribution of a halo are not the same as the projection of the axes computed from the 3D mass distribution. Rather, since the line of sight information is lost to projection prior to the calculation of the principal axes, the axes identified from the projected mass are generally different in both direction and length ratio from those in the 3D case.

We expect axis ratios to be an indicator of dynamical state since young and unrelaxed systems may be elongated along the axis of recent mergers. Meanwhile, older systems are more spherical due to their dynamical relaxation.

\subsubsection{Concentration}

The concentration $c$ of a halo is defined in terms of the Navarro-Frenk-White (NFW) density profile of the halo \citep{Navarro1996,Navarro1997}. In particular, it is defined as $c=r_\text{vir}/r_s$, where $r_\text{vir}$ is the virial radius of the halo, and $r_s$ is the scale radius as defined by \citet{Navarro1996,Navarro1997}:
\begin{equation}
    \rho(r)=\frac{\rho_0}{\frac{r}{r_s}\left(1+\frac{r}{r_s}\right)^2}\,.
    \label{eq:NFW}
\end{equation}
The virial radius indicates the boundary of the spherical region within which the kinetic energy of the mass reaches equilibrium under gravitational collapse \citep{Mo2010}. The scale radius is computed by fitting an NFW profile with free parameter $r_s$ to the mass distribution of a halo.

\begin{figure} 
    \centering
    \includegraphics[width=\linewidth]{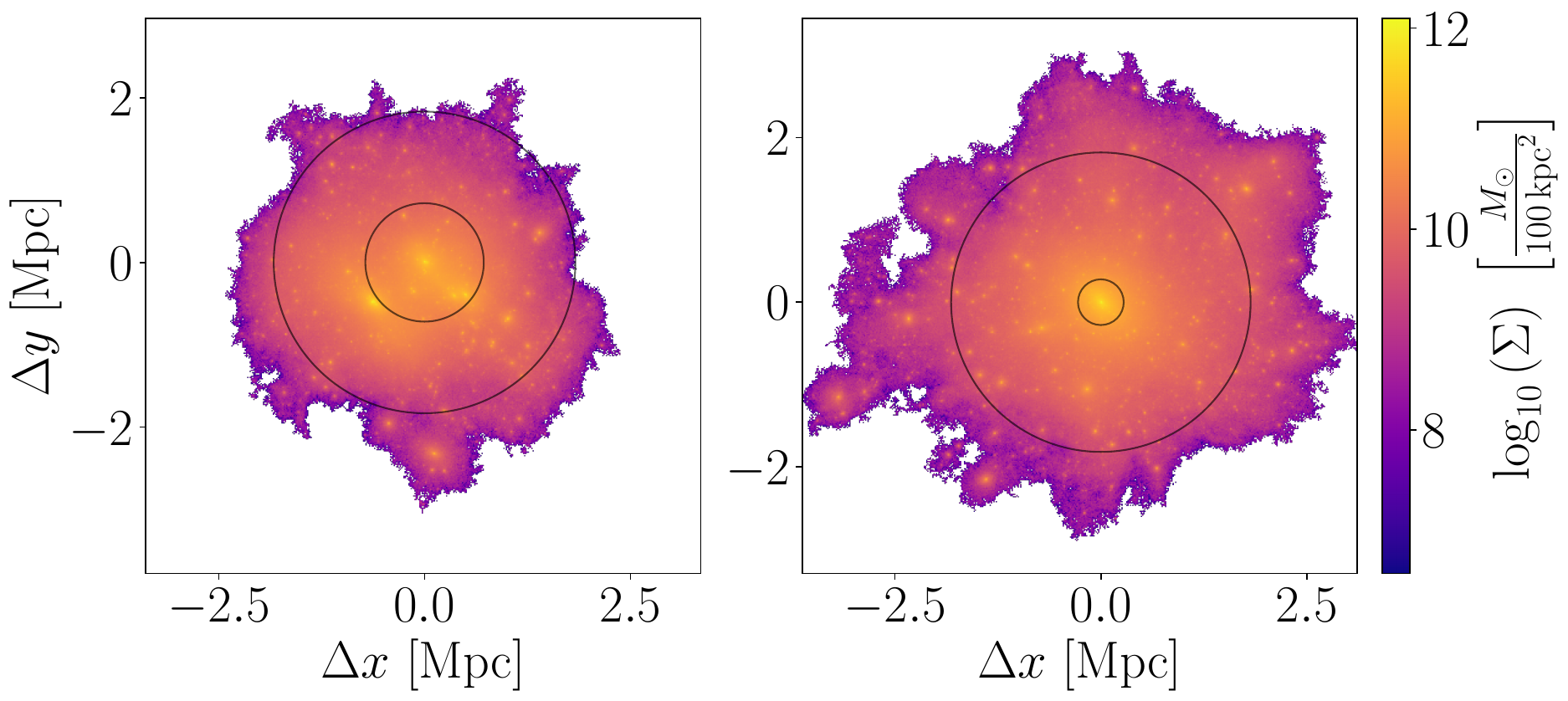}
    \caption{The surface mass densities $\Sigma$ and concentrations of two massive halos at $z=0$ in TNG300-1. The inner circles are the scale radii $r_s$ of the halos, while the outer circles are $R_{200c}$. The ratio of these radii gives the concentration $c_{200c}$. The left panel shows a low-concentration halo with $c_{200c}=2.5$, while the right panel is a higher-concentration halo with $c_{200c}=6.5$.}
    \label{fig:concentration-visual}
\end{figure}

A supplementary data catalogue produced by \citet{Anbajagane2022} is provided with the IllustrisTNG simulation data and contains concentration values for all identified halos. This concentration is defined as:
\begin{equation}
    c_{200c}=R_{200c}/r_s\,.
\end{equation}
In cluster halos, $R_{200c}$ roughly corresponds to the virial radius $r_\text{vir}$ and is generally much easier to identify \citep{Navarro1997}. We make use of this catalogue of concentrations for our halo sample. Figure~\ref{fig:concentration-visual} provides a visual example of $r_s$ and $R_{200c}$ for a low-concentration halo in the left panel and of a high-concentration halo in the right panel.

We expect the concentration of a halo to be correlated with its MAH since concentration is primarily determined by the mean density of the universe when the halo formed \citep{Navarro1997,Paranjape2015}. As the universe expands and its mean density decreases, we expect more recently formed halos to be less concentrated. As a result, halos with high concentrations are generally expected to be older and dynamically relaxed systems, whereas halos with low concentrations are expected to be younger and dynamically unrelaxed systems. We note, however, that major mergers can produce significant short-term variations in concentration \citep{Wang2020}, adding scatter to the general trend with age/relaxation.

\subsubsection{Asymmetry}

\begin{figure} 
    \centering
    \includegraphics[width=0.59\linewidth]{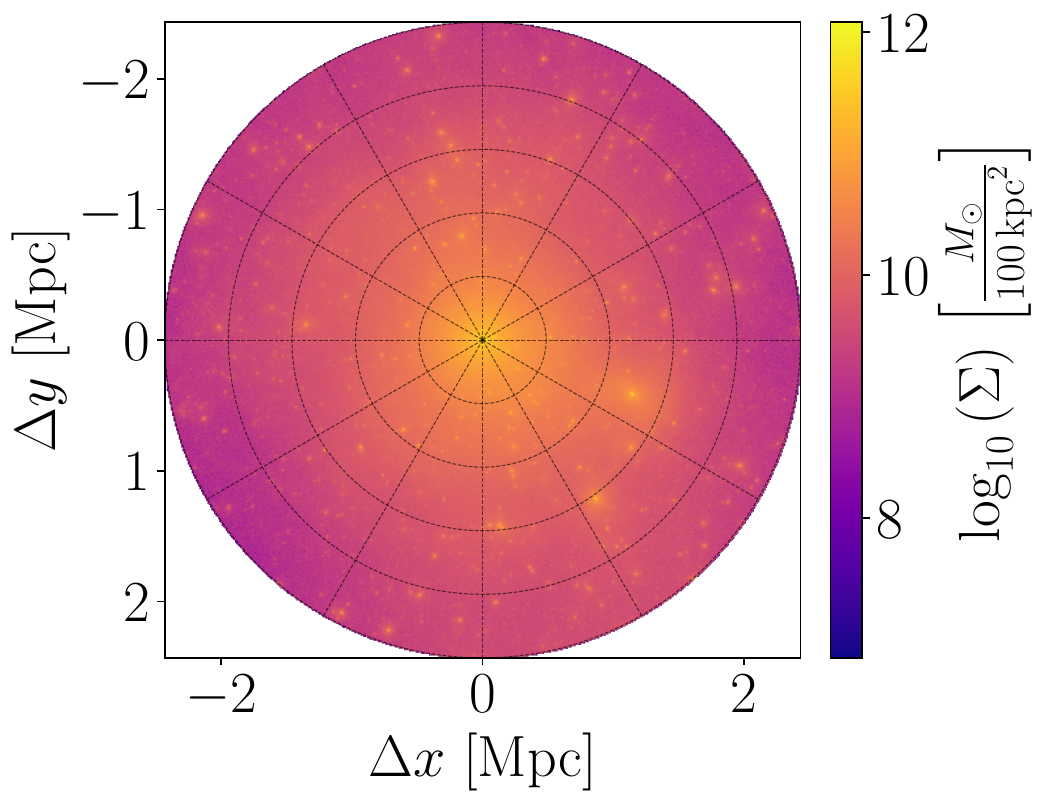}
    \caption{The surface mass density $\Sigma$ around a massive halo at $z=0$ in TNG300-1, sorted into evenly spaced angular and radial bins. The total surface mass density is shown in colour scale. The dashed lines show the divisions between bins used to compute halo asymmetry. For clarity, only a quarter of the total bins used for the calculation are shown. The total mass asymmetry of this halo is $A_m=0.252$.}
    \label{fig:asymmetry-visual}
\end{figure}

The asymmetry of a halo is a measurement of the asymmetric fluctuations in the projected mass or brightness distribution around the BCG. Asymmetry is quantified by the symmetry number $A$, which is defined by \citet{Conselice1997} as:
\begin{equation}
    A^2=\frac{\sum1/2\left(I_0-I_{180}\right)^2}{\sum I_0^2}\,,
    \label{eq:old-asymmetry}
\end{equation}
where $I_0$ is the image of a cluster and $I_{180}$ is the image rotated by $180^\circ$. This expression quantifies how much signal from the image of a halo is not equally represented on the other side of the halo centre. Motivated by this definition of asymmetry $A$, we adopt the following measurement for mass asymmetry from a simulated particle distribution:
\begin{equation}
    A_m^2=\frac{\sum_i1/2\left(m_i-m_{i,180}\right)^2}{\sum_i m_i^2}\,,
    \label{eqn:halo-asymmetry}
\end{equation}
where $m_i$ is the mass in a 2D bin $i$ of the projected particle distribution, and $m_{i,180}$ is the mass in the bin $180^\circ$ around the halo at the same radius. Figure~\ref{fig:asymmetry-visual} provides a visual example of the division of a halo's projected mass distribution into radial and angular bins used for the calculation of asymmetry.

In this work, we compute asymmetry by dividing the projected mass distribution into equally spaced radial and angular bins around the most bound particle of a halo, then applying Equation~\ref{eqn:halo-asymmetry} to the mass in the bins. As noted above, the most bound particle generally lies in the centre of the BCG, which provides a clear shared reference point between simulations and observational data. For the total mass asymmetry, we consider all particles within $R_{200c}$ tangent to the line of sight and $\pm2$ cMpc along the line of sight for each halo. This depth includes all of the particles in a halo while limiting projection effects so that the total mass asymmetry can act as a measurement of intrinsic halo properties. For the stellar mass asymmetry, we consider only star particles within a 0.5 cMpc radius tangent to the line of sight and within the full simulation box length along the line of sight. This depth provides an analogue to observations with poor constraints on line of sight distances, as with the stellar offset and galaxy offset discussed in Section~\ref{subsec:offset-explanation}. The impact of better line of sight constraints is discussed briefly in Section~\ref{subsec:structure-asymmetry}. In each case, we split the halo surface mass distributions into 10 radial bins and 24 angular bins. We tested the robustness of the asymmetry measurements to resolution changes by doubling and halving the number of radial and angular bins around these fiducial values. The total mass asymmetry is very robust to changes in the bin numbers. The stellar mass asymmetry distribution skews more toward values of 1 for high bin numbers, but the MAHs of subsamples presented in Section~\ref{subsec:structure-asymmetry} are fairly robust to bin number changes.

Halos with massive galaxies that are not symmetrically distributed around the centre of the cluster will have high asymmetry numbers $A$. We expect asymmetry to be correlated with recent mass accretion through major mergers since infalling material to a halo should produce a large mass component that is offset to one side of the BCG and not equally represented on the other side of the halo. Systems with large asymmetries should thus be young and dynamically unrelaxed. Conversely, systems with small asymmetries should have little infalling material, which enables the mass distribution to settle symmetrically around the halo centre. Systems with small asymmetries are thus expected to be old and dynamically relaxed systems.

\subsubsection{Magnitude gap \& stellar mass ratio}
\label{subsec:mag-gap-explanation}

The magnitude gap of a galaxy cluster is defined as the difference between the magnitudes of two bright galaxies in the cluster in a given spectral band \citep{Jones2003,Dariush2010,More2012}. Magnitude gaps are often taken between the two most luminous galaxies within some radius of the BCG \citep{Jones2003}. They may also be computed between the brightest galaxy and the third or fourth brightest galaxy in a cluster, which may correlate well with the hierarchical growth of the BCG \citep{Dariush2010,Golden-Marx2018,Golden-Marx2025}. A similar property of galaxy clusters is the stellar mass ratio, which is the logarithm of the ratio between the stellar masses of two massive galaxies in a cluster \citep{Golden-Marx2018,ahad2025arXiv}.

In this work, we compute one magnitude gap and stellar mass ratio for each galaxy cluster. We consider the r-band magnitude gap between the two brightest galaxies in a cluster, which we denote $r_{12}$. We also consider the stellar mass ratio between the first and fourth most massive galaxies in a cluster, which we denote $m_{14}$. The r-band magnitudes and stellar masses of galaxies are computed from the star particles within a 30 kpc radius of the galaxy centre, as are the galaxy offsets described in Section~\ref{subsec:offset-explanation}. The 3D magnitude gaps and stellar mass ratios consider only galaxies within $R_{500c}$ around the halo centre in 3D. The projected magnitude gaps and stellar mass ratios consider galaxies that are projected into $R_{500c}$ across the full simulation box along the line of sight. We chose to only include galaxies within this radius so as to easily compare the simulated cluster sample to optically selected cluster samples that identify member galaxies within $R_{500c}$, such as those presented in, e.g., \cite{WenHan2024,ahad2025arXiv}.

We expect the magnitude gap of a cluster to be correlated with its recent MAH since major mergers insert massive galaxies into the cluster, thus decreasing its magnitude gap. As a result, clusters with small magnitude gaps should be young and dynamically unrelaxed systems. Conversely, the central galaxy of a cluster that formed earlier will have cannibalised other massive galaxies, reducing the magnitude gap \citep{Farahi2020}. Consequently, systems with large magnitude gaps should be old and dynamically relaxed.

\subsection{Mass accretion history parameters}

The primary MAH parameters used to characterise the growth histories of halos in this work are $z_{50}$, $z_{75}$ and $z_{90}$. These parameters represent the redshifts $z_x$ at which the main progenitor branch of each halo crossed a threshold of $x$\% of its final mass. More precisely, the redshift $z_x$ is taken to be the redshift of the first simulation snapshot after a halo crossed the mass threshold $x$. In the case of non-monotonic MAHs, $z_x$ is the lowest-redshift crossing of the mass threshold $x$. Among the mass history parameters $z_{50}$, $z_{75}$ and $z_{90}$, we found that $z_{75}$ presents the best correlations with most structural parameters described in this section. As a result, MAHs in this work are generally parametrised using $z_{75}$.

Another common measure of MAH is the accretion rate $\Gamma_\mathrm{dyn}$ within the last dynamical time $t_\mathrm{dyn}$. This accretion rate is defined by \citet{Diemer2017} as:
\begin{equation}
    \Gamma_\mathrm{dyn}(t)=\frac{\log[M_{200m}(t)]-\log[M_{200m}(t-t_\mathrm{dyn})]}{\log[a(t)]-\log[a(t-t_\mathrm{dyn})]}\,,
    \label{eq:gamma}
\end{equation}
where $a$ is the scale factor and $M_{200m}$ is the mass of the halo within a radius where the mean density is 200 times the mean density of the universe. \citet{Diemer2017} defines $t_\mathrm{dyn}$ as the time for a particle to cross a halo with mean density 200 times that of the universe. This can be expressed as:
\begin{equation}
    t_\mathrm{dyn}=\frac{1}{5\sqrt{\Omega_m}}\cdot\frac{1}{H(z)}\,,
    \label{eq:dynamical-time}
\end{equation}
where $\Omega_m$ is the matter fraction and $H(z)$ is the Hubble parameter. At redshift $z=0$, the last dynamical time is $t_\mathrm{dyn}=5.19\,\mathrm{Gyr}$. Motivated by \cite{Xhakaj2022}, we used the parameter $\Gamma_\mathrm{dyn}$ in previous work \citep{ahad2025arXiv} to compare the mass accretion rates (MAR) of structurally selected cluster samples. In this work, we find that $\Gamma_\mathrm{dyn}$ is more weakly correlated with most structural parameters than $z_{75}$. Consequently, we only present its correlations with other parameters in Section~\ref{subsec:correlations}.

\section{Correlations between structure and growth history} 
\label{sec:3correlations}

In this section, we explore simple correlations between the centre of mass offsets of clusters and their MAHs to see how halos may be grouped into dynamically relaxed and unrelaxed subsamples. In order to increase the sample size of halos and reduce shot noise in the parameter distributions, we consider all halos with masses $1\times10^{13}\leq M_{200c}/\mathrm{M}_\odot\leq5\times10^{13}$ at redshift $z=0$. This mass range produces 3031 halos with measurements of $z_{75}$. While these are group-mass halos rather than large clusters, we generalise the results to more massive systems in Section~\ref{sec:4structure-selection}.

\begin{figure*} 
    \centering
    \includegraphics[width=0.78\linewidth]{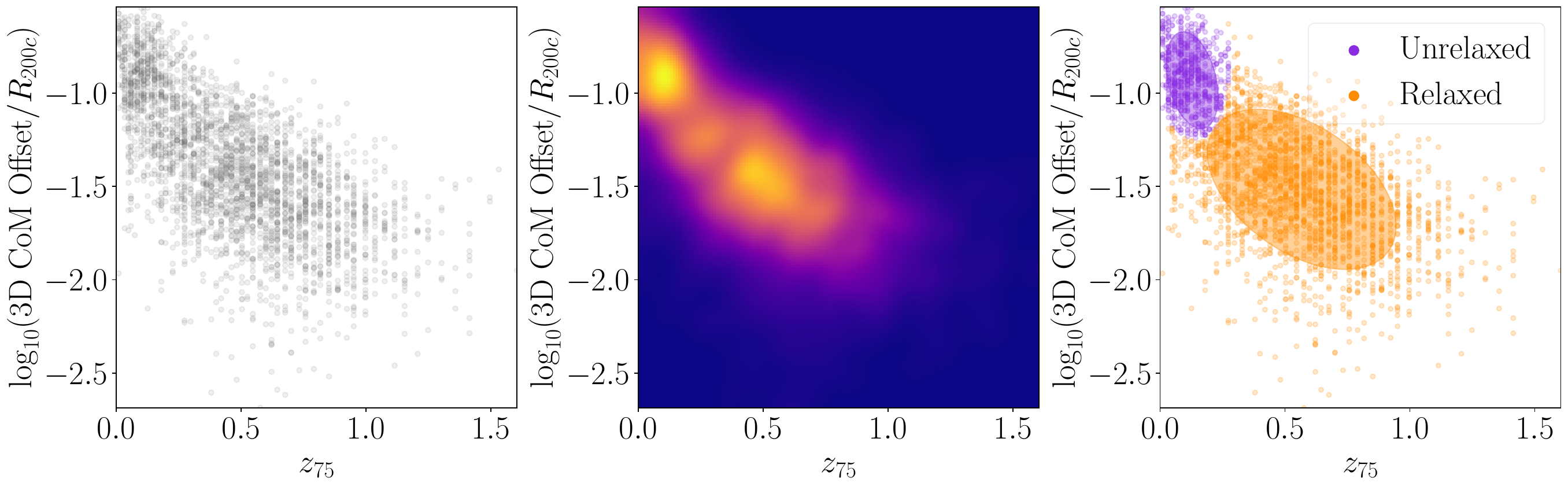}
    \caption{The left panel shows the 3D centre of mass offset versus the MAH parameter $z_{75}$ for group-mass halos. The middle panel is a heat map of the distribution with a Gaussian smoothing kernel applied. The right panel splits the distribution using a two-component Gaussian mixture model. The shaded regions in the right panel are the $1\sigma$ boundaries for the Gaussian components, and the colours of the dots represent the group membership of the halos.}
    \label{fig:com-offset-groups}
\end{figure*}

As an initial indication of how well 3D centre of mass offset correlates with recent MAH, we compare it to the MAH parameter $z_{75}$ in Figure~\ref{fig:com-offset-groups}. It is clear from the distribution of offsets and $z_{75}$ in the left panel that the two parameters are well correlated despite the scatter in the relationship. We see that halos with large offsets are almost exclusively systems with low $z_{75}$, indicating that they have likely experienced recent mergers. However, it is notable that not all systems with low $z_{75}$ have large offsets. This is likely due to the oscillation of the centre of mass offset during mergers. An infalling group repeatedly crosses through the centre of a halo and out the other side as the system gradually relaxes. This causes the centre of mass offset to oscillate with the phase of the infall. As a result, while systems with large offsets are generally experiencing mergers, systems experiencing mergers do not necessarily have large offsets. As reflected in the distribution in the left panel of Figure~\ref{fig:com-offset-groups}, this suggests that splitting a halo sample based on 3D centre of mass offset would produce a sample of young, dynamically unrelaxed clusters with low contamination, but also with low completeness. In order to produce a sample of old clusters with low contamination, we would have to take only halos with very small centre of mass offsets, excluding those with middling offsets which may be currently experiencing mergers.

To better understand the distribution of 3D centre of mass offsets against $z_{75}$, we plot a heat map of the values in the middle panel of Figure~\ref{fig:com-offset-groups}. The heat map was constructed by applying a Gaussian smoothing kernel to the distribution in the left panel of the figure. We see from the heat map that there are two distinct peaks in the distribution representing two populations of halos. There is a population of halos with recent mass accretion and large centre of mass offsets, as well as a population without recent mass accretion and with small offsets. There is a fainter peak between these two distributions as well, though it is less pronounced. Characterising the two prominent populations of halos in the middle panel of Figure~\ref{fig:com-offset-groups} grants insight into the properties and MAHs of halos with different physical structures.

To characterise the two main populations of halos identifiable by eye in the middle panel of Figure~\ref{fig:com-offset-groups}, we applied a Gaussian mixture model to the distribution. In this context, a Gaussian mixture model is a fit of several Gaussian distributions to some data set where we assume that each point is randomly drawn from the mixture of those Gaussians. In the right panel of Figure~\ref{fig:com-offset-groups}, a Gaussian mixture model is applied to the offset and $z_{75}$ distribution. We see that the model does indeed pick out the two groups that are identified by eye. Thus, by fitting a two-component Gaussian mixture model to the cluster data, we split the halos into two groups --- relaxed and unrelaxed --- based on their structures and MAHs. We explored adding more Gaussian components to fit the distribution and made use of the Akaike and Bayesian information criteria to judge the goodness of fit against the number of new parameters. In doing so, we found that a 2-component Gaussian mixture model is strongly favoured over a single-component model, but we did not find sufficient evidence to support more than two populations of halos. These results are discussed further in Appendix~\ref{appx:gaussian-groups}.

\begin{figure} 
    \centering
    \includegraphics[width=\linewidth]{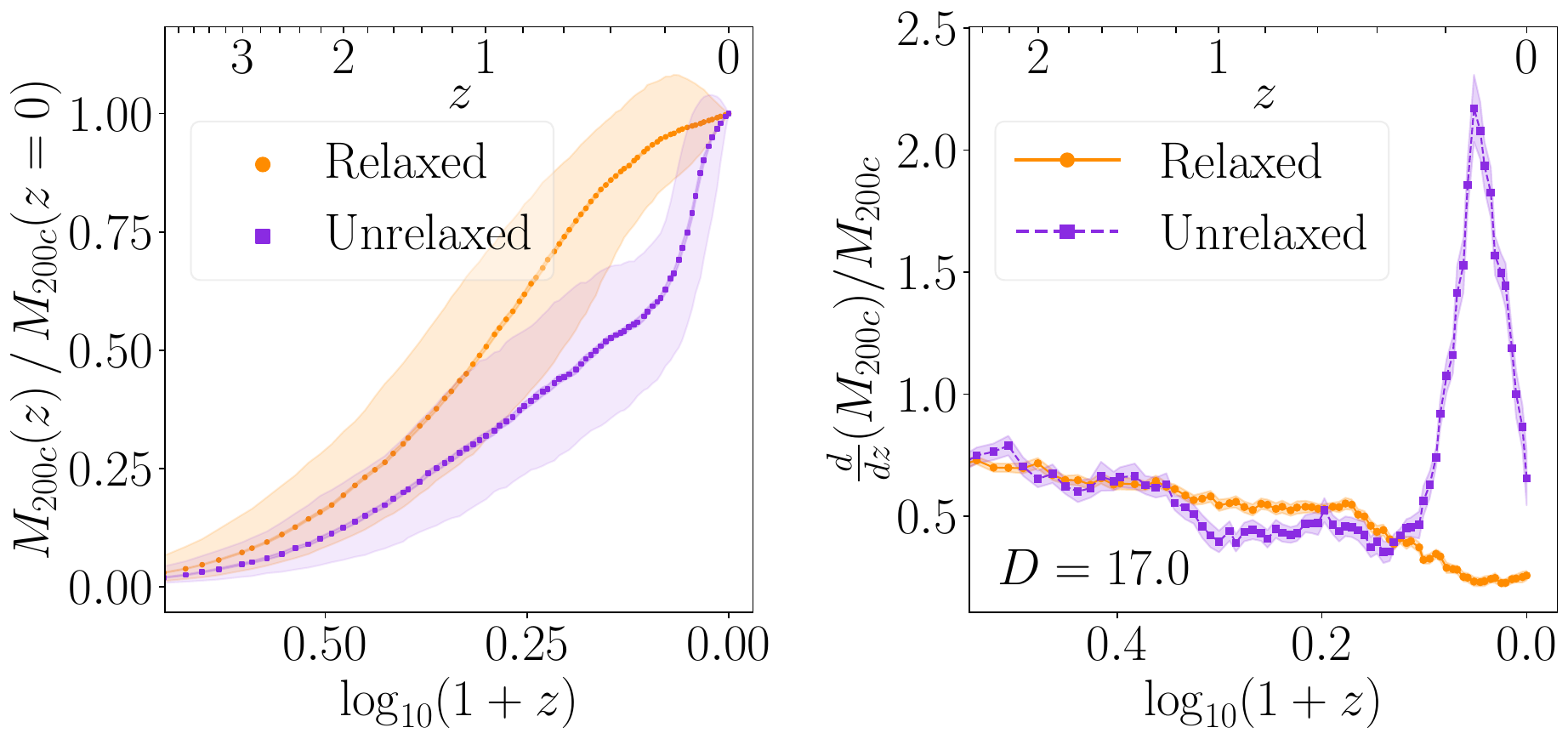}
    \caption{The median MAHs and MARs of the relaxed and unrelaxed halos as defined in Figure~\ref{fig:com-offset-groups}. The left panel shows the MAHs normalised relative to their masses at $z=0$. The points represent the median masses of halos in each group at the available snapshots. The thin shaded regions around the points are the uncertainties on the medians. The wide shaded regions around the points are the $1\sigma$ variations around the medians. The right panel shows the MARs, which are the derivatives of the MAHs with respect to redshift, divided by the median mass at each redshift. The points represent the median accretion rates of halos in each group at the available snapshots. The shaded regions around the points are the uncertainties on the medians. The value of $D$ in the right panel is the maximum normalised difference between the mass median accretion rates, as defined in Equation~\ref{eq:maximum-difference}.}
    \label{fig:com-offset-history}
\end{figure}

To more comprehensively compare the growth histories of the halos split into groups based on 3D centre of mass offset and $z_{75}$, we plot the median normalised MAHs of the two groups as a function of redshift in the left panel of Figure~\ref{fig:com-offset-history}. We see that the median MAHs of the relaxed and unrelaxed groups as defined by the Gaussian mixture model differ significantly at low redshift. The thin shaded regions that represent the uncertainty on the median accretion history are extremely well separated. The large shaded regions representing the $1\sigma$ scatter in the MAHs are also well separated at low redshift, indicating a statistically significant difference in the MAHs of the two subsamples. The relaxed sample experiences a gradual mass increase with very little recent accretion, whereas the unrelaxed group experiences a slow mass increase at high redshift, followed by sudden accretion at low redshift. This behaviour for the unrelaxed systems is consistent with recent major mergers. In the right panel of Figure~\ref{fig:com-offset-history}, we plot the median MAR of the relaxed and unrelaxed groups. This is the derivative of the MAH with respect to redshift $z$, normalised by the halo mass at each redshift. We see that both subsamples of clusters have comparable accretion rates at high redshift, but that the unrelaxed group experiences a large and sudden spike in MAR. This is again consistent with recent major mergers.

To quantify the difference in the MARs between the relaxed and unrelaxed subsamples in the right panel of Figure~\ref{fig:com-offset-history}, we introduce the maximum normalised difference:
\begin{equation}
    D=\max_z\left(\frac{\left|f_\mathrm{rel}(z)-f_\mathrm{unrel}(z)\right|}{\sqrt{\sigma_\mathrm{rel}^2(z)+\sigma_\mathrm{unrel}^2(z)}}\right)\,,
    \label{eq:maximum-difference}
\end{equation}
where $\sigma_\mathrm{rel}$ and $\sigma_\mathrm{unrel}$ are the uncertainties on the median MARs of the relaxed and unrelaxed groups, respectively, and where:
\begin{equation}
    f(z)=\frac{1}{M_{200c}(z)}\frac{dM_{200c}}{dz}
\end{equation}
is the median MAR of a subsample. As reported in the bottom left of the right panel in Figure~\ref{fig:com-offset-history}, we see that the maximum difference between the median MARs is $D=17$ times the uncertainty in the difference. This large difference highlights the significance of the disparity between the formation histories of the cluster samples. The maximum normalised difference metric is used throughout Section~\ref{sec:4structure-selection} to quantify the difference in growth histories between structurally selected samples.

\begin{figure} 
    \centering
    \includegraphics[width=\linewidth]{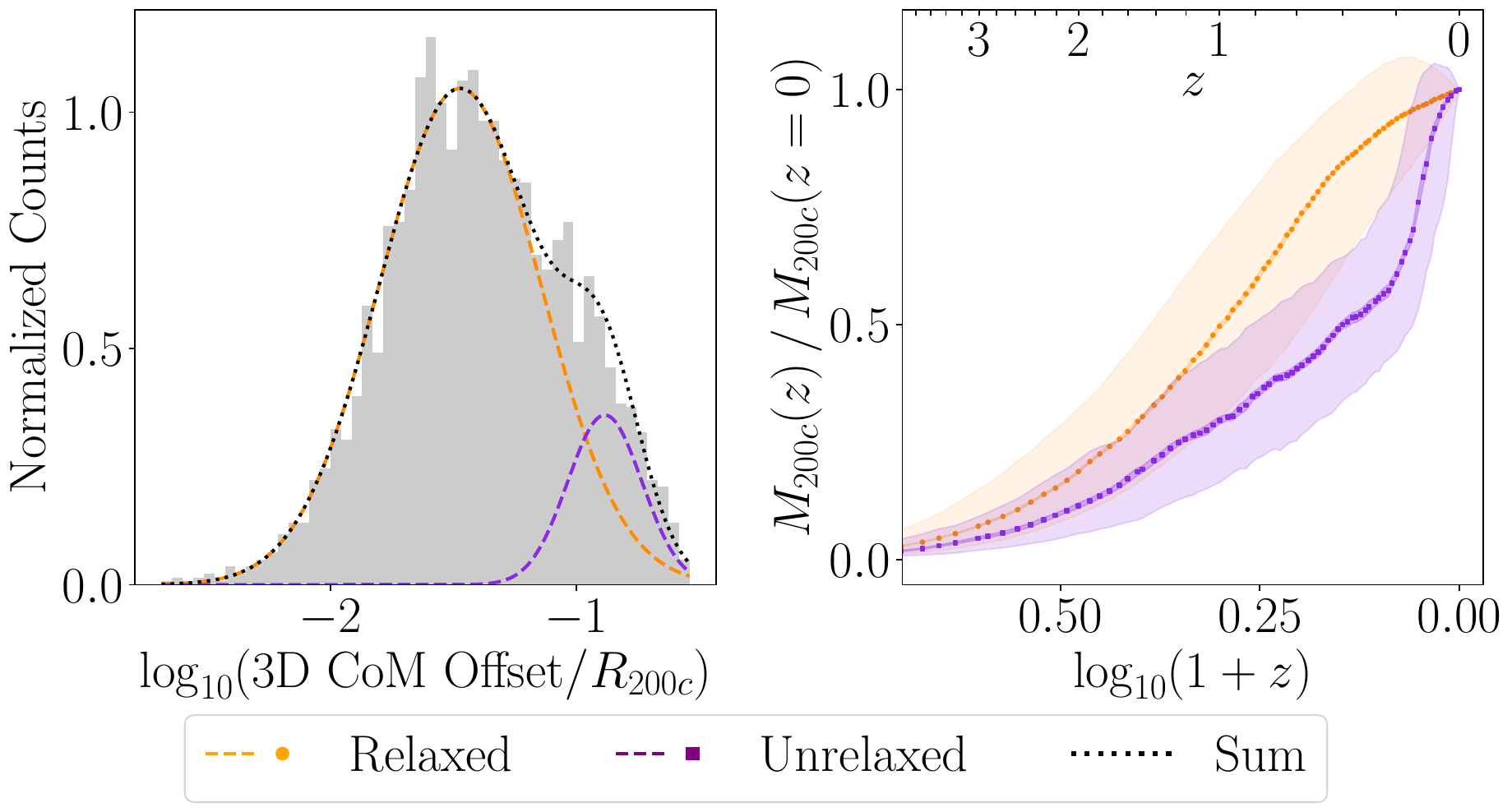}
    \caption{The left panel shows the 3D centre of mass offsets of group-mass halos fitted with a Gaussian mixture model that splits the distribution into relaxed and unrelaxed subsamples. The right panel shows the corresponding MAHs of the subsamples. MAHs are formatted as in Figure~\ref{fig:com-offset-history}.}
    \label{fig:com-offset-1D-gaussian}
\end{figure}

While we have found that halos can be effectively classified as dynamically relaxed or unrelaxed based on 3D centre of mass offsets and the $z_{75}$ parameter, this classification requires knowledge of the MAH, which is not directly observable. In Figure~\ref{fig:com-offset-1D-gaussian}, we use a one-dimensional Gaussian mixture model to classify halos as relaxed and unrelaxed based only on the 3D centre of mass offset. In contrast to Figure~\ref{fig:com-offset-history}, this includes no explicit reference to the MAH. We see in the left panel of Figure~\ref{fig:com-offset-1D-gaussian} that the Gaussian mixture model picks out a broad population of halos with small offsets and a smaller population of halos with large offsets. These are the dynamically relaxed and unrelaxed groups, respectively. Though there is significant overlap between the two Gaussian groups, we choose to split the halos based on the probability of membership in each group as determined by the Gaussian mixture model. The transition point between the groups is reflected in the left panel of Figure~\ref{fig:com-offset-1D-gaussian} as the intersection of the two Gaussian curves. In the right panel of Figure~\ref{fig:com-offset-1D-gaussian}, we plot the median MAHs of these two groups. The difference between the median MAHs is not as large as in Figure~\ref{fig:com-offset-history}, where we fit the Gaussian mixture model to $z_{75}$, but we still see a significant difference between the histories of the two groups. The median mass histories are very different at low redshift compared to their uncertainties, and the $1\sigma$ scatters in the accretion histories of the two groups do not quite overlap at the point of maximum separation.

In practice, 3D centre of mass offsets cannot be measured for real systems due to poor constraints on line of sight distances. We can perform the same exercise as in Figure~\ref{fig:com-offset-1D-gaussian} for projected centre of mass offsets, but the Gaussian mixture model fails to pick out two meaningful groups. Since a mixture model based on realistic observational measurements cannot reliably classify clusters as relaxed or unrelaxed, we elect to compare cluster subsamples at the two extreme ends of the distribution for all measured parameters. The mixture model presented in Figure~\ref{fig:com-offset-groups} identified $\sim$20$\%$ of halos as unrelaxed. On this basis, in Section~\ref{sec:4structure-selection} we will take the top and bottom 20\% of each parameter distribution to be our relaxed and unrelaxed samples, respectively.

\section{Structurally selected samples} 
\label{sec:4structure-selection}

In this section, we examine the correlation between a number of different structural measurements and the MAHs of halos. This enables us to determine what observable properties of clusters are effective indicators of the formation histories and dynamical states of halos. We include measurements of every halo in TNG300-1 with mass $5\times10^{13}\leq M_{200c}/\mathrm{M}_\odot\leq1\times10^{14}$ at redshift $z=0$. This selection includes 420 halos, which allows us to explore a sufficiently large population of systems within a reasonably narrow mass range. Each relaxed/unrelaxed split then extracts from the total distribution of structural parameter values the lowest/highest 20\% of clusters, producing two subsamples of 84 clusters each.

\subsection{Centre of mass offset}
\label{subsec:structure-offsets}

\begin{figure*} 
    \centering
    \includegraphics[width=\linewidth]{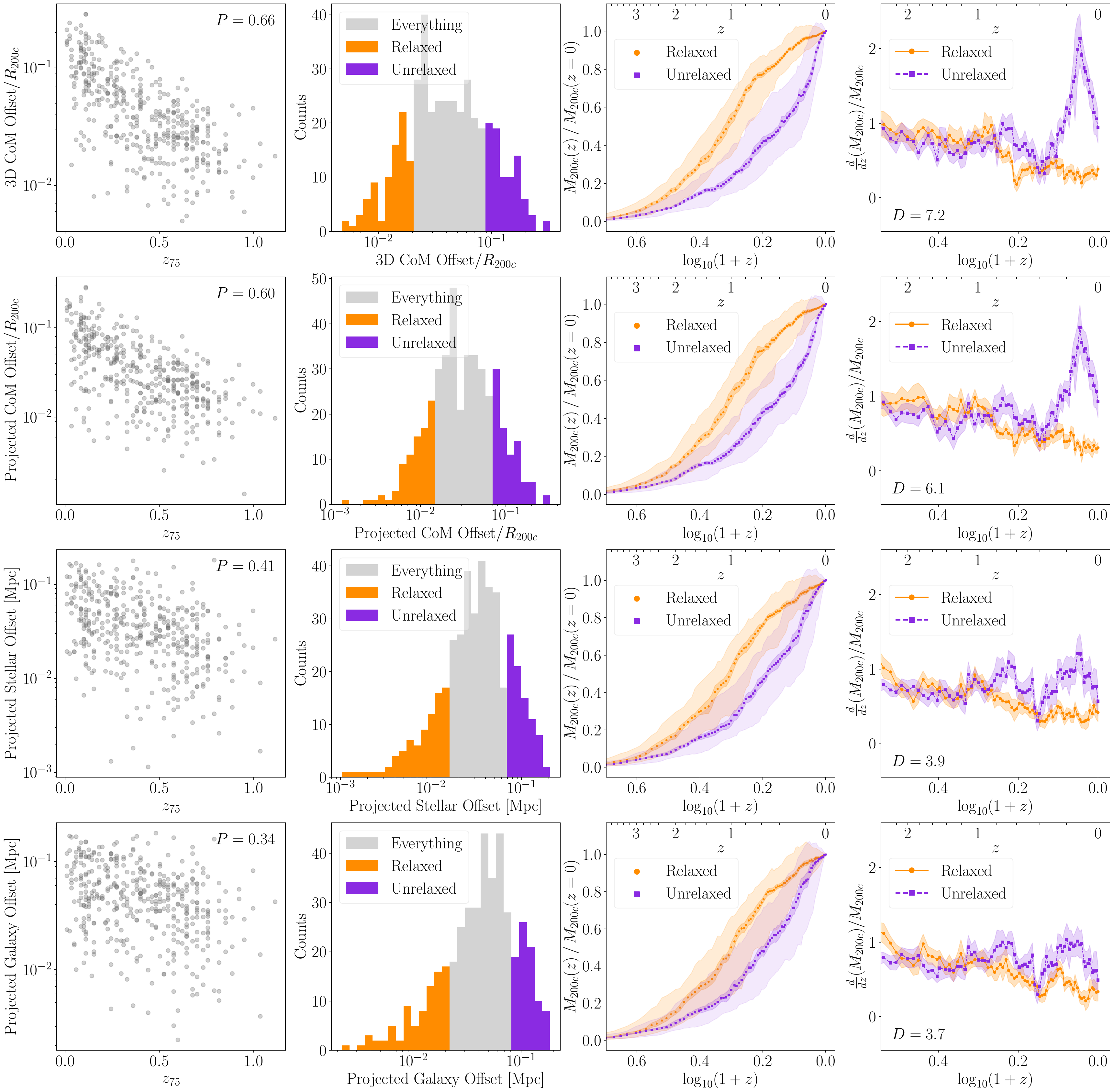}
    \caption{The leftmost column of panels shows the distributions of 3D, projected, stellar and galaxy centre of mass offsets against the MAH parameter $z_{75}$. The values of $P$ listed in the leftmost column are the absolute Pearson correlation coefficients between the plotted quantities, as further discussed in Section~\ref{subsec:correlations}. The second column shows the offset distributions split into relaxed and unrelaxed subsamples. The third column shows the median MAHs of the subsamples, while the rightmost column shows the relative accretion rates. The values of $D$ listed in the rightmost column are the greatest difference between the two accretion rates in units of its uncertainty, as defined in Equation~\ref{eq:maximum-difference}. Mass accretion histories and rates are formatted as in Figure~\ref{fig:com-offset-history}.}
    \label{fig:offsets-panels}
\end{figure*}

The 3D centre of mass offset is the offset measurement that is most intrinsically tied to the dynamical state of a halo. In the top row of panels in Figure~\ref{fig:offsets-panels}, we explore the 3D offsets of our halos and their relation to MAH. The leftmost panel shows that the centre of mass offset is strongly correlated with $z_{75}$ with fairly minimal scatter. The third panel shows that the median MAHs of the relaxed and unrelaxed subsamples split by 3D offset are quite different at low redshift. In the rightmost panel, we compare the MARs of the subsamples as a function of redshift. As expected, the unrelaxed population experiences rapid mass accretion consistent with recent major mergers. As indicated in the rightmost panel, the maximum difference between the median accretion rates is $D=7.2$ times the uncertainty in the difference. This difference is much less than that shown in Section~\ref{sec:3correlations}, primarily due to the greater uncertainties on the medians due to the smaller sample size of halos. Despite the greater uncertainties, the relaxed and unrelaxed groups of halos still have significantly different median MARs at low redshift. 

The projected centre of mass offset is a slightly more practical measurement since it does not rely on line of sight distances, which are poorly constrained in observed systems. It is presented in the second row of Figure~\ref{fig:offsets-panels}. In the leftmost panel, we see that the projected offset is slightly less well correlated with $z_{75}$ than the 3D offset, but still with reasonable scatter. In the rightmost panel, we see that the MARs of the two subsamples are still quite different at low redshift despite the loss of line of sight information. This demonstrates that centre of mass offset measurements are sufficiently robust to projection so as to preserve mass history information.

While both the 3D and projected centre of mass offsets offer strong discrimination between dynamically relaxed and unrelaxed systems, the detailed mass distributions of individual galaxy clusters are not easily observable. As a result, neither of these offsets is an ideal indicator of halo mass history. The projected stellar offset is much more observationally achievable since it relies only on the distribution of stellar mass within a large projected volume. This measurement is presented in the third row of Figure~\ref{fig:offsets-panels}. In the leftmost panel, while there is still a correlation between the offset and $z_{75}$, it is weaker and shows more scatter than seen previously. In the third panel, the MAHs of the two subsamples are still somewhat well separated. However, the $1\sigma$ scatters around the medians overlap at all redshifts. In the rightmost panel, the difference between the median MARs is statistically significant but much noisier and less pronounced than in previous cases. This is reflected in the maximum normalised difference, which drops to $D=3.9$. The loss of the difference between the median MARs of the two subsamples is due to the inclusion of more early-forming halos in the unrelaxed subsample. As can be seen in the leftmost panel of the figure, an appreciable fraction of halos with large stellar offsets also have high $z_{75}$ values. This results in contamination of the unrelaxed subsample, which suppresses the jump in the median MAR. If we impose strong line of sight constraints on the measurement and include only stars within the cluster, the difference between the median MARs increases but remains significantly less pronounced than in the projected offset case. This result establishes a pattern throughout this section: measurements that are more intrinsic to halo structure --- such as 3D offset --- trace the MAH with little contamination, while more observationally accessible measurements --- such as stellar offset --- introduce scatter and contamination that reduce the signal in the MAH.

Lastly, we consider the projected galaxy offset as a very observationally accessible measurement of dynamical state, as it relies only on the positions and stellar masses of the projected galaxies in a cluster. This is presented in the bottom row of Figure~\ref{fig:offsets-panels}. In the leftmost panel, the correlation between the offset and $z_{75}$ is weaker and more scattered than all other offset measurements. In the third panel, while the median MARs are still significantly different, there is extensive overlap between their $1\sigma$ scatters. In the rightmost panel, the unrelaxed group has largely lost the characteristic jump in mass accretion rate at low redshift that indicates recent major mergers. As with the stellar offset, the difference in the MARs can be increased by reducing line of sight projection. However, the signal always remains weaker than that from all other offset measurements. This further confirms that observational measurements of halos which trace the total mass distribution less directly are weaker probes of MAH than more intrinsic measurements.

\subsection{Axis ratio}

\begin{figure*} 
    \centering
    \includegraphics[width=\linewidth]{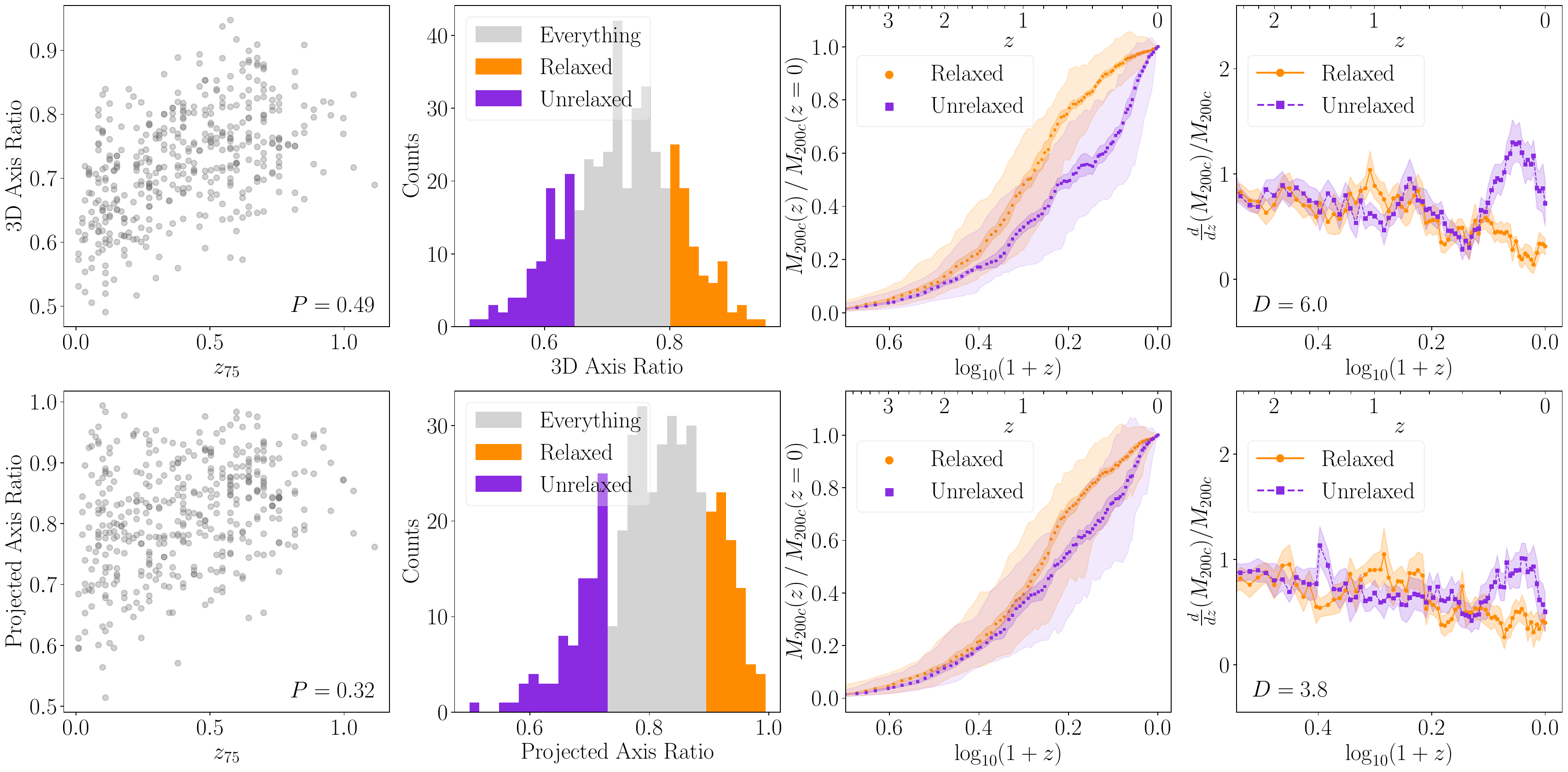}
    \caption{Parameter correlations and split sample properties, as in Figure~\ref{fig:offsets-panels}, but for samples split based on the values of the 3D and projected axis ratios.}
    \label{fig:axis-ratios-panels}
\end{figure*}

Given the triaxial shapes of dark matter halos, 3D axis ratios are more closely related to the dynamical state of cluster halos than projected axis ratios. We find that the major-to-minor axis ratio is the best correlated with MAH, and we present it in the top row of Figure~\ref{fig:axis-ratios-panels}. We see in the leftmost panel that the 3D axis ratio is well correlated with $z_{75}$, but that there is significant scatter in the relationship. In the third panel, we see that the relaxed and unrelaxed samples split by 3D axis ratio have quite different MAHs. The rightmost panel shows that the unrelaxed group experiences a recent jump in mass accretion relative to the relaxed group. However, the two curves are not as significantly different as those split using either the 3D or projected centre of mass offset, as reflected in the maximum normalised difference $D=6.0$. This indicates that even considering the full mass distributions of halos, axis ratios contain less information about MAH than centre of mass offsets.

The bottom row of Figure~\ref{fig:axis-ratios-panels} shows how robust axis ratio measurements are to projection. We see in the leftmost panel that the projected axis ratio is much more weakly correlated with $z_{75}$ and that halos with large ratios span a wide range of redshifts. In the two right panels, we see little difference between the MAHs of the relaxed and unrelaxed subsamples compared to other parameters. This shows that axis ratios are less robust to projection than other structural measurements that depend on the same information about the overall mass distribution.

\subsection{Concentration}
\label{subsec:concentration}

\begin{figure*} 
    \centering
    \includegraphics[width=\linewidth]{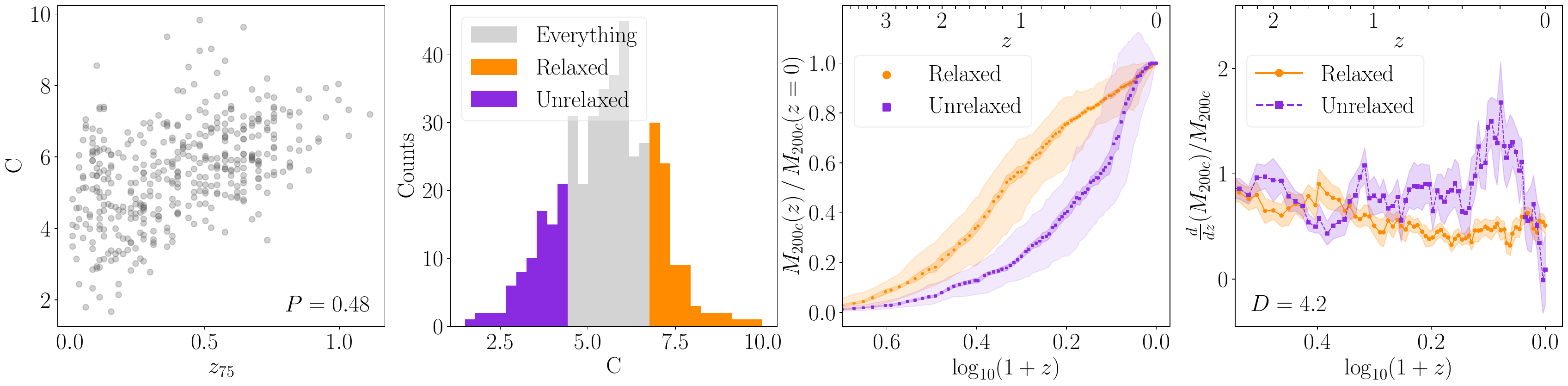}
    \caption{Parameter correlations and split sample properties, as in Figure~\ref{fig:offsets-panels}, but for samples split based on the value of the 3D concentration, $c_{200c}$.}
\label{fig:concentrations-panels}
\end{figure*}

Figure~\ref{fig:concentrations-panels} shows the correlation between 3D concentration $c_{200c}$ and the MAHs of cluster halos. In the leftmost panel, we see that concentration is correlated with $z_{75}$, though there is significant scatter in the relationship. While halos with low concentrations tend to have low $z_{75}$, there is also a population of young halos with very low $z_{75}$ but high concentrations. This highly concentrated but young population of halos likely represents systems undergoing mergers. Concentration has been shown to oscillate during mergers due to the infall and splashback of merging material \citep{Ludlow2012,Lee2018,Wang2020}. As a result, as reflected in the leftmost panel of Figure~\ref{fig:concentrations-panels}, a sample of clusters selected to have high concentrations may include some young and very unrelaxed systems, as well as older and more relaxed ones. On the other hand, a sample of very low concentrations would represent a pure but incomplete sample of young systems.

Another potential source of scatter in the concentrations of young halos is the quality of their fits to NFW profiles. Work by \citet{Drakos2019b} shows that merger remnants are not generally well described by NFW profiles. Notably, they show that the scale radius $r_s$ is a function of the energy of a merger. This can then affect the measured concentration value for the system $c_{200c}=R_{200c}/r_s$. While other profile metrics --- e.g., sparsity \citep{Balmes14} --- or other definitions of concentration \citep[e.g.][]{Klypin2016} may better describe young halos, we present the usual NFW definition since it is the most common in the literature.

Despite these complications, we see in the third panel of Figure~\ref{fig:concentrations-panels} that the MAHs of the relaxed and unrelaxed subsamples are quite different. The median histories are maximally different at a higher redshift than subsamples constructed using other structural measurements. This suggests that concentration probes earlier mass accretion rather than recent mergers. In the rightmost panel, we see that the relaxed halos are accreting mass more rapidly at very low redshift. We find this feature to be more significant at lower halo masses. Concentration probes differences in early mass accretion, but it is not the most effective indicator of recent mergers and dynamical state, especially given that it depends on a full mass distribution and is thus challenging to measure for individual halos.

\subsection{Asymmetry}
\label{subsec:structure-asymmetry}

\begin{figure*} 
    \centering
    \includegraphics[width=\linewidth]{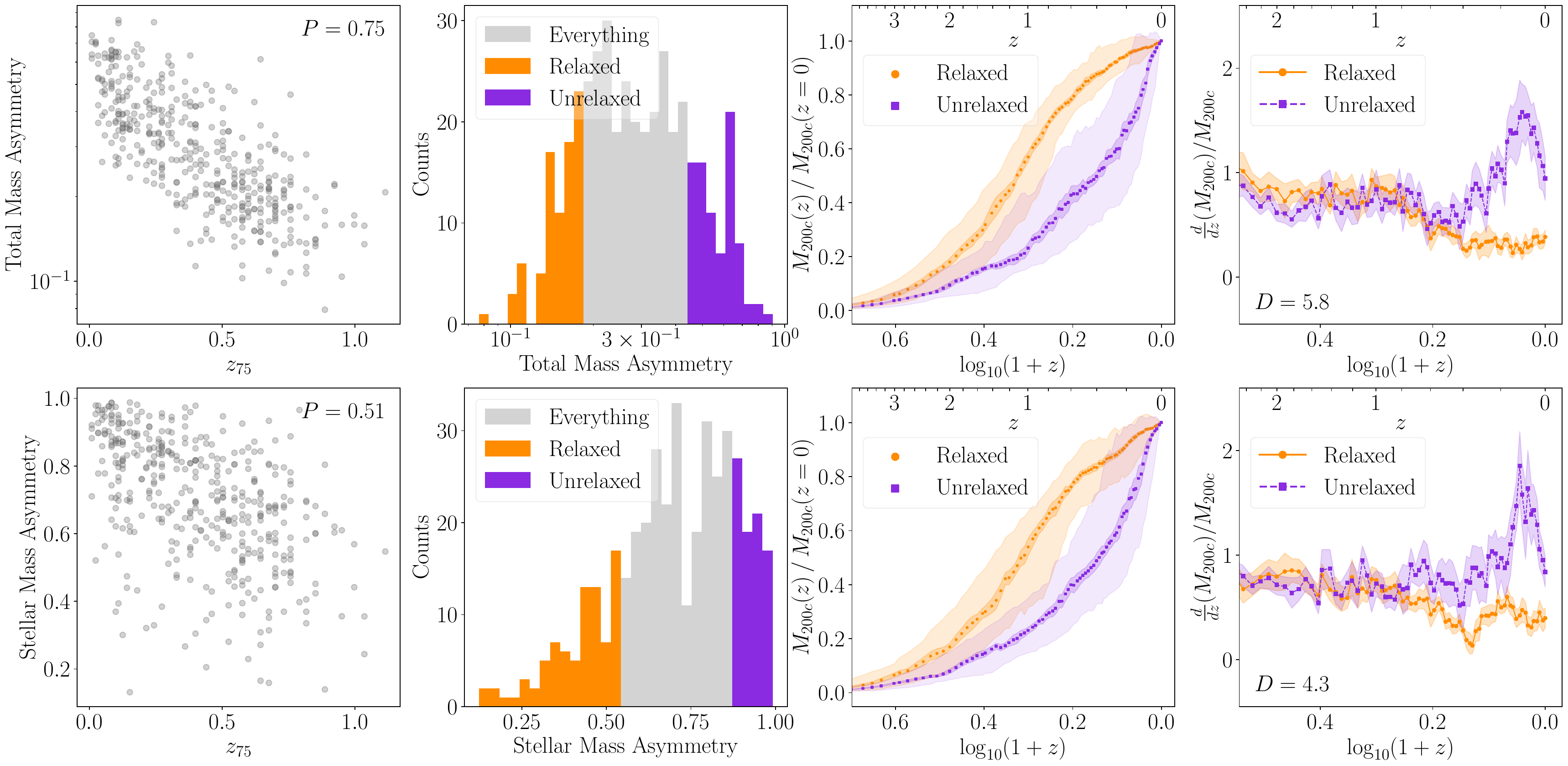}
    \caption{Parameter correlations and split sample properties, as in Figure~\ref{fig:offsets-panels}, but for samples split based on the value of the total and stellar mass asymmetry.}
    \label{fig:asymmetry-panels}
\end{figure*}

The total mass asymmetry of a halo is the asymmetry measurement that is most intrinsically tied to the dynamical state of a halo since it traces the total mass distribution around a cluster. In the top row of Figure~\ref{fig:asymmetry-panels}, we explore how total mass asymmetry relates to MAH. In the leftmost panel, we see that total mass asymmetry is strongly correlated with $z_{75}$. It has the tightest relationship with this MAH parameter of any of the structural measurements considered in this work ($P = 0.75$). In the two right panels, we see that the MAHs of relaxed and unrelaxed halo subsamples are very different. The relaxed sample has little mass accretion at low redshift, while the unrelaxed sample has a median increase in mass consistent with recent major mergers. The total mass asymmetry-based MARs are about as well differentiated as those classified by projected centre of mass offset, as indicated by the similar value of $D=5.8$. Asymmetry is also an inherently projected quantity and is thus more observationally accessible than 3D measurements.

Although it is a projected quantity and thus easier to measure observationally, the total mass asymmetry does require knowledge of the total mass distribution of a halo. In contrast, stellar mass asymmetry depends only on the projected stellar distribution, which is much more observationally tractable. We explore stellar mass asymmetry in the bottom row of Figure~\ref{fig:asymmetry-panels}. In the leftmost panel, we see that stellar mass asymmetry is correlated with $z_{75}$, though with much more scatter at low asymmetries than we saw previously. In the two right panels, we see that the MAHs of the relaxed and unrelaxed subsamples are well separated. Though the maximum difference between the MAHs occurs at a slightly higher redshift than with most other structurally split samples, the relaxed and unrelaxed groups maintain different accretion rates down to low redshift. Despite being a projected stellar quantity, the stellar mass asymmetry produces split samples with median MAHs that are nearly as distinct as those produced using the total mass asymmetry. The maximum normalised difference $D=4.3$ outperforms other optical measurements, such as the stellar offset. The difference between the median MARs of the subsamples increases very slightly if we match the line of sight constraints of the total mass asymmetry case, but the results are largely consistent. Of all the observationally accessible measurements presented in this work, the stellar mass asymmetry is that which best separates the recent MAHs of structural subsamples while being very robust to projection effects.

\subsection{Magnitude map \& stellar mass ratio}
\label{subsec:mag-gap}

In previous work \citep{ahad2025arXiv}, we considered samples split by a combination of two different parameters: $r_{12}$, the absolute r-band magnitude gap between the BCG and the second brightest galaxy, and $M^*_{14}$, the stellar mass ratio between the BCG and the fourth most massive galaxy. We will include this two-parameter split here for comparison, considering both the original 3D version and a projected version, which are calculated as described in Section~\ref{subsec:mag-gap-explanation}.

\begin{figure*} 
    \centering
    \includegraphics[width=\linewidth]{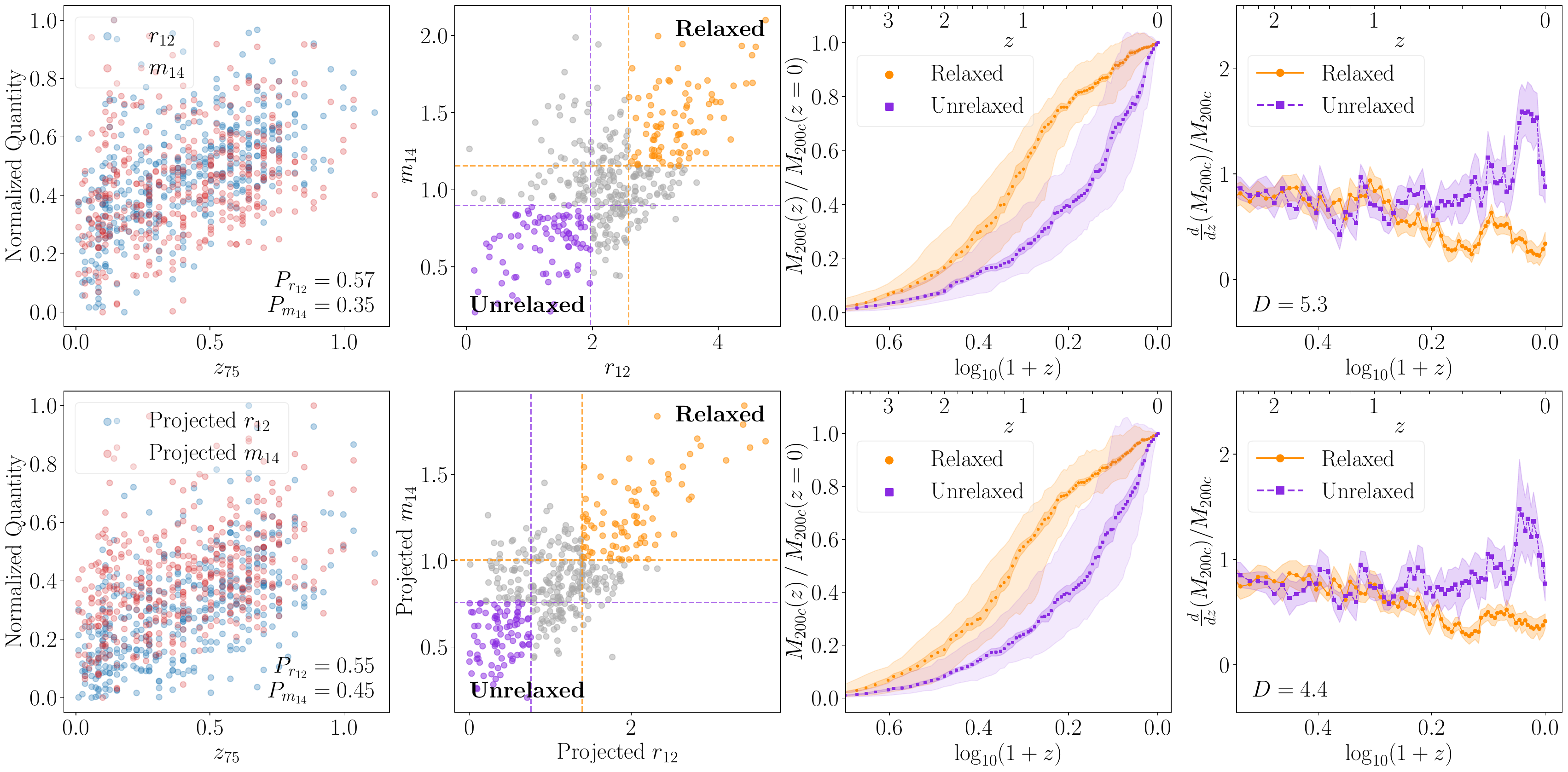}
    \caption{The leftmost panels show the distributions of magnitude gap and stellar mass ratio versus the MAH parameter $z_{75}$. The values of $P_{r_{12}}$ and $P_{m_{14}}$ listed in the leftmost column are the absolute Pearson correlation coefficients between $z_{75}$ and $r_{12}$ and between $z_{75}$ and $m_{14}$, respectively. The second column of panels shows the magnitude gap and stellar mass ratio distributions split into relaxed and unrelaxed subsamples. The third column shows the median MAHs of the subsamples, while the rightmost column shows their relative accretion rates. In the top row we include only galaxies within each halo in 3D, while in the bottom row we include projected galaxies as well. Mass accretion histories and rates are formatted as in Figure~\ref{fig:com-offset-history}.}
    \label{fig:mag-gap-panels}
\end{figure*}

In the top row of Figure~\ref{fig:mag-gap-panels}, we relate the 3D magnitude gaps and stellar mass ratios to halo MAHs. In the leftmost panel, we see that both parameters are well correlated with $z_{75}$. However, $r_{12}$ has less scatter in the relationship, as indicated by their Pearson correlation coefficients, $P_{r_{12}} = 0.57$ and $P_{m_{14}} = 0.35$. In the two right panels, we see a strong difference between the MAHs of the relaxed and unrelaxed subsamples. Despite being based solely on simple optical properties of galaxies, the magnitude gap and stellar mass ratio perform nearly as well as other structural parameters that trace the full mass distributions of halos, such as the projected centre of mass offset and the total mass asymmetry.

Given that many galaxy clusters are identified using photometric redshifts that have poor line of sight distance constraints, we also consider the projected magnitude gaps and mass ratios in the bottom row of Figure~\ref{fig:mag-gap-panels}. In the left panel, we see that both distributions are slightly less well correlated with $z_{75}$ than in the 3D case, but that a clear trend remains. In the two right panels, we see that the mass accretion histories are only slightly less well separated than in the 3D case, indicating that these measurements are robust to projection and are generally effective. The projected magnitude gap and mass ratio are easily observable and are nearly as effective at distinguishing between relaxed and unrelaxed subsamples of halos as the stellar mass asymmetry, as indicated by the similar maximum normalised separation $D=4.4$.

\section{Discussion} 
\label{sec:5discussion}

\subsection{Correlation between structural parameters}
\label{subsec:correlations}

In the previous section, we have shown that several structural parameters accessible in large optical samples, notably stellar mass asymmetry, magnitude gap and stellar mass ratio, have potential for classifying the dynamical states of clusters. Here we explore the correlations between different measures to understand which are closely related and which are relatively independent.

\begin{figure*} 
    \centering
    \includegraphics[width=0.75\linewidth]{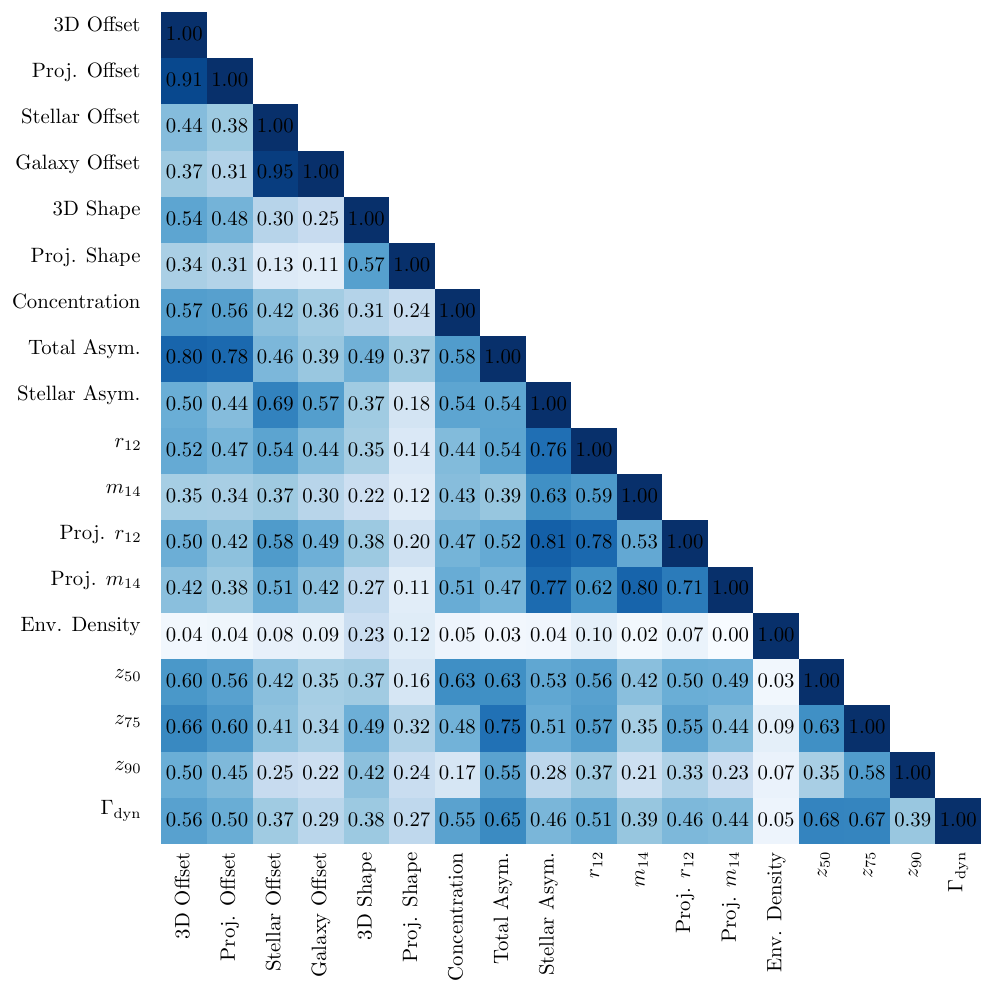}
    \caption{The Pearson correlation coefficients for the structural parameters and formation history parameters of halos. Some of these corrections are negative; we omit that information in this figure for clarity. Structural parameters that are presented in a $\log_{10}$ scale in previous figures have been converted to $\log_{10}$-scale values before computing these correlations.}
    \label{fig:corner-plot}
\end{figure*}

Figure~\ref{fig:corner-plot} shows the Pearson correlation coefficient for each of the structural parameters measured for the halo sample in Section~\ref{sec:4structure-selection}. This coefficient expresses the covariance of two parameters divided by the product of their variances. This quantity can be negative, but we present the absolute value of the coefficient to most clearly convey the strength of the correlations. As such, a high absolute value of the Pearson correlation coefficient implies a tight linear relationship between two measurements. Non-linear correlations are not necessarily well captured, however, so the coefficient only provides a partial indication of the full relationship between any two parameters.

There are several notable features in Figure~\ref{fig:corner-plot} that highlight relationships between our chosen structural parameters. First, we see strong correlations between the total mass asymmetry and the total centre of mass offsets. This result is expected intuitively, as both parameters trace the asymmetric distribution of mass in a halo. Second, the stellar mass asymmetry is found to be correlated with the magnitude gap and the stellar mass ratio. Each of these parameters is affected by large stellar components offset from the BCG. Third, we see that the projected shape parameter is poorly correlated with most other parameters. This is not surprising, as we found in Section~\ref{sec:4structure-selection} that the projected axis ratio carries little dynamical information about clusters. For instance, when comparing relaxed and unrelaxed subsamples split using magnitude gaps and stellar mass ratios, the mean 3D axis ratios of these subsamples differ from the mean of the total sample by only $\sim\!6\%$. Similarly, the 2D axis ratios differ by only $\sim\!3\%$.

Lastly, we find that local environmental density --- defined as the total mass within a 10 Mpc radius around a cluster centre, minus $M_{200c}$ for the cluster --- is poorly correlated with most parameters, except for halo shape. We tested this conclusion for different volumes and halo mass definitions and found consistent results. This is in agreement with cluster-mass halos exhibiting weak environment-dependent assembly bias signals in simulations \citep[e.g.][]{Mao2018}, though other works find the contrary \citep[e.g.][]{Cadiou2021}, and there is some evidence of stronger assembly bias effects than found here in observations \cite[e.g.][]{Liu2024}.

\subsection{Comparison to the literature}

Correlations between structural parameters and growth history have been considered previously by a number of authors. \cite{Haggar2024a} used principal component analysis (PCA) and uniform manifold approximation and projection (UMAP) to study the correlations between multiple measures of the dynamical state of galaxy clusters, taken from The Three Hundred hydrodynamical simulations \citep{Cui2018}. They find a two-dimensional decomposition provides a good description of the dynamical state. The primary component reflects structural parameters such as the centre of mass offset and substructure fraction, while a secondary component reflects the virialisation of the diffuse cluster halo. By studying MAHs, they show that the first component can be used to separate clusters into early-forming and late-forming, in agreement with our work. They also show that morphological measurements of clusters from mock X-ray and SZ maps \citep{Deluca2021} can trace this component as well. The very recent growth history of clusters is traced by their secondary component, which separates clusters by their growth rates in the last $\sim\!1\,$Gyr. Similarly to the results shown in Figure~\ref{fig:corner-plot}, they find a weak correlation between cluster dynamical state and environment. They define environment based on the number of cosmic filaments connected to a cluster, and they find that this is only weakly correlated with most other structural parameters.

\cite{Kim2024} explored whether combinations of structural parameters could accurately determine the merger status of a cluster and distinguish between recent and ancient mergers. They find that the stellar mass gap between their brightest and second-brightest galaxy is a powerful tracer of cluster merger history, as we also show in Figure~\ref{fig:mag-gap-panels}. Additionally, they demonstrate that the centre of mass offset of clusters provides additional information on recent merger activity ($\sim\!1\,$Gyr ago), while a higher satellite stellar mass fraction indicates an ancient merger event ($3-5\,$Gyr ago). \cite{Valles2025} find similar results, showing that centre of mass offset and substructure fraction of simulated clusters correlate well with their accretion histories at $z=0.2$ and $z=0.4$, respectively.

\cite{Kimmig2025} studied correlations between assembly history, structural parameters, and the intracluster light (ICL) across four different hydrodynamical simulations --- Magneticum \citep{dolag2025}, TNG100 of IllustrisTNG \citep{Nelson2018,Pillepich2018b}, Horizon-AGN \citep{dubois2014}, and Hydrangea \citep{bahe2017}. They defined the formation redshift $z_{\rm form}$ as the redshift where the cluster assembled half of its halo mass ($M_{200c}$ at $z=0$), equivalent to our $z_{50}$, and used this as a proxy for the cluster mass assembly history. They found that the mass fraction in BCG+ICL, as well as the stellar mass ratio between the BCG and the 2$^{nd}$ and 4$^{th}$ most massive galaxies in the cluster, $m_{12}$ and $m_{14}$, are strongly correlated with $z_{\rm form}$ across all simulations. They also found several other dynamical state indicators that are strongly correlated with $z_{\rm form}$ but which can only be measured in a robust way in simulations. These include the fraction of total cluster mass in all of the subhalos and the fraction of total cluster mass in the $8^{th}$ most massive satellite. We find strong correlations between $m_{14}$ and the redshifts $z_{50}$, $z_{75}$ and $z_{90}$, consistent with their results.

Finally, \cite{Magnus2026} studied the connection between gas or dark matter profiles and the accretion rate in the FLAMINGO hydrodynamical simulations \citep{Schaye2023}. They also tested for dependence on numerical effects, such as resolution and the assumed feedback model. They found the X-ray centroid shift to be a reasonably robust indicator of recent accretion. While we do not calculate this quantity directly in our work, the parallel to our total and stellar centre of mass offsets is interesting. They also found that dark matter profiles are more concentrated for relaxed systems, consistent with our results in Section~\ref{subsec:concentration}. Lastly, they measured a similar or stronger correlation between the stellar mass ratio (equivalent to our ratio $m_{14}$) and the accretion rate. The absolute Pearson coefficients ranged from 0.35 to 0.55 at low redshift, similar to our value of 0.39. However, they found across the range of FLAMINGO simulations that this correlation is more sensitive to numerical effects than that of the X-ray centroid offset. We will explore this possible sensitivity across other simulations in future work.

\subsection{Application to scaling relations}

The mass-richness scaling relation connects the total mass of a cluster to the number of member galaxies within some radius and above some mass or luminosity threshold \citep[e.g.][]{Andreon2010}. The relation is usually defined as a power law of the form:
\begin{equation}
    n_{200c}=A\left(\frac{M_{200c}}{M_\odot}\right)^k\,,
    \label{eq:mass-richness}
\end{equation}
where $n_{200c}$ and $M_{200c}$ are the number of galaxies and the total mass within $R_{200c}$, respectively, $A$ is the normalisation, and $k$ is the power law index. For this application, we consider all member galaxies with stellar masses $M_*\geq10^9M_\odot$. To determine whether the scaling relation depends on dynamical state, we fit this power law to our samples of relaxed and unrelaxed clusters separately. We considered dynamical splits based on the structural parameters that we determined to be the most practical tracers of recent MAH in Section~\ref{sec:4structure-selection}, namely asymmetry, magnitude gap and stellar mass ratio. Each scaling relation was fit using 84 halos per subsample, which we note limits the statistical significance of the trends observed.

\begin{figure} 
    \centering
    \includegraphics[width=\linewidth]{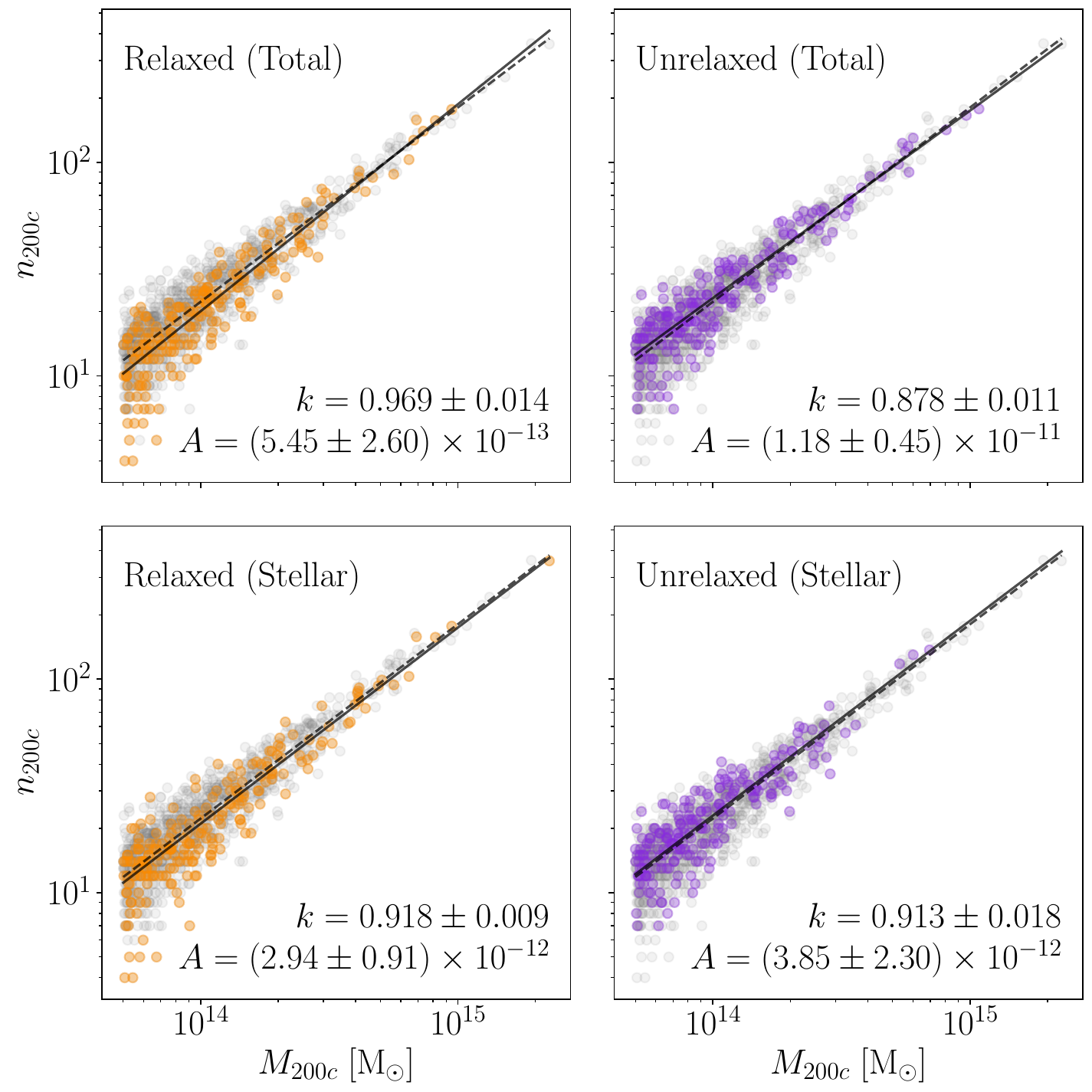}
    \caption{The mass-richness relations for relaxed and unrelaxed subsamples split by total mass asymmetry (top row) or stellar mass asymmetry (bottom row). The grey dots represent the total halo sample. The coloured dots are those included in the panel's relevant subsample. The solid black lines represent the best fit power law from Equation~\ref{eq:mass-richness} to the subsample. The dashed black lines are the fit to the total distribution.}
    \label{fig:asymmetry-mass-richness}
\end{figure}

In the top row of Figure~\ref{fig:asymmetry-mass-richness}, we compare the mass-richness relation for relaxed and unrelaxed halos split by their total mass asymmetry. We see that the scaling relation for relaxed clusters has a lower normalisation and a greater power-law index than the relation for unrelaxed clusters. Quantitatively, despite the small sample size, there is a $2.4\sigma$ difference between the fitted values of $A$ and a $3.6\sigma$ difference between the fitted values for $k$. In the bottom row of Figure~\ref{fig:asymmetry-mass-richness}, we perform the same analysis for halos split by stellar mass asymmetry. We see the same trends, though the differences in $A$ and $k$ are both less than $0.3\sigma$, making them much less statistically significant than in the total mass asymmetry case.

\begin{figure} 
    \centering
    \includegraphics[width=\linewidth]{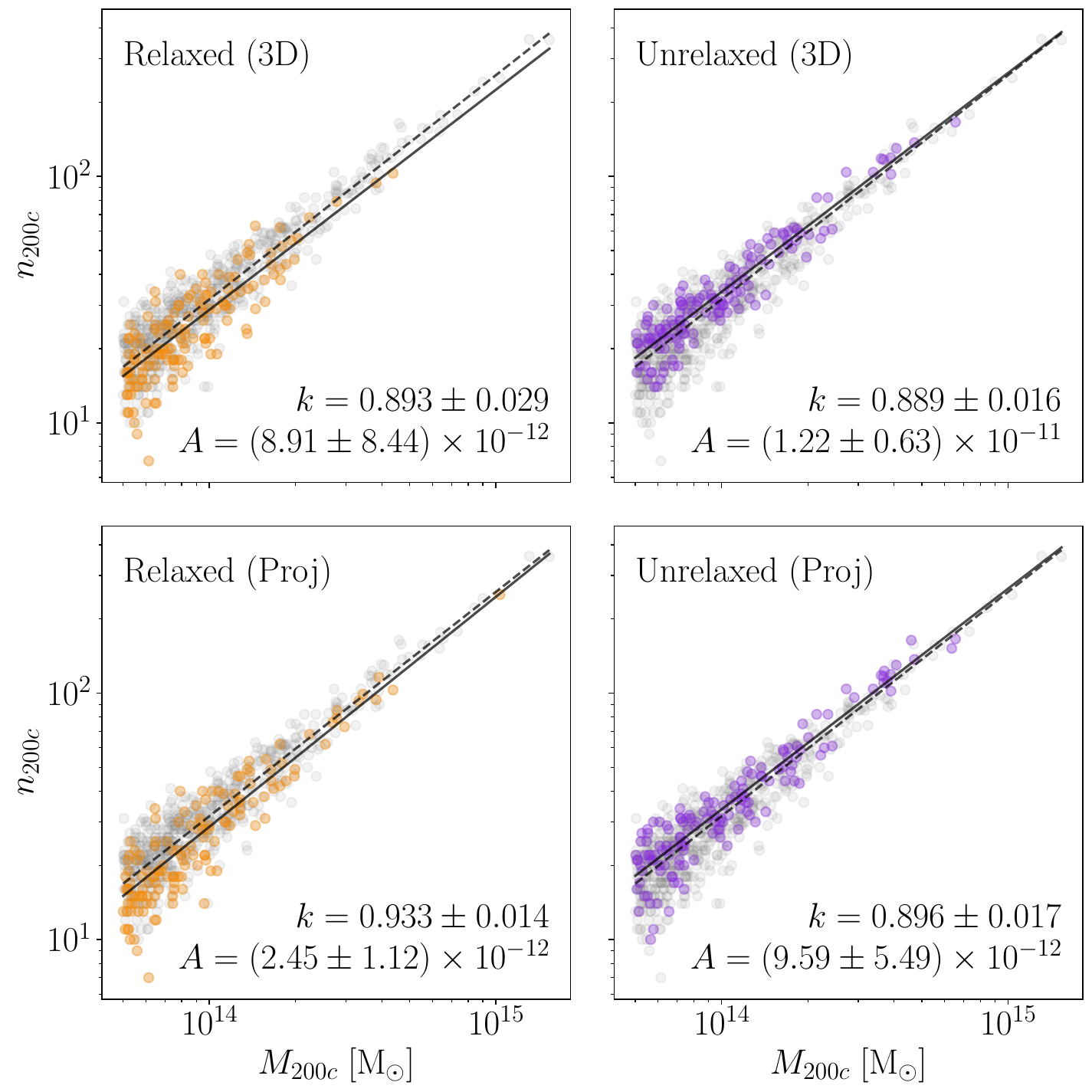}
    \caption{The mass-richness relations for relaxed and unrelaxed subsamples split by magnitude gap and stellar mass ratio using 3D cluster members (top row) or projected member galaxies (bottom row). Formatting is as in Figure~\ref{fig:asymmetry-mass-richness}.}
    \label{fig:mag-gap-mass-richness}
\end{figure}

In the top row of Figure~\ref{fig:mag-gap-mass-richness}, we compare the scaling relations for relaxed and unrelaxed halos split using $r_{12}$ and $m_{14}$ as in Section~\ref{subsec:mag-gap}. Once again, the scaling relation for relaxed halos has a smaller normalisation and a larger power-law index. However, the statistical significance of these differences is again quite small --- less than $0.3\sigma$. In the bottom row of Figure~\ref{fig:mag-gap-mass-richness}, we perform the analysis using projected magnitude gaps and stellar mass ratios. In this case, we see the same trends in the best fit parameters with a slightly greater statistical significance of $\sim\!1\sigma$ for each parameter.

These results indicate a consistent difference in the mass-richness scaling relations for relaxed and unrelaxed clusters. In particular, the relation for relaxed clusters has a lower normalisation and steeper slope than the relation for unrelaxed clusters. This appears to be consistent with \cite{Mulroy2019}, who found that relaxed, low-central-entropy clusters have higher stellar fractions, and also with \cite{Farahi2020}, who found that systems with larger magnitude gaps have a smaller number of satellites overall. The significance of the trends in IllustrisTNG is weak, however, due to the small sample size. In future work, we will seek further evidence for these trends in larger simulated or observed samples.

For each of these structural selections and scaling relation fits, we also compared the scatter of the residuals around the best fit power laws. For the total and stellar mass asymmetry, the relative difference in the standard deviations of the relaxed and unrelaxed subsamples is no more than 10\%. For the magnitude gap and mass ratio, the relative difference in the standard deviations of the subsamples is less than 20\%. Across these different structural selections, there is no consistent pattern as to whether the scatter is greater for the relaxed or the unrelaxed subsample. Given these results and the small sample size, we conclude that the difference in the scatter around the scaling relations between the subsamples is not significant.

\section{Conclusions} 
\label{sec:6conclusions}

In this work, we have considered practical ways of classifying individual dynamical states for large, optically selected samples of galaxy clusters in order to create subsets of clusters with similar growth histories. To compare different metrics, we have used an approach we refer to as the ``split sample'' method. An initial sample is sorted according to the value of a structural parameter; then ``relaxed'' and ``unrelaxed'' subsamples at either end of the distribution are compared. We look specifically at the median growth history of systems in each subsample, as indicated by the mass accretion history. From these comparisons, we select the structural parameters that produce the largest difference in median mass accretion history as the most effective metrics for characterising the dynamical state. We find that several structural parameters accessible in large optical samples, notably stellar asymmetry, magnitude gap and stellar mass ratio, can produce subsamples with significantly different median mass accretion histories. Thus, these parameters have potential for classifying the dynamical states of clusters.

With poor line of sight constraints, relaxed and unrelaxed cluster samples selected based on stellar mass asymmetry have median mass accretion rates with a maximum difference of $D=4.3$ times the uncertainty in the difference. Samples selected on projected magnitude gaps and stellar mass ratios similarly achieve $D=4.4$. With strong discrimination between cluster members and background galaxies, this maximum difference is improved to $D=5.3$. Detailed analyses of galaxy clusters (e.g.~strong lensing models) can to some degree trace their total projected mass distributions. Such analyses enable better structural measurements of halos, such as the projected centre of mass offset and the total mass asymmetry. Both of these measurements outperform basic optical splits, producing maximum differences between the mass accretion rates of $D=6.1$ and $D=5.8$, respectively. Lastly, among measurements that trace the full 3D mass distributions of halos, the 3D centre of mass offset produces the greatest difference between the mass accretion rates of the relaxed and unrelaxed subsamples, $D=7.2$.

While we find trends and correlations between different parameters that are consistent with a number of previous studies in the literature, many of the parameters considered in these studies would be difficult or impossible to determine for large optical samples. The projected stellar metrics considered here have the advantage that they can be applied directly to these samples without ancillary data or measurements. 

To illustrate a possible application of split samples, we determine richness-lensing mass scaling relations for samples split by dynamical state, revealing a small but consistent difference between the relations for relaxed and unrelaxed systems. More generally, we can use the split sample approach to study problems in sample selection, bias in mass estimates, and the impact of dynamical state on cosmological tests. In particular, we have already applied the approach to a UNIONS $\times$ DESI sample \citep{ahad2025arXiv}, showing that the resulting relaxed and unrelaxed cluster samples differ in mean galaxy luminosity function and weak lensing mass profile.

We note several shortcomings of the split sample approach. First, we have focused only on limited aspects of the ``dynamical state'' and ``growth history'' of clusters. Our dynamical state indicators all relate to the overall mass distribution or stellar mass distribution in each cluster. These quantities are mainly sensitive to the spatial distribution of collisionless (dark matter or stellar) material. Optical or X-ray spectroscopy, as well as imaging at X-ray or sub-millimetre wavelengths, can provide much more detailed information about the kinematics of cluster components or the state of the ICM. They are much more expensive to obtain observationally, however, and are thus unlikely to be available for the largest samples in any near future. Similarly, we have only considered the simplest aspects of the growth history, representing this by a few simple moments of the mass accretion history. The details of each major merger event, such as its orbital energy, angular momentum and line of sight orientation, can have a strong effect on the overall appearance of the final system. Here too, however, modelling this dependence in detail is only realistic for small samples of well-studied objects.

Second, our assessment of the relative effectiveness of different splits is based on a particular set of simulations, the IllustrisTNG simulations, which include a specific implementation of baryonic physics. While some structural indicators are expected to be relatively independent of this assumed physics, others, such as the magnitude gap or stellar mass ratio, may be sensitive to it and should be tested across multiple simulations. From this point of view, the study by \cite{Kimmig2025}, which compared parameter correlations across four different simulations, provides an interesting point of reference for some of our indicators. 

Finally, another shortcoming of the split sample approach is that most of the indicators we have considered are only sensitive to fairly recent accretion onto the cluster. The properties of cluster galaxies may depend more on the early history of the cluster, on the properties of their original host halos, and/or on the extent to which they experienced preprocessing \citep[e.g.][]{werner2022,Ahad2024}. In future work, we will search for additional structural indicators that can trace these earlier phases of growth.

\section*{Acknowledgements}

We thank Susmita Adhikari, Niayesh Afshordi, Camille Avestruz and Mike Hudson for useful discussions. CTM is supported by an appointment to the NASA Postdoctoral Program at the NASA Goddard Space Flight Center, administered by Oak Ridge Associated Universities under contract with NASA. JET acknowledges support from the Natural Sciences and Engineering Research Council of Canada (NSERC) through a Discovery Grant. This research was enabled in part by support provided by Compute Ontario (www.computeontario.ca) and the Digital Research Alliance of Canada (alliancecan.ca). 

\section*{Data availability}

The data generated for this paper are available on request from the authors. Data from the IllustrisTNG simulations are available at \url{https://www.tng-project.org/data/}.


\bibliographystyle{mnras}
\bibliography{references} 

@ARTICLE{dolag2025,
       author = {{Dolag}, Klaus and {Remus}, Rhea-Silvia and {Valenzuela}, Lucas M. and {Kimmig}, Lucas C. and {Seidel}, Benjamin and {Fortune}, Silvio and {Stoiber}, Johannes and {Ivleva}, Anna and {Hoffmann}, Tadziu and {Biffi}, Veronica and {Marini}, Ilaria and {Popesso}, Paola and {Vladutescu-Zopp}, Stephan},
        title = "{Encyclopedia Magneticum: Scaling Relations from Cosmic Dawn to Present Day}",
      journal = {arXiv e-prints},
         year = 2025,
        month = apr,
          eid = {arXiv:2504.01061},
        pages = {arXiv:2504.01061},
          doi = {10.48550/arXiv.2504.01061},
archivePrefix = {arXiv},
       eprint = {2504.01061},
 primaryClass = {astro-ph.CO},
       adsurl = {https://ui.adsabs.harvard.edu/abs/2025arXiv250401061D}
}

@ARTICLE{ahad2025arXiv,
       author = {{Ahad}, Syeda Lammim and {Reid}, Rashaad and {Mpetha}, Charlie T. and {Taylor}, James E. and {Hildebrandt}, Hendrik and {Hudson}, Michael J. and {Chambers}, Kenneth C. and {de Boer}, Thomas and {Guerrini}, Sacha and {Guinot}, Axel and {Gwyn}, Stephen and {Kilbinger}, Martin and {Van Waerbeke}, Ludovic},
        title = "{Cluster properties as a function of dynamical state in the DESI Legacy x UNIONS surveys}",
      journal = {arXiv e-prints},
         year = 2025,
        month = dec,
          eid = {arXiv:2512.14636},
        pages = {arXiv:2512.14636},
          doi = {10.48550/arXiv.2512.14636},
archivePrefix = {arXiv},
       eprint = {2512.14636},
 primaryClass = {astro-ph.GA},
       adsurl = {https://ui.adsabs.harvard.edu/abs/2025arXiv251214636A}
}

@ARTICLE{werner2022,
       author = {{Werner}, S.~V. and {Hatch}, N.~A. and {Muzzin}, A. and {van der Burg}, R.~F.~J. and {Balogh}, M.~L. and {Rudnick}, G. and {Wilson}, G.},
        title = "{Satellite quenching was not important for z {\ensuremath{\sim}} 1 clusters: most quenching occurred during infall}",
      journal = {\mnras},
         year = 2022,
        month = feb,
       volume = {510},
       number = {1},
        pages = {674-686},
          doi = {10.1093/mnras/stab3484},
archivePrefix = {arXiv},
       eprint = {2111.14624},
 primaryClass = {astro-ph.GA},
       adsurl = {https://ui.adsabs.harvard.edu/abs/2022MNRAS.510..674W}
}

@ARTICLE{Ahad2024,
       author = {{Ahad}, Syeda Lammim and {Muzzin}, Adam and {Bah{\'e}}, Yannick M. and {Hoekstra}, Henk},
        title = "{An environment-dependent halo mass function as a driver for the early quenching of z {\ensuremath{\geq}} 1.5 cluster galaxies}",
      journal = {\mnras},
         year = 2024,
        month = mar,
       volume = {528},
       number = {4},
        pages = {6329-6339},
          doi = {10.1093/mnras/stae341},
archivePrefix = {arXiv},
       eprint = {2307.01147},
 primaryClass = {astro-ph.GA},
       adsurl = {https://ui.adsabs.harvard.edu/abs/2024MNRAS.528.6329A}
}

@ARTICLE{bahe2017,
       author = {{Bah{\'e}}, Yannick M. and {Barnes}, David J. and {Dalla Vecchia}, Claudio and {Kay}, Scott T. and {White}, Simon D.~M. and {McCarthy}, Ian G. and {Schaye}, Joop and {Bower}, Richard G. and {Crain}, Robert A. and {Theuns}, Tom and {Jenkins}, Adrian and {McGee}, Sean L. and {Schaller}, Matthieu and {Thomas}, Peter A. and {Trayford}, James W.},
        title = "{The Hydrangea simulations: galaxy formation in and around massive clusters}",
      journal = {\mnras},
         year = 2017,
        month = oct,
       volume = {470},
       number = {4},
        pages = {4186-4208},
          doi = {10.1093/mnras/stx1403},
archivePrefix = {arXiv},
       eprint = {1703.10610},
 primaryClass = {astro-ph.GA},
       adsurl = {https://ui.adsabs.harvard.edu/abs/2017MNRAS.470.4186B}
}

@ARTICLE{dubois2014,
       author = {{Dubois}, Y. and {Pichon}, C. and {Welker}, C. and {Le Borgne}, D. and {Devriendt}, J. and {Laigle}, C. and {Codis}, S. and {Pogosyan}, D. and {Arnouts}, S. and {Benabed}, K. and {Bertin}, E. and {Blaizot}, J. and {Bouchet}, F. and {Cardoso}, J.-F. and {Colombi}, S. and {de Lapparent}, V. and {Desjacques}, V. and {Gavazzi}, R. and {Kassin}, S. and {Kimm}, T. and {McCracken}, H. and {Milliard}, B. and {Peirani}, S. and {Prunet}, S. and {Rouberol}, S. and {Silk}, J. and {Slyz}, A. and {Sousbie}, T. and {Teyssier}, R. and {Tresse}, L. and {Treyer}, M. and {Vibert}, D. and {Volonteri}, M.},
        title = "{Dancing in the dark: galactic properties trace spin swings along the cosmic web}",
      journal = {\mnras},
         year = 2014,
        month = oct,
       volume = {444},
       number = {2},
        pages = {1453-1468},
          doi = {10.1093/mnras/stu1227},
archivePrefix = {arXiv},
       eprint = {1402.1165},
 primaryClass = {astro-ph.CO},
       adsurl = {https://ui.adsabs.harvard.edu/abs/2014MNRAS.444.1453D}
}

@ARTICLE{Zhang2022,
    author = {{Zhang}, Bowei and {Cui}, Weiguang and {Wang}, Yuhuan and {Dave}, Romeel and {De Petris}, Marco},
    title = "{THE THREE HUNDRED: cluster dynamical states and relaxation period}",
    journal = {\mnras},
    year = 2022,
    month = oct,
    volume = {516},
    number = {1},
    pages = {26-38},
    doi = {10.1093/mnras/stac2171}
}

@ARTICLE{Kelkar2023,
    author = {{Kelkar}, K. and {Jaff{\'e}}, Y.~L. and {Louren{\c{c}}o}, A.~C.~C. and {P{\'e}rez-Mill{\'a}n}, D. and {Fritz}, J. and {Vulcani}, B. and {Crossett}, J.~P. and {Poggianti}, B. and {Moretti}, A.},
    title = "{Post-processing of galaxies due to major cluster mergers. I. Hints from galaxy colours and morphologies}",
    journal = {\aap},
    year = 2023,
    month = dec,
    volume = {680},
    eid = {A54},
    pages = {A54},
    doi = {10.1051/0004-6361/202347660}
}

@ARTICLE{Amoura2021,
    author = {{Amoura}, Yuba and {Drakos}, Nicole E. and {Berrouet}, Anael and {Taylor}, James E.},
    title = "{Cluster assembly times as a cosmological test}",
    journal = {\mnras},
    year = 2021,
    month = nov,
    volume = {508},
    number = {1},
    pages = {100-117},
    doi = {10.1093/mnras/stab2467}
}

@ARTICLE{Euclid2019,
    author  = {{Euclid Collaboration} and {Adam}, R. and {Vannier}, M. and {Maurogordato}, S. and {Biviano}, A. and {Adami}, C. and {Ascaso}, B. and {Bellagamba}, F. and {Benoist}, C. and {Cappi}, A. and {D{\'\i}az-S{\'a}nchez}, A. and {Durret}, F. and {Farrens}, S. and {Gonzalez}, A.~H. and {Iovino}, A. and {Licitra}, R. and {Maturi}, M. and {Mei}, S. and {Merson}, A. and {Munari}, E. and {Pell{\'o}}, R. and {Ricci}, M. and {Rocci}, P.~F. and {Roncarelli}, M. and {Sarron}, F. and {Amoura}, Y. and {Andreon}, S. and {Apostolakos}, N. and {Arnaud}, M. and {Bardelli}, S. and {Bartlett}, J. and {Baugh}, C.~M. and {Borgani}, S. and {Brodwin}, M. and {Castander}, F. and {Castignani}, G. and {Cucciati}, O. and {De Lucia}, G. and {Dubath}, P. and {Fosalba}, P. and {Giocoli}, C. and {Hoekstra}, H. and {Mamon}, G.~A. and {Melin}, J.~B. and {Moscardini}, L. and {Paltani}, S. and {Radovich}, M. and {Sartoris}, B. and {Schultheis}, M. and {Sereno}, M. and {Weller}, J. and {Burigana}, C. and {Carvalho}, C.~S. and {Corcione}, L. and {Kurki-Suonio}, H. and {Lilje}, P.~B. and {Sirri}, G. and {Toledo-Moreo}, R. and {Zamorani}, G.},
    title   = {Euclid preparation. {III}. {G}alaxy cluster detection in the wide photometric survey, performance and algorithm selection},
    journal = {\aap},
    volume  = {627},
    pages   = {A23},
    year    = {2019},
    doi     = {10.1051/0004-6361/201935088}
}

@ARTICLE{Ivezic2019,
    author = {{Ivezi{\'c}}, {\v{Z}}eljko and {Kahn}, Steven M. and {Tyson}, J. Anthony and {Abel}, Bob and {Acosta}, Emily and {Allsman}, Robyn and {Alonso}, David and {AlSayyad}, Yusra and {Anderson}, Scott F. and {Andrew}, John and {Angel}, James Roger P. and {Angeli}, George Z. and {Ansari}, Reza and {Antilogus}, Pierre and {Araujo}, Constanza and {Armstrong}, Robert and {Arndt}, Kirk T. and {Astier}, Pierre and {Aubourg}, {\'E}ric and {Auza}, Nicole and {Axelrod}, Tim S. and {Bard}, Deborah J. and {Barr}, Jeff D. and {Barrau}, Aurelian and {Bartlett}, James G. and {Bauer}, Amanda E. and {Bauman}, Brian J. and {Baumont}, Sylvain and {Bechtol}, Ellen and {Bechtol}, Keith and {Becker}, Andrew C. and {Becla}, Jacek and {Beldica}, Cristina and {Bellavia}, Steve and {Bianco}, Federica B. and {Biswas}, Rahul and {Blanc}, Guillaume and {Blazek}, Jonathan and {Blandford}, Roger D. and {Bloom}, Josh S. and {Bogart}, Joanne and {Bond}, Tim W. and {Booth}, Michael T. and {Borgland}, Anders W. and {Borne}, Kirk and {Bosch}, James F. and {Boutigny}, Dominique and {Brackett}, Craig A. and {Bradshaw}, Andrew and {Brandt}, William Nielsen and {Brown}, Michael E. and {Bullock}, James S. and {Burchat}, Patricia and {Burke}, David L. and {Cagnoli}, Gianpietro and {Calabrese}, Daniel and {Callahan}, Shawn and {Callen}, Alice L. and {Carlin}, Jeffrey L. and {Carlson}, Erin L. and {Chandrasekharan}, Srinivasan and {Charles-Emerson}, Glenaver and {Chesley}, Steve and {Cheu}, Elliott C. and {Chiang}, Hsin-Fang and {Chiang}, James and {Chirino}, Carol and {Chow}, Derek and {Ciardi}, David R. and {Claver}, Charles F. and {Cohen-Tanugi}, Johann and {Cockrum}, Joseph J. and {Coles}, Rebecca and {Connolly}, Andrew J. and {Cook}, Kem H. and {Cooray}, Asantha and {Covey}, Kevin R. and {Cribbs}, Chris and {Cui}, Wei and {Cutri}, Roc and {Daly}, Philip N. and {Daniel}, Scott F. and {Daruich}, Felipe and {Daubard}, Guillaume and {Daues}, Greg and {Dawson}, William and {Delgado}, Francisco and {Dellapenna}, Alfred and {de Peyster}, Robert and {de Val-Borro}, Miguel and {Digel}, Seth W. and {Doherty}, Peter and {Dubois}, Richard and {Dubois-Felsmann}, Gregory P. and {Durech}, Josef and {Economou}, Frossie and {Eifler}, Tim and {Eracleous}, Michael and {Emmons}, Benjamin L. and {Fausti Neto}, Angelo and {Ferguson}, Henry and {Figueroa}, Enrique and {Fisher-Levine}, Merlin and {Focke}, Warren and {Foss}, Michael D. and {Frank}, James and {Freemon}, Michael D. and {Gangler}, Emmanuel and {Gawiser}, Eric and {Geary}, John C. and {Gee}, Perry and {Geha}, Marla and {Gessner}, Charles J.~B. and {Gibson}, Robert R. and {Gilmore}, D. Kirk and {Glanzman}, Thomas and {Glick}, William and {Goldina}, Tatiana and {Goldstein}, Daniel A. and {Goodenow}, Iain and {Graham}, Melissa L. and {Gressler}, William J. and {Gris}, Philippe and {Guy}, Leanne P. and {Guyonnet}, Augustin and {Haller}, Gunther and {Harris}, Ron and {Hascall}, Patrick A. and {Haupt}, Justine and {Hernandez}, Fabio and {Herrmann}, Sven and {Hileman}, Edward and {Hoblitt}, Joshua and {Hodgson}, John A. and {Hogan}, Craig and {Howard}, James D. and {Huang}, Dajun and {Huffer}, Michael E. and {Ingraham}, Patrick and {Innes}, Walter R. and {Jacoby}, Suzanne H. and {Jain}, Bhuvnesh and {Jammes}, Fabrice and {Jee}, M. James and {Jenness}, Tim and {Jernigan}, Garrett and {Jevremovi{\'c}}, Darko and {Johns}, Kenneth and {Johnson}, Anthony S. and {Johnson}, Margaret W.~G. and {Jones}, R. Lynne and {Juramy-Gilles}, Claire and {Juri{\'c}}, Mario and {Kalirai}, Jason S. and {Kallivayalil}, Nitya J. and {Kalmbach}, Bryce and {Kantor}, Jeffrey P. and {Karst}, Pierre and {Kasliwal}, Mansi M. and {Kelly}, Heather and {Kessler}, Richard and {Kinnison}, Veronica and {Kirkby}, David and {Knox}, Lloyd and {Kotov}, Ivan V. and {Krabbendam}, Victor L. and {Krughoff}, K. Simon and {Kub{\'a}nek}, Petr and {Kuczewski}, John and {Kulkarni}, Shri and {Ku}, John and {Kurita}, Nadine R. and {Lage}, Craig S. and {Lambert}, Ron and {Lange}, Travis and {Langton}, J. Brian and {Le Guillou}, Laurent and {Levine}, Deborah and {Liang}, Ming and {Lim}, Kian-Tat and {Lintott}, Chris J. and {Long}, Kevin E. and {Lopez}, Margaux and {Lotz}, Paul J. and {Lupton}, Robert H. and {Lust}, Nate B. and {MacArthur}, Lauren A. and {Mahabal}, Ashish and {Mandelbaum}, Rachel and {Markiewicz}, Thomas W. and {Marsh}, Darren S. and {Marshall}, Philip J. and {Marshall}, Stuart and {May}, Morgan and {McKercher}, Robert and {McQueen}, Michelle and {Meyers}, Joshua and {Migliore}, Myriam and {Miller}, Michelle and {Mills}, David J.},
    title = "{LSST: From Science Drivers to Reference Design and Anticipated Data Products}",
    journal = {\apj},
    year = 2019,
    month = mar,
    volume = {873},
    number = {2},
    eid = {111},
    pages = {111},
    doi = {10.3847/1538-4357/ab042c}
}

@ARTICLE{Naiman2018,
    author = {{Naiman}, Jill P. and {Pillepich}, Annalisa and {Springel}, Volker and {Ramirez-Ruiz}, Enrico and {Torrey}, Paul and {Vogelsberger}, Mark and {Pakmor}, R{\"u}diger and {Nelson}, Dylan and {Marinacci}, Federico and {Hernquist}, Lars and {Weinberger}, Rainer and {Genel}, Shy},
    title = "{First results from the IllustrisTNG simulations: a tale of two elements - chemical evolution of magnesium and europium}",
    journal = {\mnras},
    year = 2018,
    month = jun,
    volume = {477},
    number = {1},
    pages = {1206-1224},
    doi = {10.1093/mnras/sty618}
}

@ARTICLE{Marinacci2018,
    author = {{Marinacci}, Federico and {Vogelsberger}, Mark and {Pakmor}, R{\"u}diger and {Torrey}, Paul and {Springel}, Volker and {Hernquist}, Lars and {Nelson}, Dylan and {Weinberger}, Rainer and {Pillepich}, Annalisa and {Naiman}, Jill and {Genel}, Shy},
    title = "{First results from the IllustrisTNG simulations: radio haloes and magnetic fields}",
    journal = {\mnras},
    year = 2018,
    month = nov,
    volume = {480},
    number = {4},
    pages = {5113-5139},
    doi = {10.1093/mnras/sty2206}
}

@ARTICLE{Pillepich2018b,
    author = {{Pillepich}, Annalisa and {Nelson}, Dylan and {Hernquist}, Lars and {Springel}, Volker and {Pakmor}, R{\"u}diger and {Torrey}, Paul and {Weinberger}, Rainer and {Genel}, Shy and {Naiman}, Jill P. and {Marinacci}, Federico and {Vogelsberger}, Mark},
    title = "{First results from the IllustrisTNG simulations: the stellar mass content of groups and clusters of galaxies}",
    journal = {\mnras},
    year = 2018,
    month = mar,
    volume = {475},
    number = {1},
    pages = {648-675},
    doi = {10.1093/mnras/stx3112}
}

@ARTICLE{Nelson2018,
    author = {{Nelson}, Dylan and {Pillepich}, Annalisa and {Springel}, Volker and {Weinberger}, Rainer and {Hernquist}, Lars and {Pakmor}, R{\"u}diger and {Genel}, Shy and {Torrey}, Paul and {Vogelsberger}, Mark and {Kauffmann}, Guinevere and {Marinacci}, Federico and {Naiman}, Jill},
    title = "{First results from the IllustrisTNG simulations: the galaxy colour bimodality}",
    journal = {\mnras},
    year = 2018,
    month = mar,
    volume = {475},
    number = {1},
    pages = {624-647},
    doi = {10.1093/mnras/stx3040}
}

@ARTICLE{Springel2018,
    author = {{Springel}, Volker and {Pakmor}, R{\"u}diger and {Pillepich}, Annalisa and {Weinberger}, Rainer and {Nelson}, Dylan and {Hernquist}, Lars and {Vogelsberger}, Mark and {Genel}, Shy and {Torrey}, Paul and {Marinacci}, Federico and {Naiman}, Jill},
    title = "{First results from the IllustrisTNG simulations: matter and galaxy clustering}",
    journal = {\mnras},
    year = 2018,
    month = mar,
    volume = {475},
    number = {1},
    pages = {676-698},
    doi = {10.1093/mnras/stx3304}
}

@ARTICLE{Planck2015,
    author = {{Planck Collaboration} and {Ade}, P.~A.~R. and {Aghanim}, N. and {Arnaud}, M. and {Ashdown}, M. and {Aumont}, J. and {Baccigalupi}, C. and {Banday}, A.~J. and {Barreiro}, R.~B. and {Bartlett}, J.~G. and {Bartolo}, N. and {Battaner}, E. and {Battye}, R. and {Benabed}, K. and {Beno{\^\i}t}, A. and {Benoit-L{\'e}vy}, A. and {Bernard}, J.-P. and {Bersanelli}, M. and {Bielewicz}, P. and {Bock}, J.~J. and {Bonaldi}, A. and {Bonavera}, L. and {Bond}, J.~R. and {Borrill}, J. and {Bouchet}, F.~R. and {Boulanger}, F. and {Bucher}, M. and {Burigana}, C. and {Butler}, R.~C. and {Calabrese}, E. and {Cardoso}, J.-F. and {Catalano}, A. and {Challinor}, A. and {Chamballu}, A. and {Chary}, R.-R. and {Chiang}, H.~C. and {Chluba}, J. and {Christensen}, P.~R. and {Church}, S. and {Clements}, D.~L. and {Colombi}, S. and {Colombo}, L.~P.~L. and {Combet}, C. and {Coulais}, A. and {Crill}, B.~P. and {Curto}, A. and {Cuttaia}, F. and {Danese}, L. and {Davies}, R.~D. and {Davis}, R.~J. and {de Bernardis}, P. and {de Rosa}, A. and {de Zotti}, G. and {Delabrouille}, J. and {D{\'e}sert}, F.-X. and {Di Valentino}, E. and {Dickinson}, C. and {Diego}, J.~M. and {Dolag}, K. and {Dole}, H. and {Donzelli}, S. and {Dor{\'e}}, O. and {Douspis}, M. and {Ducout}, A. and {Dunkley}, J. and {Dupac}, X. and {Efstathiou}, G. and {Elsner}, F. and {En{\ss}lin}, T.~A. and {Eriksen}, H.~K. and {Farhang}, M. and {Fergusson}, J. and {Finelli}, F. and {Forni}, O. and {Frailis}, M. and {Fraisse}, A.~A. and {Franceschi}, E. and {Frejsel}, A. and {Galeotta}, S. and {Galli}, S. and {Ganga}, K. and {Gauthier}, C. and {Gerbino}, M. and {Ghosh}, T. and {Giard}, M. and {Giraud-H{\'e}raud}, Y. and {Giusarma}, E. and {Gjerl{\o}w}, E. and {Gonz{\'a}lez-Nuevo}, J. and {G{\'o}rski}, K.~M. and {Gratton}, S. and {Gregorio}, A. and {Gruppuso}, A. and {Gudmundsson}, J.~E. and {Hamann}, J. and {Hansen}, F.~K. and {Hanson}, D. and {Harrison}, D.~L. and {Helou}, G. and {Henrot-Versill{\'e}}, S. and {Hern{\'a}ndez-Monteagudo}, C. and {Herranz}, D. and {Hildebrandt}, S.~R. and {Hivon}, E. and {Hobson}, M. and {Holmes}, W.~A. and {Hornstrup}, A. and {Hovest}, W. and {Huang}, Z. and {Huffenberger}, K.~M. and {Hurier}, G. and {Jaffe}, A.~H. and {Jaffe}, T.~R. and {Jones}, W.~C. and {Juvela}, M. and {Keih{\"a}nen}, E. and {Keskitalo}, R. and {Kisner}, T.~S. and {Kneissl}, R. and {Knoche}, J. and {Knox}, L. and {Kunz}, M. and {Kurki-Suonio}, H. and {Lagache}, G. and {L{\"a}hteenm{\"a}ki}, A. and {Lamarre}, J.-M. and {Lasenby}, A. and {Lattanzi}, M. and {Lawrence}, C.~R. and {Leahy}, J.~P. and {Leonardi}, R. and {Lesgourgues}, J. and {Levrier}, F. and {Lewis}, A. and {Liguori}, M. and {Lilje}, P.~B. and {Linden-V{\o}rnle}, M. and {L{\'o}pez-Caniego}, M. and {Lubin}, P.~M. and {Mac{\'\i}as-P{\'e}rez}, J.~F. and {Maggio}, G. and {Maino}, D. and {Mandolesi}, N. and {Mangilli}, A. and {Marchini}, A. and {Maris}, M. and {Martin}, P.~G. and {Martinelli}, M. and {Mart{\'\i}nez-Gonz{\'a}lez}, E. and {Masi}, S. and {Matarrese}, S. and {McGehee}, P. and {Meinhold}, P.~R. and {Melchiorri}, A. and {Melin}, J.-B. and {Mendes}, L. and {Mennella}, A. and {Migliaccio}, M. and {Millea}, M. and {Mitra}, S. and {Miville-Desch{\^e}nes}, M.-A. and {Moneti}, A. and {Montier}, L. and {Morgante}, G. and {Mortlock}, D. and {Moss}, A. and {Munshi}, D. and {Murphy}, J.~A. and {Naselsky}, P. and {Nati}, F. and {Natoli}, P. and {Netterfield}, C.~B. and {N{\o}rgaard-Nielsen}, H.~U. and {Noviello}, F. and {Novikov}, D. and {Novikov}, I. and {Oxborrow}, C.~A. and {Paci}, F. and {Pagano}, L. and {Pajot}, F. and {Paladini}, R. and {Paoletti}, D. and {Partridge}, B. and {Pasian}, F. and {Patanchon}, G. and {Pearson}, T.~J. and {Perdereau}, O. and {Perotto}, L. and {Perrotta}, F. and {Pettorino}, V. and {Piacentini}, F. and {Piat}, M. and {Pierpaoli}, E. and {Pietrobon}, D. and {Plaszczynski}, S. and {Pointecouteau}, E. and {Polenta}, G. and {Popa}, L. and {Pratt}, G.~W. and {Pr{\'e}zeau}, G.},
    title = "{Planck 2015 results. XIII. Cosmological parameters}",
    journal = {\aap},
    year = 2016,
    month = sep,
    volume = {594},
    eid = {A13},
    pages = {A13},
    doi = {10.1051/0004-6361/201525830}
}

@ARTICLE{Springel2001,
    author = {{Springel}, Volker and {White}, Simon D.~M. and {Tormen}, Giuseppe and {Kauffmann}, Guinevere},
    title = "{Populating a cluster of galaxies - I. Results at z=0}",
    journal = {\mnras},
    year = 2001,
    month = dec,
    volume = {328},
    number = {3},
    pages = {726-750},
    doi = {10.1046/j.1365-8711.2001.04912.x}
}

@ARTICLE{More2011,
    author = {{More}, Surhud and {Kravtsov}, Andrey V. and {Dalal}, Neal and {Gottl{\"o}ber}, Stefan},
    title = "{The Overdensity and Masses of the Friends-of-friends Halos and Universality of Halo Mass Function}",
    journal = {\apjs},
    year = 2011,
    month = jul,
    volume = {195},
    number = {1},
    eid = {4},
    pages = {4},
    doi = {10.1088/0067-0049/195/1/4}
}

@ARTICLE{Onions2012,
    author = {{Onions}, Julian and {Knebe}, Alexander and {Pearce}, Frazer R. and {Muldrew}, Stuart I. and {Lux}, Hanni and {Knollmann}, Steffen R. and {Ascasibar}, Yago and {Behroozi}, Peter and {Elahi}, Pascal and {Han}, Jiaxin and {Maciejewski}, Michal and {Merch{\'a}n}, Manuel E. and {Neyrinck}, Mark and {Ruiz}, Andr{\'e}s. N. and {Sgr{\'o}}, Mario A. and {Springel}, Volker and {Tweed}, Dylan},
    title = "{Subhaloes going Notts: the subhalo-finder comparison project}",
    journal = {\mnras},
    year = 2012,
    month = jun,
    volume = {423},
    number = {2},
    pages = {1200-1214},
    doi = {10.1111/j.1365-2966.2012.20947.x}
}

@ARTICLE{Press1982,
    author = {{Press}, W.~H. and {Davis}, M.},
    title = "{How to identify and weigh virialized clusters of galaxies in a complete redshift catalog}",
    journal = {\apj},
    year = 1982,
    month = aug,
    volume = {259},
    pages = {449-473},
    doi = {10.1086/160183}
}

@phdthesis{Amoura2023,
    author       = {Yuba Amoura},
    title        = {Cosmology with Cluster Structural Properties},
    year         = {2023},
    school       = {University of Waterloo},
    type         = {{PhD} thesis}
}

@ARTICLE{Schaye2015,
    author = {{Schaye}, Joop and {Crain}, Robert A. and {Bower}, Richard G. and {Furlong}, Michelle and {Schaller}, Matthieu and {Theuns}, Tom and {Dalla Vecchia}, Claudio and {Frenk}, Carlos S. and {McCarthy}, I.~G. and {Helly}, John C. and {Jenkins}, Adrian and {Rosas-Guevara}, Y.~M. and {White}, Simon D.~M. and {Baes}, Maarten and {Booth}, C.~M. and {Camps}, Peter and {Navarro}, Julio F. and {Qu}, Yan and {Rahmati}, Alireza and {Sawala}, Till and {Thomas}, Peter A. and {Trayford}, James},
    title = "{The EAGLE project: simulating the evolution and assembly of galaxies and their environments}",
    journal = {\mnras},
    year = 2015,
    month = jan,
    volume = {446},
    number = {1},
    pages = {521-554},
    doi = {10.1093/mnras/stu2058}
}

@ARTICLE{Engler2021,
    author = {{Engler}, Christoph and {Pillepich}, Annalisa and {Joshi}, Gandhali D. and {Nelson}, Dylan and {Pasquali}, Anna and {Grebel}, Eva K. and {Lisker}, Thorsten and {Zinger}, Elad and {Donnari}, Martina and {Marinacci}, Federico and {Vogelsberger}, Mark and {Hernquist}, Lars},
    title = "{The distinct stellar-to-halo mass relations of satellite and central galaxies: insights from the IllustrisTNG simulations}",
    journal = {\mnras},
    year = 2021,
    month = jan,
    volume = {500},
    number = {3},
    pages = {3957-3975},
    doi = {10.1093/mnras/staa3505}
}

@ARTICLE{Power2012,
    author = {{Power}, Chris and {Knebe}, Alexander and {Knollmann}, Steffen R.},
    title = "{The dynamical state of dark matter haloes in cosmological simulations - I. Correlations with mass assembly history}",
    journal = {\mnras},
    year = 2012,
    month = jan,
    volume = {419},
    number = {2},
    pages = {1576-1587},
    doi = {10.1111/j.1365-2966.2011.19820.x}
}

@ARTICLE{Maccio2007,
    author = {{Macci{\`o}}, Andrea V. and {Dutton}, Aaron A. and {van den Bosch}, Frank C. and {Moore}, Ben and {Potter}, Doug and {Stadel}, Joachim},
    title = "{Concentration, spin and shape of dark matter haloes: scatter and the dependence on mass and environment}",
    journal = {\mnras},
    year = 2007,
    month = jun,
    volume = {378},
    number = {1},
    pages = {55-71},
    doi = {10.1111/j.1365-2966.2007.11720.x}
}

@ARTICLE{Anbajagane2022,
    author = {{Anbajagane}, Dhayaa and {Evrard}, August E. and {Farahi}, Arya},
    title = "{Baryonic imprints on DM haloes: population statistics from dwarf galaxies to galaxy clusters}",
    journal = {\mnras},
    year = 2022,
    month = jan,
    volume = {509},
    number = {3},
    pages = {3441-3461},
    doi = {10.1093/mnras/stab3177}
}

@ARTICLE{Bett2012,
    author = {{Bett}, Philip},
    title = "{Halo shapes from weak lensing: the impact of galaxy-halo misalignment}",
    journal = {\mnras},
    year = 2012,
    month = mar,
    volume = {420},
    number = {4},
    pages = {3303-3323},
    doi = {10.1111/j.1365-2966.2011.20258.x}
}

@ARTICLE{Wong2012,
    author = {{Wong}, Anson W.~C. and {Taylor}, James E.},
    title = "{What Do Dark Matter Halo Properties Tell Us about Their Mass Assembly Histories?}",
    journal = {\apj},
    year = 2012,
    month = sep,
    volume = {757},
    number = {1},
    eid = {102},
    pages = {102},
    doi = {10.1088/0004-637X/757/1/102}
}

@ARTICLE{Navarro1996,
    author = {{Navarro}, Julio F. and {Frenk}, Carlos S. and {White}, Simon D.~M.},
    title = "{The Structure of Cold Dark Matter Halos}",
    journal = {\apj},
    year = 1996,
    month = may,
    volume = {462},
    pages = {563},
    doi = {10.1086/177173}
}

@ARTICLE{Navarro1997,
    author = {{Navarro}, Julio F. and {Frenk}, Carlos S. and {White}, Simon D.~M.},
    title = "{A Universal Density Profile from Hierarchical Clustering}",
    journal = {\apj},
    year = 1997,
    month = dec,
    volume = {490},
    number = {2},
    pages = {493-508},
    doi = {10.1086/304888}
}

@ARTICLE{Conselice1997,
    author = {{Conselice}, C.~J.},
    title = "{The Symmetry, Color, and Morphology of Galaxies}",
    journal = {\pasp},
    year = 1997,
    month = nov,
    volume = {109},
    pages = {1251-1255},
    doi = {10.1086/134004}
}

@ARTICLE{Dariush2010,
    author = {{Dariush}, Ali A. and {Raychaudhury}, Somak and {Ponman}, Trevor J. and {Khosroshahi}, Habib G. and {Benson}, Andrew J. and {Bower}, Richard G. and {Pearce}, Frazer},
    title = "{The mass assembly of galaxy groups and the evolution of the magnitude gap}",
    journal = {\mnras},
    year = 2010,
    month = jul,
    volume = {405},
    number = {3},
    pages = {1873-1887},
    doi = {10.1111/j.1365-2966.2010.16569.x},
}

@ARTICLE{More2012,
    author = {Surhud More},
    title = {MAGNITUDE GAP STATISTICS AND THE CONDITIONAL LUMINOSITY FUNCTION},
    journal = {\apj},
    year = {2012},
    month = {dec},
    publisher = {The American Astronomical Society},
    volume = {761},
    number = {2},
    pages = {127},
    doi = {10.1088/0004-637X/761/2/127}
}

@article{Jones2003,
    author = {{Jones}, L.~R. and {Ponman}, T.~J. and {Horton}, A. and {Babul}, A. and {Ebeling}, H. and {Burke}, D.~J.},
    title = "{The nature and space density of fossil groups of galaxies}",
    journal = {\mnras},
    year = 2003,
    month = aug,
    volume = {343},
    number = {2},
    pages = {627-638},
    doi = {10.1046/j.1365-8711.2003.06702.x},
}

@ARTICLE{Golden-Marx2025,
    author = {{Golden-Marx}, Jesse B. and {Zhang}, Y. and {Ogando}, R.~L.~C. and {Yanny}, B. and {da Silva Pereira}, M.~E. and {Hilton}, M. and {Aguena}, M. and {Allam}, S. and {Andrade-Oliveira}, F. and {Bacon}, D. and {Brooks}, D. and {Carnero Rosell}, A. and {Carretero}, J. and {Cheng}, T.-Y. and {da Costa}, L.~N. and {De Vicente}, J. and {Desai}, S. and {Doel}, P. and {Everett}, S. and {Ferrero}, I. and {Frieman}, J. and {Garc{\'\i}a-Bellido}, J. and {Gatti}, M. and {Giannini}, G. and {Gruen}, D. and {Gruendl}, R.~A. and {Gutierrez}, G. and {Hinton}, S.~R. and {Hollowood}, D.~L. and {Honscheid}, K. and {James}, D.~J. and {Kuehn}, K. and {Lee}, S. and {Mena-Fern{\'a}ndez}, J. and {Menanteau}, F. and {Miquel}, R. and {Mohr}, J. and {Palmese}, A. and {Pieres}, A. and {Plazas Malag{\'o}n}, A.~A. and {Samuroff}, S. and {Sanchez}, E. and {Schubnell}, M. and {Sevilla-Noarbe}, I. and {Smith}, M. and {Suchyta}, E. and {Tarle}, G. and {Vikram}, V. and {Walker}, A.~R. and {Weaverdyck}, N. and {Wiseman}, P.},
    title = "{The hierarchical growth of bright central galaxies and intracluster light as traced by the magnitude gap}",
    journal = {\mnras},
    year = 2025,
    month = apr,
    volume = {538},
    number = {2},
    pages = {622-638},
    doi = {10.1093/mnras/staf277}
}

@BOOK{Mo2010,
    author = {{Mo}, Houjun and {van den Bosch}, Frank C. and {White}, Simon},
    title = "{Galaxy Formation and Evolution}",
    year = 2010,
    publisher = {Cambridge University Press},
    doi = {10.1017/CBO9780511807244}
}

@ARTICLE{Paranjape2015,
    author = {{Paranjape}, Aseem and {Kova{\v{c}}}, Katarina and {Hartley}, William G. and {Pahwa}, Isha},
    title = "{Correlating galaxy colour and halo concentration: a tunable halo model of galactic conformity}",
    journal = {\mnras},
    year = 2015,
    month = dec,
    volume = {454},
    number = {3},
    pages = {3030-3048},
    doi = {10.1093/mnras/stv2137}
}

@ARTICLE{Ludlow2012,
    author = {{Ludlow}, Aaron D. and {Navarro}, Julio F. and {Li}, Ming and {Angulo}, Raul E. and {Boylan-Kolchin}, Michael and {Bett}, Philip E.},
    title = "{The dynamical state and mass-concentration relation of galaxy clusters}",
    journal = {\mnras},
    year = 2012,
    month = dec,
    volume = {427},
    number = {2},
    pages = {1322-1328},
    doi = {10.1111/j.1365-2966.2012.21892.x}
}

@ARTICLE{Lee2018,
    author = {{Lee}, Christoph T. and {Primack}, Joel R. and {Behroozi}, Peter and {Rodr{\'\i}guez-Puebla}, Aldo and {Hellinger}, Doug and {Dekel}, Avishai},
    title = "{Tidal stripping and post-merger relaxation of dark matter haloes: causes and consequences of mass-loss}",
    journal = {\mnras},
    year = 2018,
    month = dec,
    volume = {481},
    number = {3},
    pages = {4038-4057},
    doi = {10.1093/mnras/sty2538}
}

@ARTICLE{Wang2020,
    author = {{Wang}, Kuan and {Mao}, Yao-Yuan and {Zentner}, Andrew R. and {Lange}, Johannes U. and {van den Bosch}, Frank C. and {Wechsler}, Risa H.},
    title = "{Concentrations of dark haloes emerge from their merger histories}",
    journal = {\mnras},
    year = 2020,
    month = nov,
    volume = {498},
    number = {3},
    pages = {4450-4464},
    doi = {10.1093/mnras/staa2733}
}

@ARTICLE{Farahi2020,
    author = {{Farahi}, Arya and {Ho}, Matthew and {Trac}, Hy},
    title = "{Aging haloes: implications of the magnitude gap on conditional statistics of stellar and gas properties of massive haloes}",
    journal = {\mnras},
    year = 2020,
    month = mar,
    volume = {493},
    number = {1},
    pages = {1361-1374},
    doi = {10.1093/mnras/staa291}
}

@ARTICLE{Andreon2010,
    author = {{Andreon}, S. and {Hurn}, M.~A.},
    title = "{The scaling relation between richness and mass of galaxy clusters: a Bayesian approach}",
    journal = {\mnras},
    year = 2010,
    month = jun,
    volume = {404},
    number = {4},
    pages = {1922-1937},
    doi = {10.1111/j.1365-2966.2010.16406.x}
}

@ARTICLE{Allen2011,
       author = {{Allen}, Steven W. and {Evrard}, August E. and {Mantz}, Adam B.},
        title = "{Cosmological Parameters from Observations of Galaxy Clusters}",
      journal = {\araa},
         year = 2011,
        month = sep,
       volume = {49},
       number = {1},
        pages = {409-470},
          doi = {10.1146/annurev-astro-081710-102514},
archivePrefix = {arXiv},
       eprint = {1103.4829},
 primaryClass = {astro-ph.CO},
       adsurl = {https://ui.adsabs.harvard.edu/abs/2011ARA&A..49..409A}
}

@ARTICLE{Ghirardini2024,
       author = {{Ghirardini}, V. and {Bulbul}, E. and {Artis}, E. and {Clerc}, N. and {Garrel}, C. and {Grandis}, S. and {Kluge}, M. and {Liu}, A. and {Bahar}, Y.~E. and {Balzer}, F. and {Chiu}, I. and {Comparat}, J. and {Gruen}, D. and {Kleinebreil}, F. and {Krippendorf}, S. and {Merloni}, A. and {Nandra}, K. and {Okabe}, N. and {Pacaud}, F. and {Predehl}, P. and {Ramos-Ceja}, M.~E. and {Reiprich}, T.~H. and {Sanders}, J.~S. and {Schrabback}, T. and {Seppi}, R. and {Zelmer}, S. and {Zhang}, X. and {Bornemann}, W. and {Brunner}, H. and {Burwitz}, V. and {Coutinho}, D. and {Dennerl}, K. and {Freyberg}, M. and {Friedrich}, S. and {Gaida}, R. and {Gueguen}, A. and {Haberl}, F. and {Kink}, W. and {Lamer}, G. and {Li}, X. and {Liu}, T. and {Maitra}, C. and {Meidinger}, N. and {Mueller}, S. and {Miyatake}, H. and {Miyazaki}, S. and {Robrade}, J. and {Schwope}, A. and {Stewart}, I.},
        title = "{The SRG/eROSITA all-sky survey: Cosmology constraints from cluster abundances in the western Galactic hemisphere}",
      journal = {\aap},
         year = 2024,
        month = sep,
       volume = {689},
          eid = {A298},
        pages = {A298},
          doi = {10.1051/0004-6361/202348852},
archivePrefix = {arXiv},
       eprint = {2402.08458},
 primaryClass = {astro-ph.CO},
       adsurl = {https://ui.adsabs.harvard.edu/abs/2024A&A...689A.298G}
}

@INCOLLECTION{Lovisari2022,
       author = {{Lovisari}, Lorenzo and {Maughan}, Ben J.},
        title = "{Scaling Relations of Clusters and Groups and Their Evolution}",
    booktitle = {Handbook of X-ray and Gamma-ray Astrophysics},
    publisher = {Springer},
         year = 2022,
       editor = {{Bambi}, Cosimo and {Sangangelo}, Andrea},
          eid = {65},
        pages = {65},
          doi = {10.1007/978-981-16-4544-0_118-1},
       adsurl = {https://ui.adsabs.harvard.edu/abs/2022hxga.book...65L}
}

@ARTICLE{Giodini2013,
       author = {{Giodini}, S. and {Lovisari}, L. and {Pointecouteau}, E. and {Ettori}, S. and {Reiprich}, T.~H. and {Hoekstra}, H.},
        title = "{Scaling Relations for Galaxy Clusters: Properties and Evolution}",
      journal = {\ssr},
         year = 2013,
        month = aug,
       volume = {177},
       number = {1-4},
        pages = {247-282},
          doi = {10.1007/s11214-013-9994-5},
archivePrefix = {arXiv},
       eprint = {1305.3286},
 primaryClass = {astro-ph.CO},
       adsurl = {https://ui.adsabs.harvard.edu/abs/2013SSRv..177..247G}
}

@ARTICLE{Haggar2024b,
       author = {{Haggar}, Roan and {Amoura}, Yuba and {Mpetha}, Charlie T. and {Taylor}, James E. and {Walker}, Kris and {Power}, Chris},
        title = "{Constraining Cosmological Parameters Using the Splashback Radius of Galaxy Clusters}",
      journal = {\apj},
         year = 2024,
        month = sep,
       volume = {972},
       number = {1},
          eid = {28},
        pages = {28},
          doi = {10.3847/1538-4357/ad5cee},
archivePrefix = {arXiv},
       eprint = {2406.17849},
 primaryClass = {astro-ph.CO},
       adsurl = {https://ui.adsabs.harvard.edu/abs/2024ApJ...972...28H}
}

@ARTICLE{Mpetha2024,
       author = {{Mpetha}, C.~T. and {Taylor}, J.~E. and {Amoura}, Y. and {Haggar}, R.},
        title = "{The infall region as a complementary probe to cluster abundance}",
      journal = {\mnras},
         year = 2024,
        month = aug,
       volume = {532},
       number = {2},
        pages = {2521-2533},
          doi = {10.1093/mnras/stae1637},
archivePrefix = {arXiv},
       eprint = {2407.01661},
 primaryClass = {astro-ph.CO},
       adsurl = {https://ui.adsabs.harvard.edu/abs/2024MNRAS.532.2521M}
}

@ARTICLE{Richstone1992,
       author = {{Richstone}, D. and {Loeb}, A. and {Turner}, E.~L.},
        title = "{A Lower Limit on the Cosmic Mean Density from the Ages of Clusters of Galaxies}",
      journal = {\apj},
         year = 1992,
        month = jul,
       volume = {393},
        pages = {477},
          doi = {10.1086/171521},
       adsurl = {https://ui.adsabs.harvard.edu/abs/1992ApJ...393..477R}
}

@ARTICLE{Evrard1993,
       author = {{Evrard}, August E. and {Mohr}, Joseph J. and {Fabricant}, Daniel G. and {Geller}, Margaret J.},
        title = "{A Morphology-Cosmology Connection for X-Ray Clusters}",
      journal = {\apjl},
         year = 1993,
        month = dec,
       volume = {419},
        pages = {L9},
          doi = {10.1086/187124},
archivePrefix = {arXiv},
       eprint = {astro-ph/9310002},
 primaryClass = {astro-ph},
       adsurl = {https://ui.adsabs.harvard.edu/abs/1993ApJ...419L...9E}
}

@ARTICLE{Mohr1995,
       author = {{Mohr}, Joseph J. and {Evrard}, August E. and {Fabricant}, Daniel G. and {Geller}, Margaret J.},
        title = "{Cosmological Constraints from Observed Cluster X-Ray Morphologies}",
      journal = {\apj},
         year = 1995,
        month = jul,
       volume = {447},
        pages = {8},
          doi = {10.1086/175852},
archivePrefix = {arXiv},
       eprint = {astro-ph/9501011},
 primaryClass = {astro-ph},
       adsurl = {https://ui.adsabs.harvard.edu/abs/1995ApJ...447....8M}
}

@ARTICLE{Karim2025,
       author = {{Karim}, Tanveer and {Singh}, Sukhdeep and {Rezaie}, Mehdi and {Eisenstein}, Daniel and {Hadzhiyska}, Boryana and {Speagle}, Joshua S. and {Aguilar}, Jessica Nicole and {Ahlen}, Steven and {Brooks}, David and {Claybaugh}, Todd and {de la Macorra}, Axel and {Ferraro}, Simone and {Forero-Romero}, Jaime E. and {Gazta{\~n}aga}, Enrique and {Gontcho}, Satya Gontcho A. and {Gutierrez}, Gaston and {Guy}, Julien and {Honscheid}, Klaus and {Juneau}, Stephanie and {Kirkby}, David and {Krolewski}, Alex and {Lambert}, Andrew and {Landriau}, Martin and {Levi}, Michael and {Meisner}, Aaron and {Miquel}, Ramon and {Moustakas}, John and {Mu{\~n}oz-Guti{\'e}rrez}, Andrea and {Myers}, Adam and {Niz}, Gustavo and {Palanque-Delabrouille}, Nathalie and {Percival}, Will and {Prada}, Francisco and {Rossi}, Graziano and {Sanchez}, Eusebio and {Schlafly}, Edward and {Schlegel}, David and {Schubnell}, Michael and {Sprayberry}, David and {Tarl{\'e}}, Gregory and {Weaver}, Benjamin Alan and {Zou}, Hu},
        title = "{Measuring {\ensuremath{\sigma}} $_{8}$ using DESI Legacy Imaging Surveys Emission-Line galaxies and Planck CMB lensing, and the impact of dust on parameter inference}",
      journal = {\jcap},
         year = 2025,
        month = feb,
       volume = {2025},
       number = {2},
          eid = {045},
        pages = {045},
          doi = {10.1088/1475-7516/2025/02/045},
archivePrefix = {arXiv},
       eprint = {2408.15909},
 primaryClass = {astro-ph.CO},
       adsurl = {https://ui.adsabs.harvard.edu/abs/2025JCAP...02..045K}
}

@ARTICLE{Lesci2022,
       author = {{Lesci}, G.~F. and {Marulli}, F. and {Moscardini}, L. and {Sereno}, M. and {Veropalumbo}, A. and {Maturi}, M. and {Giocoli}, C. and {Radovich}, M. and {Bellagamba}, F. and {Roncarelli}, M. and {Bardelli}, S. and {Contarini}, S. and {Covone}, G. and {Ingoglia}, L. and {Nanni}, L. and {Puddu}, E.},
        title = "{AMICO galaxy clusters in KiDS-DR3: Cosmological constraints from counts and stacked weak lensing}",
      journal = {\aap},
         year = 2022,
        month = mar,
       volume = {659},
          eid = {A88},
        pages = {A88},
          doi = {10.1051/0004-6361/202040194},
archivePrefix = {arXiv},
       eprint = {2012.12273},
 primaryClass = {astro-ph.CO},
       adsurl = {https://ui.adsabs.harvard.edu/abs/2022A&A...659A..88L}
}

@ARTICLE{Aymerich2024,
       author = {{Aymerich}, G. and {Douspis}, M. and {Pratt}, G.~W. and {Salvati}, L. and {Soubri{\'e}}, E. and {Andrade-Santos}, F. and {Forman}, W.~R. and {Jones}, C. and {Aghanim}, N. and {Kraft}, R. and {van Weeren}, R.~J.},
        title = "{Cosmological constraints from the Planck cluster catalogue with new multi-wavelength mass calibration from Chandra and CFHT}",
      journal = {\aap},
         year = 2024,
        month = oct,
       volume = {690},
          eid = {A238},
        pages = {A238},
          doi = {10.1051/0004-6361/202449513},
archivePrefix = {arXiv},
       eprint = {2402.04006},
 primaryClass = {astro-ph.CO},
       adsurl = {https://ui.adsabs.harvard.edu/abs/2024A&A...690A.238A}
}

@ARTICLE{Abdalla2022,
       author = {{Abdalla}, Elcio and {Abell{\'a}n}, Guillermo Franco and {Aboubrahim}, Amin and {Agnello}, Adriano and {Akarsu}, {\"O}zg{\"u}r and {Akrami}, Yashar and {Alestas}, George and {Aloni}, Daniel and {Amendola}, Luca and {Anchordoqui}, Luis A. and {Anderson}, Richard I. and {Arendse}, Nikki and {Asgari}, Marika and {Ballardini}, Mario and {Barger}, Vernon and {Basilakos}, Spyros and {Batista}, Ronaldo C. and {Battistelli}, Elia S. and {Battye}, Richard and {Benetti}, Micol and {Benisty}, David and {Berlin}, Asher and {de Bernardis}, Paolo and {Berti}, Emanuele and {Bidenko}, Bohdan and {Birrer}, Simon and {Blakeslee}, John P. and {Boddy}, Kimberly K. and {Bom}, Clecio R. and {Bonilla}, Alexander and {Borghi}, Nicola and {Bouchet}, Fran{\c{c}}ois R. and {Braglia}, Matteo and {Buchert}, Thomas and {Buckley-Geer}, Elizabeth and {Calabrese}, Erminia and {Caldwell}, Robert R. and {Camarena}, David and {Capozziello}, Salvatore and {Casertano}, Stefano and {Chen}, Geoff C.-F. and {Chluba}, Jens and {Chen}, Angela and {Chen}, Hsin-Yu and {Chudaykin}, Anton and {Cicoli}, Michele and {Copi}, Craig J. and {Courbin}, Fred and {Cyr-Racine}, Francis-Yan and {Czerny}, Bo{\.z}ena and {Dainotti}, Maria and {D'Amico}, Guido and {Davis}, Anne-Christine and {de Cruz P{\'e}rez}, Javier and {de Haro}, Jaume and {Delabrouille}, Jacques and {Denton}, Peter B. and {Dhawan}, Suhail and {Dienes}, Keith R. and {Di Valentino}, Eleonora and {Du}, Pu and {Eckert}, Dominique and {Escamilla-Rivera}, Celia and {Fert{\'e}}, Agn{\`e}s and {Finelli}, Fabio and {Fosalba}, Pablo and {Freedman}, Wendy L. and {Frusciante}, Noemi and {Gazta{\~n}aga}, Enrique and {Giar{\`e}}, William and {Giusarma}, Elena and {G{\'o}mez-Valent}, Adri{\`a} and {Handley}, Will and {Harrison}, Ian and {Hart}, Luke and {Hazra}, Dhiraj Kumar and {Heavens}, Alan and {Heinesen}, Asta and {Hildebrandt}, Hendrik and {Hill}, J. Colin and {Hogg}, Natalie B. and {Holz}, Daniel E. and {Hooper}, Deanna C. and {Hosseininejad}, Nikoo and {Huterer}, Dragan and {Ishak}, Mustapha and {Ivanov}, Mikhail M. and {Jaffe}, Andrew H. and {Jang}, In Sung and {Jedamzik}, Karsten and {Jimenez}, Raul and {Joseph}, Melissa and {Joudaki}, Shahab and {Kamionkowski}, Marc and {Karwal}, Tanvi and {Kazantzidis}, Lavrentios and {Keeley}, Ryan E. and {Klasen}, Michael and {Komatsu}, Eiichiro and {Koopmans}, L{\'e}on V.~E. and {Kumar}, Suresh and {Lamagna}, Luca and {Lazkoz}, Ruth and {Lee}, Chung-Chi and {Lesgourgues}, Julien and {Levi Said}, Jackson and {Lewis}, Tiffany R. and {L'Huillier}, Benjamin and {Lucca}, Matteo and {Maartens}, Roy and {Macri}, Lucas M. and {Marfatia}, Danny and {Marra}, Valerio and {Martins}, Carlos J.~A.~P. and {Masi}, Silvia and {Matarrese}, Sabino and {Mazumdar}, Arindam and {Melchiorri}, Alessandro and {Mena}, Olga and {Mersini-Houghton}, Laura and {Mertens}, James and {Milakovi{\'c}}, Dinko and {Minami}, Yuto and {Miranda}, Vivian and {Moreno-Pulido}, Cristian and {Moresco}, Michele and {Mota}, David F. and {Mottola}, Emil and {Mozzon}, Simone and {Muir}, Jessica and {Mukherjee}, Ankan and {Mukherjee}, Suvodip and {Naselsky}, Pavel and {Nath}, Pran and {Nesseris}, Savvas and {Niedermann}, Florian and {Notari}, Alessio and {Nunes}, Rafael C. and {{\'O} Colg{\'a}in}, Eoin and {Owens}, Kayla A. and {{\"O}z{\"u}lker}, Emre and {Pace}, Francesco and {Paliathanasis}, Andronikos and {Palmese}, Antonella and {Pan}, Supriya and {Paoletti}, Daniela and {Perez Bergliaffa}, Santiago E. and {Perivolaropoulos}, Leandros and {Pesce}, Dominic W. and {Pettorino}, Valeria and {Philcox}, Oliver H.~E. and {Pogosian}, Levon and {Poulin}, Vivian and {Poulot}, Gaspard and {Raveri}, Marco and {Reid}, Mark J. and {Renzi}, Fabrizio and {Riess}, Adam G. and {Sabla}, Vivian I. and {Salucci}, Paolo and {Salzano}, Vincenzo and {Saridakis}, Emmanuel N. and {Sathyaprakash}, Bangalore S. and {Schmaltz}, Martin and {Sch{\"o}neberg}, Nils and {Scolnic}, Dan and {Sen}, Anjan A. and {Sehgal}, Neelima and {Shafieloo}, Arman and {Sheikh-Jabbari}, M.~M. and {Silk}, Joseph and {Silvestri}, Alessandra and {Skara}, Foteini and {Sloth}, Martin S. and {Soares-Santos}, Marcelle and {Sol{\`a} Peracaula}, Joan and {Songsheng}, Yu-Yang and {Soriano}, Jorge F. and {Staicova}, Denitsa and {Starkman}, Glenn D. and {Szapudi}, Istv{\'a}n and {Teixeira}, Elsa M. and {Thomas}, Brooks and {Treu}, Tommaso and {Trott}, Emery and {van de Bruck}, Carsten and {Vazquez}, J. Alberto and {Verde}, Licia and {Visinelli}, Luca and {Wang}, Deng and {Wang}, Jian-Min and {Wang}, Shao-Jiang and {Watkins}, Richard and {Watson}, Scott and {Webb}, John K. and {Weiner}, Neal and {Weltman}, Amanda and {Witte}, Samuel J. and {Wojtak}, Rados{\l}aw and {Yadav}, Anil Kumar},
        title = "{Cosmology intertwined: A review of the particle physics, astrophysics, and cosmology associated with the cosmological tensions and anomalies}",
      journal = {Journal of High Energy Astrophysics},
         year = 2022,
        month = jun,
       volume = {34},
        pages = {49-211},
          doi = {10.1016/j.jheap.2022.04.002},
archivePrefix = {arXiv},
       eprint = {2203.06142},
 primaryClass = {astro-ph.CO},
       adsurl = {https://ui.adsabs.harvard.edu/abs/2022JHEAp..34...49A}
}

@ARTICLE{Kravtsov2012,
       author = {{Kravtsov}, Andrey V. and {Borgani}, Stefano},
        title = "{Formation of Galaxy Clusters}",
      journal = {\araa},
         year = 2012,
        month = sep,
       volume = {50},
        pages = {353-409},
          doi = {10.1146/annurev-astro-081811-125502},
archivePrefix = {arXiv},
       eprint = {1205.5556},
 primaryClass = {astro-ph.CO},
       adsurl = {https://ui.adsabs.harvard.edu/abs/2012ARA&A..50..353K}
}

@ARTICLE{Kaiser1986,
       author = {{Kaiser}, N.},
        title = "{Evolution and clustering of rich clusters.}",
      journal = {\mnras},
         year = 1986,
        month = sep,
       volume = {222},
        pages = {323-345},
          doi = {10.1093/mnras/222.2.323},
       adsurl = {https://ui.adsabs.harvard.edu/abs/1986MNRAS.222..323K}
}

@ARTICLE{Jeon2022,
       author = {{Jeon}, Seyoung and {Yi}, Sukyoung K. and {Dubois}, Yohan and {Chung}, Aeree and {Devriendt}, Julien and {Han}, San and {Jackson}, Ryan A. and {Kimm}, Taysun and {Pichon}, Christophe and {Rhee}, Jinsu},
        title = "{Star Formation History and Transition Epoch of Cluster Galaxies Based on the Horizon-AGN Simulation}",
      journal = {\apj},
         year = 2022,
        month = dec,
       volume = {941},
       number = {1},
          eid = {5},
        pages = {5},
          doi = {10.3847/1538-4357/ac9d8c},
archivePrefix = {arXiv},
       eprint = {2210.05285},
 primaryClass = {astro-ph.GA},
       adsurl = {https://ui.adsabs.harvard.edu/abs/2022ApJ...941....5J}
}

@ARTICLE{Oxland2024,
       author = {{Oxland}, M. and {Parker}, L.~C. and {de Carvalho}, R.~R. and {Sampaio}, V.~M.},
        title = "{Satellite quenching and morphological transformation of galaxies in groups and clusters}",
      journal = {\mnras},
         year = 2024,
        month = apr,
       volume = {529},
       number = {4},
        pages = {3651-3665},
          doi = {10.1093/mnras/stae747},
archivePrefix = {arXiv},
       eprint = {2403.07742},
 primaryClass = {astro-ph.GA},
       adsurl = {https://ui.adsabs.harvard.edu/abs/2024MNRAS.529.3651O}
}

@ARTICLE{Eckert2022,
       author = {{Eckert}, D. and {Ettori}, S. and {Robertson}, A. and {Massey}, R. and {Pointecouteau}, E. and {Harvey}, D. and {McCarthy}, I.~G.},
        title = "{Constraints on dark matter self-interaction from the internal density profiles of X-COP galaxy clusters}",
      journal = {\aap},
         year = 2022,
        month = oct,
       volume = {666},
          eid = {A41},
        pages = {A41},
          doi = {10.1051/0004-6361/202243205},
archivePrefix = {arXiv},
       eprint = {2205.01123},
 primaryClass = {astro-ph.CO},
       adsurl = {https://ui.adsabs.harvard.edu/abs/2022A&A...666A..41E}
}

@ARTICLE{Vogt2024,
       author = {{Vogt}, Sophie M.~L. and {Bocquet}, Sebastian and {Davies}, Christopher T. and {Mohr}, Joseph J. and {Schmidt}, Fabian},
        title = "{Constraining f (R ) gravity using future galaxy cluster abundance and weak-lensing mass calibration datasets}",
      journal = {\prd},
         year = 2024,
        month = jun,
       volume = {109},
       number = {12},
          eid = {123503},
        pages = {123503},
          doi = {10.1103/PhysRevD.109.123503},
archivePrefix = {arXiv},
       eprint = {2401.09959},
 primaryClass = {astro-ph.CO},
       adsurl = {https://ui.adsabs.harvard.edu/abs/2024PhRvD.109l3503V}
}

@ARTICLE{Roche2024,
       author = {{Roche}, Cian and {McDonald}, Michael and {Borrow}, Josh and {Vogelsberger}, Mark and {Shen}, Xuejian and {Springel}, Volker and {Hernquist}, Lars and {Pakmor}, Ruediger and {Bose}, Sownak and {Kannan}, Rahul},
        title = "{Brightest Cluster Galaxy Offsets in Cold Dark Matter}",
      journal = {The Open Journal of Astrophysics},
         year = 2024,
        month = aug,
       volume = {7},
          eid = {65},
        pages = {65},
          doi = {10.33232/001c.122309},
archivePrefix = {arXiv},
       eprint = {2402.00928},
 primaryClass = {astro-ph.GA},
       adsurl = {https://ui.adsabs.harvard.edu/abs/2024OJAp....7E..65R}
}

@ARTICLE{Butt2025,
       author = {{Butt}, Minahil Adil and {Haridasu}, Sandeep and {Diaferio}, Antonaldo and {Benetti}, Francesco and {Boumechta}, Yacer and {Baccigalupi}, Carlo and {Lapi}, Andrea},
        title = "{Outer regions of galaxy clusters as a new probe to test modifications to gravity}",
      journal = {arXiv e-prints},
         year = 2025,
        month = apr,
          eid = {arXiv:2504.16685},
        pages = {arXiv:2504.16685},
          doi = {10.48550/arXiv.2504.16685},
archivePrefix = {arXiv},
       eprint = {2504.16685},
 primaryClass = {astro-ph.CO},
       adsurl = {https://ui.adsabs.harvard.edu/abs/2025arXiv250416685B}
}

@ARTICLE{Chen2025,
       author = {{Chen}, Mingjing and {Fang}, Wenjuan and {Zhang}, Yufei and {Wen}, Zhonglue and {Cui}, Weiguang},
        title = "{Constraining the neutrino mass with the CSST galaxy clusters}",
      journal = {\prd},
         year = 2025,
        month = sep,
       volume = {112},
       number = {6},
          eid = {063506},
        pages = {063506},
          doi = {10.1103/t821-6gr8},
archivePrefix = {arXiv},
       eprint = {2411.02752},
 primaryClass = {astro-ph.CO},
       adsurl = {https://ui.adsabs.harvard.edu/abs/2025PhRvD.112f3506C}
}

@ARTICLE{Supanitsky2026,
       author = {{Supanitsky}, A.~D. and {Nuza}, S.~E.},
        title = "{Exploring the role of accretion shocks in galaxy clusters as sources of ultrahigh-energy cosmic rays}",
      journal = {\prd},
         year = 2026,
        month = feb,
       volume = {113},
       number = {4},
          eid = {043008},
        pages = {043008},
          doi = {10.1103/xz32-7pk7},
archivePrefix = {arXiv},
       eprint = {2601.18411},
 primaryClass = {astro-ph.HE},
       adsurl = {https://ui.adsabs.harvard.edu/abs/2026PhRvD.113d3008S}
}

@ARTICLE{Smith2008,
       author = {{Smith}, Graham P. and {Taylor}, James E.},
        title = "{Connecting Substructure in Galaxy Cluster Cores at z = 0.2 with Cluster Assembly Histories}",
      journal = {\apjl},
         year = 2008,
        month = aug,
       volume = {682},
       number = {2},
        pages = {L73},
          doi = {10.1086/591271},
archivePrefix = {arXiv},
       eprint = {0806.3981},
 primaryClass = {astro-ph},
       adsurl = {https://ui.adsabs.harvard.edu/abs/2008ApJ...682L..73S}
}

@ARTICLE{Drakos2019a,
       author = {{Drakos}, Nicole E. and {Taylor}, James E. and {Berrouet}, Anael and {Robotham}, Aaron S.~G. and {Power}, Chris},
        title = "{Major mergers between dark matter haloes - I. Predictions for size, shape, and spin}",
      journal = {\mnras},
         year = 2019,
        month = jul,
       volume = {487},
       number = {1},
        pages = {993-1007},
          doi = {10.1093/mnras/stz1306},
archivePrefix = {arXiv},
       eprint = {1811.12839},
 primaryClass = {astro-ph.GA},
       adsurl = {https://ui.adsabs.harvard.edu/abs/2019MNRAS.487..993D}
}

@ARTICLE{Drakos2019b,
       author = {{Drakos}, Nicole E. and {Taylor}, James E. and {Berrouet}, Anael and {Robotham}, Aaron S.~G. and {Power}, Chris},
        title = "{Major mergers between dark matter haloes - II. Profile and concentration changes}",
      journal = {\mnras},
         year = 2019,
        month = jul,
       volume = {487},
       number = {1},
        pages = {1008-1024},
          doi = {10.1093/mnras/stz1307},
archivePrefix = {arXiv},
       eprint = {1811.12844},
 primaryClass = {astro-ph.GA},
       adsurl = {https://ui.adsabs.harvard.edu/abs/2019MNRAS.487.1008D}
}

@ARTICLE{Valles2025,
       author = {{Vall{\'e}s-P{\'e}rez}, David and {Planelles}, Susana and {Quilis}, Vicent},
        title = "{The eventful life journey of galaxy clusters: I. Impact of dark matter halo and intracluster medium properties on their full assembly histories}",
      journal = {\aap},
         year = 2025,
        month = jul,
       volume = {699},
          eid = {A1},
        pages = {A1},
          doi = {10.1051/0004-6361/202453483},
archivePrefix = {arXiv},
       eprint = {2505.02891},
 primaryClass = {astro-ph.CO},
       adsurl = {https://ui.adsabs.harvard.edu/abs/2025A&A...699A...1V}
}

@ARTICLE{Valles2026,
       author = {{Vall{\'e}s-P{\'e}rez}, David and {Planelles}, Susana and {Quilis}, Vicent},
        title = "{The eventful life journey of galaxy clusters: II. Impact of mass accretion on the thermodynamic structure of the ICM}",
      journal = {\aap},
         year = 2026,
        month = jan,
       volume = {706},
          eid = {A23},
        pages = {A23},
          doi = {10.1051/0004-6361/202557496},
archivePrefix = {arXiv},
       eprint = {2512.09023},
 primaryClass = {astro-ph.CO},
       adsurl = {https://ui.adsabs.harvard.edu/abs/2026A&A...706A..23V}
}

@ARTICLE{Wang2025,
       author = {{Wang}, Kuan and {Mansfield}, Philip and {Anbajagane}, Dhayaa and {Avestruz}, Camille},
        title = "{Merger Response of Halo Anisotropy Properties}",
      journal = {\apj},
         year = 2025,
        month = feb,
       volume = {979},
       number = {2},
          eid = {223},
        pages = {223},
          doi = {10.3847/1538-4357/ada4aa},
archivePrefix = {arXiv},
       eprint = {2311.08664},
 primaryClass = {astro-ph.CO},
       adsurl = {https://ui.adsabs.harvard.edu/abs/2025ApJ...979..223W}
}

@ARTICLE{Chen2020,
       author = {{Chen}, Yangyao and {Mo}, H.~J. and {Li}, Cheng and {Wang}, Huiyuan and {Yang}, Xiaohu and {Zhang}, Youcai and {Wang}, Kai},
        title = "{Relating the Structure of Dark Matter Halos to Their Assembly and Environment}",
      journal = {\apj},
         year = 2020,
        month = aug,
       volume = {899},
       number = {1},
          eid = {81},
        pages = {81},
          doi = {10.3847/1538-4357/aba597},
archivePrefix = {arXiv},
       eprint = {2003.05137},
 primaryClass = {astro-ph.GA},
       adsurl = {https://ui.adsabs.harvard.edu/abs/2020ApJ...899...81C}
}

@ARTICLE{Jeeson2011,
       author = {{Jeeson-Daniel}, Akila and {Dalla Vecchia}, Claudio and {Haas}, Marcel R. and {Schaye}, Joop},
        title = "{The correlation structure of dark matter halo properties}",
      journal = {\mnras},
         year = 2011,
        month = jul,
       volume = {415},
       number = {1},
        pages = {L69-L73},
          doi = {10.1111/j.1745-3933.2011.01081.x},
archivePrefix = {arXiv},
       eprint = {1103.5467},
 primaryClass = {astro-ph.CO},
       adsurl = {https://ui.adsabs.harvard.edu/abs/2011MNRAS.415L..69J}
}

@ARTICLE{Skibba2011,
       author = {{Skibba}, Ramin A. and {Macci{\`o}}, Andrea V.},
        title = "{Properties of dark matter haloes and their correlations: the lesson from principal component analysis}",
      journal = {\mnras},
         year = 2011,
        month = sep,
       volume = {416},
       number = {3},
        pages = {2388-2400},
          doi = {10.1111/j.1365-2966.2011.19218.x},
archivePrefix = {arXiv},
       eprint = {1103.1641},
 primaryClass = {astro-ph.CO},
       adsurl = {https://ui.adsabs.harvard.edu/abs/2011MNRAS.416.2388S}
}

@ARTICLE{Haggar2024a,
       author = {{Haggar}, Roan and {De Luca}, Federico and {De Petris}, Marco and {Sazonova}, Elizaveta and {Taylor}, James E. and {Knebe}, Alexander and {Gray}, Meghan E. and {Pearce}, Frazer R. and {Contreras-Santos}, Ana and {Cui}, Weiguang and {Kuchner}, Ulrike and {Mostoghiu Paun}, Robert A. and {Power}, Chris},
        title = "{Reconsidering the dynamical states of galaxy clusters using PCA and UMAP}",
      journal = {\mnras},
         year = 2024,
        month = jul,
       volume = {532},
       number = {1},
        pages = {1031-1048},
          doi = {10.1093/mnras/stae1566},
archivePrefix = {arXiv},
       eprint = {2406.15555},
 primaryClass = {astro-ph.GA},
       adsurl = {https://ui.adsabs.harvard.edu/abs/2024MNRAS.532.1031H}
}

@ARTICLE{Seppi2023,
       author = {{Seppi}, R. and {Comparat}, J. and {Nandra}, K. and {Dolag}, K. and {Biffi}, V. and {Bulbul}, E. and {Liu}, A. and {Ghirardini}, V. and {Ider-Chitham}, J.},
        title = "{Offset between X-ray and optical centers in clusters of galaxies: Connecting eROSITA data with simulations}",
      journal = {\aap},
         year = 2023,
        month = mar,
       volume = {671},
          eid = {A57},
        pages = {A57},
          doi = {10.1051/0004-6361/202245138},
archivePrefix = {arXiv},
       eprint = {2212.10107},
 primaryClass = {astro-ph.CO},
       adsurl = {https://ui.adsabs.harvard.edu/abs/2023A&A...671A..57S}
}

@ARTICLE{Richardson2022,
       author = {{Richardson}, T.~R.~G. and {Corasaniti}, P.-S.},
        title = "{Timing the last major merger of galaxy clusters with large halo sparsity}",
      journal = {\mnras},
         year = 2022,
        month = jul,
       volume = {513},
       number = {4},
        pages = {4951-4967},
          doi = {10.1093/mnras/stac1241},
archivePrefix = {arXiv},
       eprint = {2112.04926},
 primaryClass = {astro-ph.CO},
       adsurl = {https://ui.adsabs.harvard.edu/abs/2022MNRAS.513.4951R}
}

@ARTICLE{Zenteno2020,
       author = {{Zenteno}, A. and {Hern{\'a}ndez-Lang}, D. and {Klein}, M. and {Vergara Cervantes}, C. and {Hollowood}, D.~L. and {Bhargava}, S. and {Palmese}, A. and {Strazzullo}, V. and {Romer}, A.~K. and {Mohr}, J.~J. and {Jeltema}, T. and {Saro}, A. and {Lidman}, C. and {Gruen}, D. and {Ojeda}, V. and {Katzenberger}, A. and {Aguena}, M. and {Allam}, S. and {Avila}, S. and {Bayliss}, M. and {Bertin}, E. and {Brooks}, D. and {Buckley-Geer}, E. and {Burke}, D.~L. and {Capasso}, R. and {Carnero Rosell}, A. and {Carrasco Kind}, M. and {Carretero}, J. and {Castander}, F.~J. and {Costanzi}, M. and {da Costa}, L.~N. and {De Vicente}, J. and {Desai}, S. and {Diehl}, H.~T. and {Doel}, P. and {Eifler}, T.~F. and {Evrard}, A.~E. and {Flaugher}, B. and {Floyd}, B. and {Fosalba}, P. and {Frieman}, J. and {Garc{\'\i}a-Bellido}, J. and {Gerdes}, D.~W. and {Gonzalez}, J.~R. and {Gruendl}, R.~A. and {Gschwend}, J. and {Gutierrez}, G. and {Hartley}, W.~G. and {Hinton}, S.~R. and {Honscheid}, K. and {James}, D.~J. and {Kuehn}, K. and {Lahav}, O. and {Lima}, M. and {McDonald}, M. and {Maia}, M.~A.~G. and {March}, M. and {Melchior}, P. and {Menanteau}, F. and {Miquel}, R. and {Ogando}, R.~L.~C. and {Paz-Chinch{\'o}n}, F. and {Plazas}, A.~A. and {Roodman}, A. and {Rykoff}, E.~S. and {Sanchez}, E. and {Scarpine}, V. and {Schubnell}, M. and {Serrano}, S. and {Sevilla-Noarbe}, I. and {Smith}, M. and {Soares-Santos}, M. and {Suchyta}, E. and {Swanson}, M.~E.~C. and {Tarle}, G. and {Thomas}, D. and {Varga}, T.~N. and {Walker}, A.~R. and {Wilkinson}, R.~D. and {DES Collaboration}},
        title = "{A joint SZ-X-ray-optical analysis of the dynamical state of 288 massive galaxy clusters}",
      journal = {\mnras},
         year = 2020,
        month = jun,
       volume = {495},
       number = {1},
        pages = {705-725},
          doi = {10.1093/mnras/staa1157},
archivePrefix = {arXiv},
       eprint = {2004.01721},
 primaryClass = {astro-ph.GA},
       adsurl = {https://ui.adsabs.harvard.edu/abs/2020MNRAS.495..705Z}
}

@ARTICLE{Mulroy2019,
       author = {{Mulroy}, Sarah L. and {Farahi}, Arya and {Evrard}, August E. and {Smith}, Graham P. and {Finoguenov}, Alexis and {O'Donnell}, Christine and {Marrone}, Daniel P. and {Abdulla}, Zubair and {Bourdin}, Herv{\'e} and {Carlstrom}, John E. and {D{\'e}mocl{\`e}s}, Jessica and {Haines}, Chris P. and {Martino}, Rossella and {Mazzotta}, Pasquale and {McGee}, Sean L. and {Okabe}, Nobuhiro},
        title = "{LoCuSS: scaling relations between galaxy cluster mass, gas, and stellar content}",
      journal = {\mnras},
         year = 2019,
        month = mar,
       volume = {484},
       number = {1},
        pages = {60-80},
          doi = {10.1093/mnras/sty3484},
archivePrefix = {arXiv},
       eprint = {1901.11276},
 primaryClass = {astro-ph.CO},
       adsurl = {https://ui.adsabs.harvard.edu/abs/2019MNRAS.484...60M}
}

@ARTICLE{Cerini2025,
       author = {{Cerini}, Giulia and {Bellomi}, Elena and {Cappelluti}, Nico and {Khizroev}, Sabina and {Lau}, Erwin T. and {Natarajan}, Priyamvada and {ZuHone}, John},
        title = "{Revisiting Galaxy Cluster Scaling Relations through Dark Matter{\textendash}Gas Coherence: Scatter Dependence on Dynamical State}",
      journal = {\apj},
         year = 2025,
        month = sep,
       volume = {991},
       number = {1},
          eid = {56},
        pages = {56},
          doi = {10.3847/1538-4357/adf744},
archivePrefix = {arXiv},
       eprint = {2509.18240},
 primaryClass = {astro-ph.CO},
       adsurl = {https://ui.adsabs.harvard.edu/abs/2025ApJ...991...56C}
}

@ARTICLE{Yuan2022,
       author = {{Yuan}, Z.~S. and {Han}, J.~L. and {Wen}, Z.~L.},
        title = "{Dynamical state of galaxy clusters evaluated from X-ray images}",
      journal = {\mnras},
         year = 2022,
        month = jun,
       volume = {513},
       number = {2},
        pages = {3013-3021},
          doi = {10.1093/mnras/stac1037},
archivePrefix = {arXiv},
       eprint = {2204.02699},
 primaryClass = {astro-ph.GA},
       adsurl = {https://ui.adsabs.harvard.edu/abs/2022MNRAS.513.3013Y}
}

@ARTICLE{Jeltema2005,
       author = {{Jeltema}, Tesla E. and {Canizares}, Claude R. and {Bautz}, Mark W. and {Buote}, David A.},
        title = "{The Evolution of Structure in X-Ray Clusters of Galaxies}",
      journal = {\apj},
         year = 2005,
        month = may,
       volume = {624},
       number = {2},
        pages = {606-629},
          doi = {10.1086/428940},
archivePrefix = {arXiv},
       eprint = {astro-ph/0501360},
 primaryClass = {astro-ph},
       adsurl = {https://ui.adsabs.harvard.edu/abs/2005ApJ...624..606J}
}

@ARTICLE{Chen2019,
       author = {{Chen}, Huanqing and {Avestruz}, Camille and {Kravtsov}, Andrey V. and {Lau}, Erwin T. and {Nagai}, Daisuke},
        title = "{Imprints of mass accretion history on the shape of the intracluster medium and the T$_{X}$-M relation}",
      journal = {\mnras},
         year = 2019,
        month = dec,
       volume = {490},
       number = {2},
        pages = {2380-2389},
          doi = {10.1093/mnras/stz2776},
archivePrefix = {arXiv},
       eprint = {1903.08662},
 primaryClass = {astro-ph.GA},
       adsurl = {https://ui.adsabs.harvard.edu/abs/2019MNRAS.490.2380C}
}

@ARTICLE{WenHan2013,
       author = {{Wen}, Z.~L. and {Han}, J.~L.},
        title = "{Substructure and dynamical state of 2092 rich clusters of galaxies derived from photometric data}",
      journal = {\mnras},
         year = 2013,
        month = nov,
       volume = {436},
       number = {1},
        pages = {275-293},
          doi = {10.1093/mnras/stt1581},
archivePrefix = {arXiv},
       eprint = {1307.0568},
 primaryClass = {astro-ph.CO},
       adsurl = {https://ui.adsabs.harvard.edu/abs/2013MNRAS.436..275W}
}

@ARTICLE{Contreras2022,
       author = {{Contreras-Santos}, Ana and {Knebe}, Alexander and {Pearce}, Frazer and {Haggar}, Roan and {Gray}, Meghan and {Cui}, Weiguang and {Yepes}, Gustavo and {De Petris}, Marco and {De Luca}, Federico and {Power}, Chris and {Mostoghiu}, Robert and {Nuza}, Sebasti{\'a}n E. and {Hoeft}, Matthias},
        title = "{The three hundred project: galaxy cluster mergers and their impact on the stellar component of brightest cluster galaxies}",
      journal = {\mnras},
         year = 2022,
        month = apr,
       volume = {511},
       number = {2},
        pages = {2897-2913},
          doi = {10.1093/mnras/stac275},
archivePrefix = {arXiv},
       eprint = {2201.12252},
 primaryClass = {astro-ph.CO},
       adsurl = {https://ui.adsabs.harvard.edu/abs/2022MNRAS.511.2897C}
}

@ARTICLE{Yuan2020,
       author = {{Yuan}, Z.~S. and {Han}, J.~L.},
        title = "{Dynamical state for 964 galaxy clusters from Chandra X-ray images}",
      journal = {\mnras},
         year = 2020,
        month = oct,
       volume = {497},
       number = {4},
        pages = {5485-5497},
          doi = {10.1093/mnras/staa2363},
archivePrefix = {arXiv},
       eprint = {2008.01299},
 primaryClass = {astro-ph.GA},
       adsurl = {https://ui.adsabs.harvard.edu/abs/2020MNRAS.497.5485Y}
}

@ARTICLE{Dressler1988,
       author = {{Dressler}, Alan and {Shectman}, Stephen A.},
        title = "{Evidence for Substructure in Rich Clusters of Galaxies from Radial-Velocity Measurements}",
      journal = {\aj},
         year = 1988,
        month = apr,
       volume = {95},
        pages = {985},
          doi = {10.1086/114694},
       adsurl = {https://ui.adsabs.harvard.edu/abs/1988AJ.....95..985D}
}

@ARTICLE{Benavides2023,
       author = {{Benavides}, Jos{\'e} A. and {Biviano}, Andrea and {Abadi}, Mario G.},
        title = "{DS+: A method for the identification of cluster substructures}",
      journal = {\aap},
         year = 2023,
        month = jan,
       volume = {669},
          eid = {A147},
        pages = {A147},
          doi = {10.1051/0004-6361/202245422},
archivePrefix = {arXiv},
       eprint = {2212.00040},
 primaryClass = {astro-ph.GA},
       adsurl = {https://ui.adsabs.harvard.edu/abs/2023A&A...669A.147B}
}

@ARTICLE{Zarattini2016,
       author = {{Zarattini}, S. and {Girardi}, M. and {Aguerri}, J.~A.~L. and {Boschin}, W. and {Barrena}, R. and {del Burgo}, C. and {Castro-Rodriguez}, N. and {Corsini}, E.~M. and {D'Onghia}, E. and {Kundert}, A. and {M{\'e}ndez-Abreu}, J. and {S{\'a}nchez-Janssen}, R.},
        title = "{Fossil group origins. VII. Galaxy substructures in fossil systems}",
      journal = {\aap},
         year = 2016,
        month = feb,
       volume = {586},
          eid = {A63},
        pages = {A63},
          doi = {10.1051/0004-6361/201527175},
archivePrefix = {arXiv},
       eprint = {1511.02854},
 primaryClass = {astro-ph.GA},
       adsurl = {https://ui.adsabs.harvard.edu/abs/2016A&A...586A..63Z}
}

@ARTICLE{Pinkey1996,
       author = {{Pinkney}, Jason and {Roettiger}, Kurt and {Burns}, Jack O. and {Bird}, Christina M.},
        title = "{Evaluation of Statistical Tests for Substructure in Clusters of Galaxies}",
      journal = {\apjs},
         year = 1996,
        month = may,
       volume = {104},
        pages = {1},
          doi = {10.1086/192290},
       adsurl = {https://ui.adsabs.harvard.edu/abs/1996ApJS..104....1P}
}

@ARTICLE{Lovisari2017,
       author = {{Lovisari}, Lorenzo and {Forman}, William R. and {Jones}, Christine and {Ettori}, Stefano and {Andrade-Santos}, Felipe and {Arnaud}, Monique and {D{\'e}mocl{\`e}s}, Jessica and {Pratt}, Gabriel W. and {Randall}, Scott and {Kraft}, Ralph},
        title = "{X-Ray Morphological Analysis of the Planck ESZ Clusters}",
      journal = {\apj},
         year = 2017,
        month = sep,
       volume = {846},
       number = {1},
          eid = {51},
        pages = {51},
          doi = {10.3847/1538-4357/aa855f},
archivePrefix = {arXiv},
       eprint = {1708.02590},
 primaryClass = {astro-ph.CO},
       adsurl = {https://ui.adsabs.harvard.edu/abs/2017ApJ...846...51L}
}

@ARTICLE{Kimmig2025,
       author = {{Kimmig}, Lucas C. and {Brough}, Sarah and {Dolag}, Klaus and {Remus}, Rhea-Silvia and {Bah{\'e}}, Yannick M. and {Martin}, Garreth and {Pillepich}, Annalisa and {Hatch}, Nina and {Montes}, Mireia and {Lammim Ahad}, Syeda and {Bellhouse}, Callum and {Brown}, Harley J. and {Ellien}, Ama{\"e}l and {Golden-Marx}, Jesse B. and {Gonzalez}, Anthony H. and {Iodice}, Enrica and {Jim{\'e}nez-Teja}, Yolanda and {Kluge}, Matthias and {Knapen}, Johan H. and {Mihos}, J. Christopher and {Ragusa}, Rossella and {Spavone}, Marilena},
        title = "{Intra-cluster light as a dynamical clock for galaxy clusters: Insights from the MAGNETICUM, IllustrisTNG, Hydrangea, and Horizon-AGN simulations}",
      journal = {\aap},
         year = 2025,
        month = aug,
       volume = {700},
          eid = {A95},
        pages = {A95},
          doi = {10.1051/0004-6361/202554777},
archivePrefix = {arXiv},
       eprint = {2503.20857},
 primaryClass = {astro-ph.GA},
       adsurl = {https://ui.adsabs.harvard.edu/abs/2025A&A...700A..95K}
}

@ARTICLE{Kim2024,
       author = {{Kim}, Hyowon and {Smith}, Rory and {Ko}, Jongwan and {Shinn}, Jong-Ho and {Chun}, Kyungwon and {Shin}, Jihye and {Yoo}, Jaewon},
        title = "{New Observational Recipes for Measuring Dynamical States of Galaxy Clusters}",
      journal = {\apj},
         year = 2024,
        month = aug,
       volume = {970},
       number = {2},
          eid = {165},
        pages = {165},
          doi = {10.3847/1538-4357/ad4f80},
archivePrefix = {arXiv},
       eprint = {2405.06245},
 primaryClass = {astro-ph.GA},
       adsurl = {https://ui.adsabs.harvard.edu/abs/2024ApJ...970..165K}
}

@ARTICLE{Liu2024,
       author = {{Liu}, A. and {Bulbul}, E. and {Shin}, T. and {von der Linden}, A. and {Ghirardini}, V. and {Kluge}, M. and {Sanders}, J.~S. and {Grandis}, S. and {Zhang}, X. and {Artis}, E. and {Bahar}, Y.~E. and {Balzer}, F. and {Clerc}, N. and {Malavasi}, N. and {Merloni}, A. and {Nandra}, K. and {Ramos-Ceja}, M.~E. and {Zelmer}, S.},
        title = "{The SRG/eROSITA All-Sky Survey: Exploring halo assembly bias with X-ray-selected superclusters}",
      journal = {\aap},
         year = 2024,
        month = aug,
       volume = {688},
          eid = {A186},
        pages = {A186},
          doi = {10.1051/0004-6361/202450519},
archivePrefix = {arXiv},
       eprint = {2404.17345},
 primaryClass = {astro-ph.CO},
       adsurl = {https://ui.adsabs.harvard.edu/abs/2024A&A...688A.186L}
}

@ARTICLE{Mao2018,
       author = {{Mao}, Yao-Yuan and {Zentner}, Andrew R. and {Wechsler}, Risa H.},
        title = "{Beyond assembly bias: exploring secondary halo biases for cluster-size haloes}",
      journal = {\mnras},
         year = 2018,
        month = mar,
       volume = {474},
       number = {4},
        pages = {5143-5157},
          doi = {10.1093/mnras/stx3111},
archivePrefix = {arXiv},
       eprint = {1705.03888},
 primaryClass = {astro-ph.CO},
       adsurl = {https://ui.adsabs.harvard.edu/abs/2018MNRAS.474.5143M}
}

@ARTICLE{Cadiou2021,
       author = {{Cadiou}, Corentin and {Pontzen}, Andrew and {Peiris}, Hiranya V. and {Lucie-Smith}, Luisa},
        title = "{The causal effect of environment on halo mass and concentration}",
      journal = {\mnras},
         year = 2021,
        month = nov,
       volume = {508},
       number = {1},
        pages = {1189-1194},
          doi = {10.1093/mnras/stab2650},
archivePrefix = {arXiv},
       eprint = {2107.03407},
 primaryClass = {astro-ph.CO},
       adsurl = {https://ui.adsabs.harvard.edu/abs/2021MNRAS.508.1189C}
}

@ARTICLE{Cui2018,
       author = {{Cui}, Weiguang and {Knebe}, Alexander and {Yepes}, Gustavo and {Pearce}, Frazer and {Power}, Chris and {Dave}, Romeel and {Arth}, Alexander and {Borgani}, Stefano and {Dolag}, Klaus and {Elahi}, Pascal and {Mostoghiu}, Robert and {Murante}, Giuseppe and {Rasia}, Elena and {Stoppacher}, Doris and {Vega-Ferrero}, Jesus and {Wang}, Yang and {Yang}, Xiaohu and {Benson}, Andrew and {Cora}, Sof{\'\i}a A. and {Croton}, Darren J. and {Sinha}, Manodeep and {Stevens}, Adam R.~H. and {Vega-Mart{\'\i}nez}, Cristian A. and {Arthur}, Jake and {Baldi}, Anna S. and {Ca{\~n}as}, Rodrigo and {Cialone}, Giammarco and {Cunnama}, Daniel and {De Petris}, Marco and {Durando}, Giacomo and {Ettori}, Stefano and {Gottl{\"o}ber}, Stefan and {Nuza}, Sebasti{\'a}n E. and {Old}, Lyndsay J. and {Pilipenko}, Sergey and {Sorce}, Jenny G. and {Welker}, Charlotte},
        title = "{The Three Hundred project: a large catalogue of theoretically modelled galaxy clusters for cosmological and astrophysical applications}",
      journal = {\mnras},
         year = 2018,
        month = nov,
       volume = {480},
       number = {3},
        pages = {2898-2915},
          doi = {10.1093/mnras/sty2111},
archivePrefix = {arXiv},
       eprint = {1809.04622},
 primaryClass = {astro-ph.GA},
       adsurl = {https://ui.adsabs.harvard.edu/abs/2018MNRAS.480.2898C}
}

@ARTICLE{Deluca2021,
       author = {{De Luca}, Federico and {De Petris}, Marco and {Yepes}, Gustavo and {Cui}, Weiguang and {Knebe}, Alexander and {Rasia}, Elena},
        title = "{The Three Hundred project: dynamical state of galaxy clusters and morphology from multiwavelength synthetic maps}",
      journal = {\mnras},
         year = 2021,
        month = jul,
       volume = {504},
       number = {4},
        pages = {5383-5400},
          doi = {10.1093/mnras/stab1073},
archivePrefix = {arXiv},
       eprint = {2011.09002},
 primaryClass = {astro-ph.CO},
       adsurl = {https://ui.adsabs.harvard.edu/abs/2021MNRAS.504.5383D}
}

@ARTICLE{WenHan2024,
       author = {{Wen}, Z.~L. and {Han}, J.~L.},
        title = "{A Catalog of 1.58 Million Clusters of Galaxies Identified from the DESI Legacy Imaging Surveys}",
      journal = {\apjs},
         year = 2024,
        month = jun,
       volume = {272},
       number = {2},
          eid = {39},
        pages = {39},
          doi = {10.3847/1538-4365/ad409d}
}

@ARTICLE{Nelson2019,
       author = {{Nelson}, Dylan and {Springel}, Volker and {Pillepich}, Annalisa and {Rodriguez-Gomez}, Vicente and {Torrey}, Paul and {Genel}, Shy and {Vogelsberger}, Mark and {Pakmor}, Ruediger and {Marinacci}, Federico and {Weinberger}, Rainer and {Kelley}, Luke and {Lovell}, Mark and {Diemer}, Benedikt and {Hernquist}, Lars},
        title = "{The IllustrisTNG simulations: public data release}",
      journal = {Computational Astrophysics and Cosmology},
         year = 2019,
        month = may,
       volume = {6},
       number = {1},
          eid = {2},
        pages = {2},
          doi = {10.1186/s40668-019-0028-x}
}

@ARTICLE{vandenBosch2002,
       author = {{van den Bosch}, Frank C.},
        title = "{The universal mass accretion history of cold dark matter haloes}",
      journal = {\mnras},
         year = 2002,
        month = mar,
       volume = {331},
       number = {1},
        pages = {98-110},
          doi = {10.1046/j.1365-8711.2002.05171.x},
archivePrefix = {arXiv},
       eprint = {astro-ph/0105158},
 primaryClass = {astro-ph},
       adsurl = {https://ui.adsabs.harvard.edu/abs/2002MNRAS.331...98V}
}

@ARTICLE{Golden-Marx2018,
       author = {{Golden-Marx}, Jesse B. and {Miller}, Christopher J.},
        title = "{The Impact of Environment on the Stellar Mass-Halo Mass Relation}",
      journal = {\apj},
         year = 2018,
        month = jun,
       volume = {860},
       number = {1},
          eid = {2},
        pages = {2},
          doi = {10.3847/1538-4357/aac2bd},
archivePrefix = {arXiv},
       eprint = {1711.00481},
 primaryClass = {astro-ph.GA},
       adsurl = {https://ui.adsabs.harvard.edu/abs/2018ApJ...860....2G}
}

@ARTICLE{Crone1996,
    author = {{Crone}, Mary M. and {Evrard}, August E. and {Richstone}, Douglas O.},
    title = "{Substructure in Clusters as a Cosmological Test}",
    journal = {\apj},
    year = 1996,
    month = aug,
    volume = {467},
    pages = {489},
    doi = {10.1086/177626}
}

@ARTICLE{Thomas1998,
    author = {{Thomas}, Peter A. and {Colberg}, Jorg M. and {Couchman}, Hugh M.~P. and {Efstathiou}, George P. and {Frenk}, Carlos S. and {Jenkins}, Adrian R. and {Nelson}, Alistair H. and {Hutchings}, Roger M. and {Peacock}, John A. and {Pearce}, Frazer R. and {White}, Simon D.~M. and {Virgo Consortium}},
    title = "{The structure of galaxy clusters in various cosmologies}",
    journal = {\mnras},
    year = 1998,
    month = jun,
    volume = {296},
    number = {4},
    pages = {1061-1071},
    doi = {10.1046/j.1365-8711.1998.01491.x}
}

@ARTICLE{Diemer2017,
    author = {{Diemer}, Benedikt},
    title = "{The Splashback Radius of Halos from Particle Dynamics. I. The SPARTA Algorithm}",
    journal = {\apjs},
    year = 2017,
    month = jul,
    volume = {231},
    number = {1},
    eid = {5},
    pages = {5},
    doi = {10.3847/1538-4365/aa799c}
}

@ARTICLE{Xhakaj2022,
    author = {{Xhakaj}, Enia and {Leauthaud}, Alexie and {Lange}, Johannes and {Hearin}, Andrew and {Diemer}, Benedikt and {Dalal}, Neal},
    title = "{Beyond mass: detecting secondary halo properties with galaxy-galaxy lensing}",
    journal = {\mnras},
    year = 2022,
    month = aug,
    volume = {514},
    number = {2},
    pages = {2876-2890},
    doi = {10.1093/mnras/stac941}
}

@ARTICLE{Magnus2026,
    author = {{Correa Magnus}, Lilia and {Kay}, Scott T. and {Schaye}, Joop and {Schaller}, Matthieu},
    title = "{Signatures of dynamical activity in the hot gas profiles of groups and clusters in the FLAMINGO simulations}",
    journal = {\mnras},
    year = 2026,
    month = jan,
    volume = {545},
    number = {2},
    eid = {staf2073},
    pages = {staf2073},
    doi = {10.1093/mnras/staf2073}
}

@ARTICLE{Klypin2016,
       author = {{Klypin}, Anatoly and {Yepes}, Gustavo and {Gottl{\"o}ber}, Stefan and {Prada}, Francisco and {He{\ss}}, Steffen},
        title = "{MultiDark simulations: the story of dark matter halo concentrations and density profiles}",
      journal = {\mnras},
         year = 2016,
        month = apr,
       volume = {457},
       number = {4},
        pages = {4340-4359},
          doi = {10.1093/mnras/stw248},
archivePrefix = {arXiv},
       eprint = {1411.4001},
 primaryClass = {astro-ph.CO},
       adsurl = {https://ui.adsabs.harvard.edu/abs/2016MNRAS.457.4340K}
}

@ARTICLE{Balmes14,
       author = {{Balm{\`e}s}, I. and {Rasera}, Y. and {Corasaniti}, P.-S. and {Alimi}, J.-M.},
        title = "{Imprints of dark energy on cosmic structure formation - III. Sparsity of dark matter halo profiles}",
      journal = {\mnras},
         year = 2014,
        month = jan,
       volume = {437},
       number = {3},
        pages = {2328-2339},
          doi = {10.1093/mnras/stt2050},
archivePrefix = {arXiv},
       eprint = {1307.2922},
 primaryClass = {astro-ph.CO},
       adsurl = {https://ui.adsabs.harvard.edu/abs/2014MNRAS.437.2328B}
}

@ARTICLE{Schaye2023,
    author = {{Schaye}, Joop and {Kugel}, Roi and {Schaller}, Matthieu and {Helly}, John C. and {Braspenning}, Joey and {Elbers}, Willem and {McCarthy}, Ian G. and {van Daalen}, Marcel P. and {Vandenbroucke}, Bert and {Frenk}, Carlos S. and {Kwan}, Juliana and {Salcido}, Jaime and {Bah{\'e}}, Yannick M. and {Borrow}, Josh and {Chaikin}, Evgenii and {Hahn}, Oliver and {Hu{\v{s}}ko}, Filip and {Jenkins}, Adrian and {Lacey}, Cedric G. and {Nobels}, Folkert S.~J.},
    title = "{The FLAMINGO project: cosmological hydrodynamical simulations for large-scale structure and galaxy cluster surveys}",
    journal = {\mnras},
    year = 2023,
    month = dec,
    volume = {526},
    number = {4},
    pages = {4978-5020},
    doi = {10.1093/mnras/stad2419}
}



\appendix

\section{Gaussian groups}
\label{appx:gaussian-groups}

In Section~\ref{sec:3correlations}, we showed that a two-component Gaussian mixture model fit to the distribution of 3D centre of mass offsets versus $z_{75}$ produces groups of halos with very different median MAHs. However, a third peak in the halo distribution that is visible in the middle panel of Figure~\ref{fig:com-offset-groups} is not captured by the two-component mixture model. This suggests that there may be a third population of halos between the relaxed and unrelaxed groups. In this appendix, we investigate whether adding more Gaussian components is warranted to better describe the population of halos.

\begin{figure} 
    \centering
    \includegraphics[width=\linewidth]{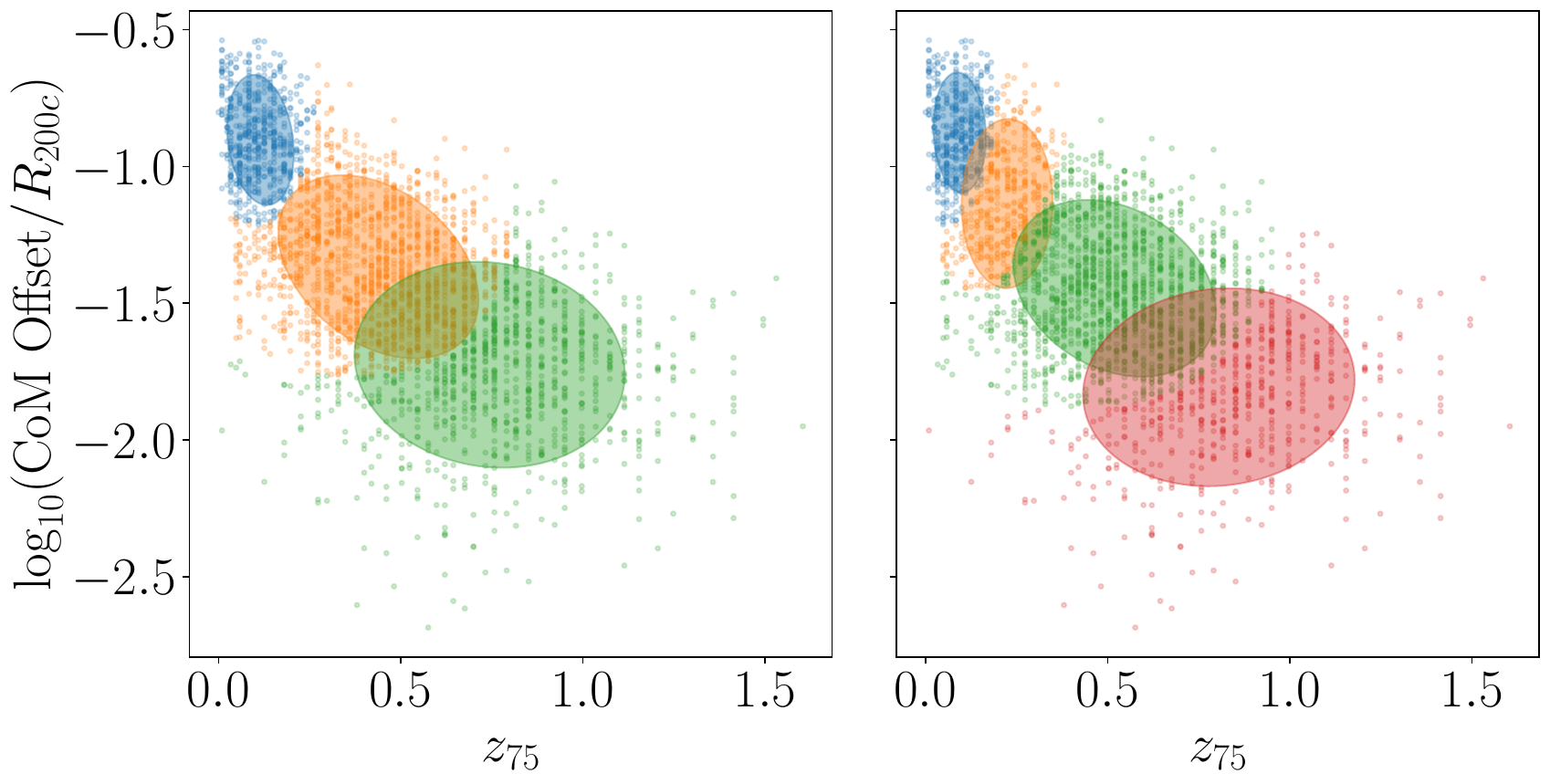}
    \caption{The 3D centre of mass offset versus $z_{75}$ for group-mass halos, split using Gaussian mixture models. The left panel shows the groups for a 3-component mixture model, while the right panel shows the same for a 4-component mixture model. Both panels are formatted as in Figure~\ref{fig:com-offset-groups}.}
    \label{fig:com-offset-multigroups}
\end{figure}

In Figure~\ref{fig:com-offset-multigroups}, we use the Gaussian mixture model fitting procedure defined in Section~\ref{sec:3correlations} to identify additional groups of halos. We see in the left panel of Figure~\ref{fig:com-offset-multigroups} that using a three-component mixture model does not capture the three peaks in the distribution that are identifiable by eye. Rather, it splits the previously defined ``relaxed'' group into two different populations. In the right panel, a four-component mixture model splits the distribution of halos into unintuitive groups along the correlation between the 3D centre of mass offset and $z_{75}$. There is also significant overlap between some of the $1\sigma$ contours in the three- and four-component models, which suggests that they do not split the population as distinctly as the two-component model.

\begin{figure} 
    \centering
    \includegraphics[width=\linewidth]{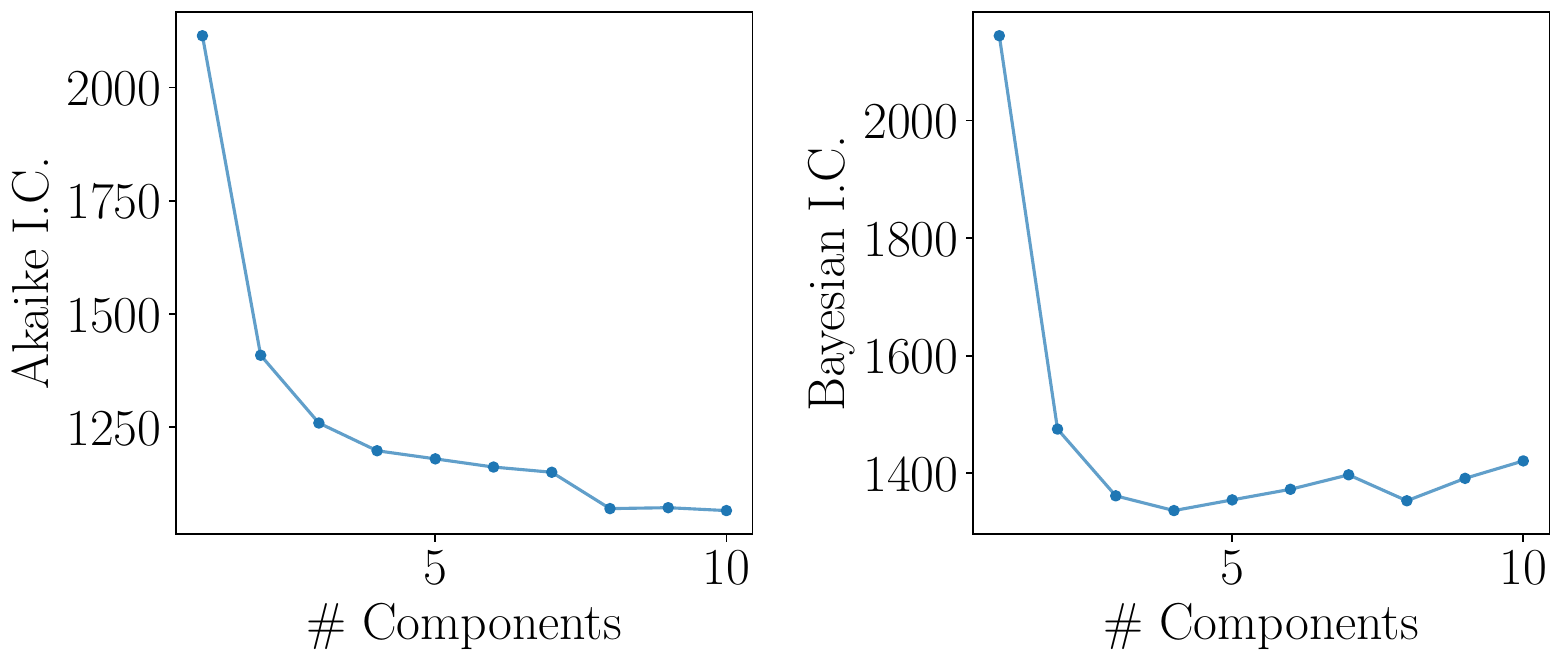}
    \caption{The BIC and AIC for Gaussian mixture models fitted to the 3D centre of mass offsets and mass history parameters $z_{75}$ of group-mass halos. The AIC and BIC are plotted as a function of the number of mixture model components.}
    \label{fig:info-criteria}
\end{figure}

To quantify how well these multi-component Gaussian mixture models fit the data without overfitting, we employ the Akaike information criterion (AIC) and the Bayesian information criterion (BIC). Each of these criteria is a numerical value that represents the goodness of fit of a model to a data set, while preferring models with fewer free parameters. The minimum value of these functions is meant to indicate the best fit for the data, weighted by the number of free parameters. Figure~\ref{fig:info-criteria} shows the values of both information criteria for our halo sample as a function of the number of Gaussian mixture model components. We see in Figure~\ref{fig:info-criteria} that the AIC decreases consistently as a function of the number of Gaussian components. Meanwhile, the BIC has a minimum at 4 Gaussian components. To test the robustness of these results, we compute the same information criteria for randomly selected halves of our data set. The AIC shows the same behaviour under this treatment, while the minimum for the BIC changes between 3--5 components, depending on the sample.

Both the continued decrease of the AIC and the minimum of the BIC suggest that the distribution of 3D centre of mass offsets and $z_{75}$ mass histories is best fit by more than two Gaussian components. However, while both information criteria decrease dramatically going from one to two Gaussian components, they only decrease marginally when adding additional components beyond the second. This provides a justification for dividing the distribution into two components, rather than three or more. Both the absence of a minimum in the AIC within a reasonable number of Gaussian components and the instability of the BIC minimum based on sample selection suggest that a mixture of Gaussian components is not a good fit to the data overall. As such, we conclude that it is appropriate to split our halo sample into only two groups --- relaxed and unrelaxed --- but that Gaussian mixture models are not ideal for decomposing this distribution.

To summarise, we have found that halos classified as ``relaxed'' or ``unrelaxed'' based on 3D centre of mass offsets and $z_{75}$ mass histories produce significantly different samples, and that there is evidence for two classes of halos based on these two parameters.

\section{Cluster-Mass Halos}
\label{appx:high-mass-halos}

In Section~\ref{sec:4structure-selection}, we present the structural properties and mass accretion histories of group-mass halos since they provide us with enough objects to reduce noise. In this section, we present some analogous results for cluster-mass halos to verify the consistency of our results for more massive systems. In particular, we consider clusters with masses $1\times10^{14}\leq M_{200c}/M_\odot\leq5\times10^{14}$, which gives us a total of 266 halos with measurements of $z_{75}$.

\begin{figure*} 
    \centering
    \includegraphics[width=\linewidth]{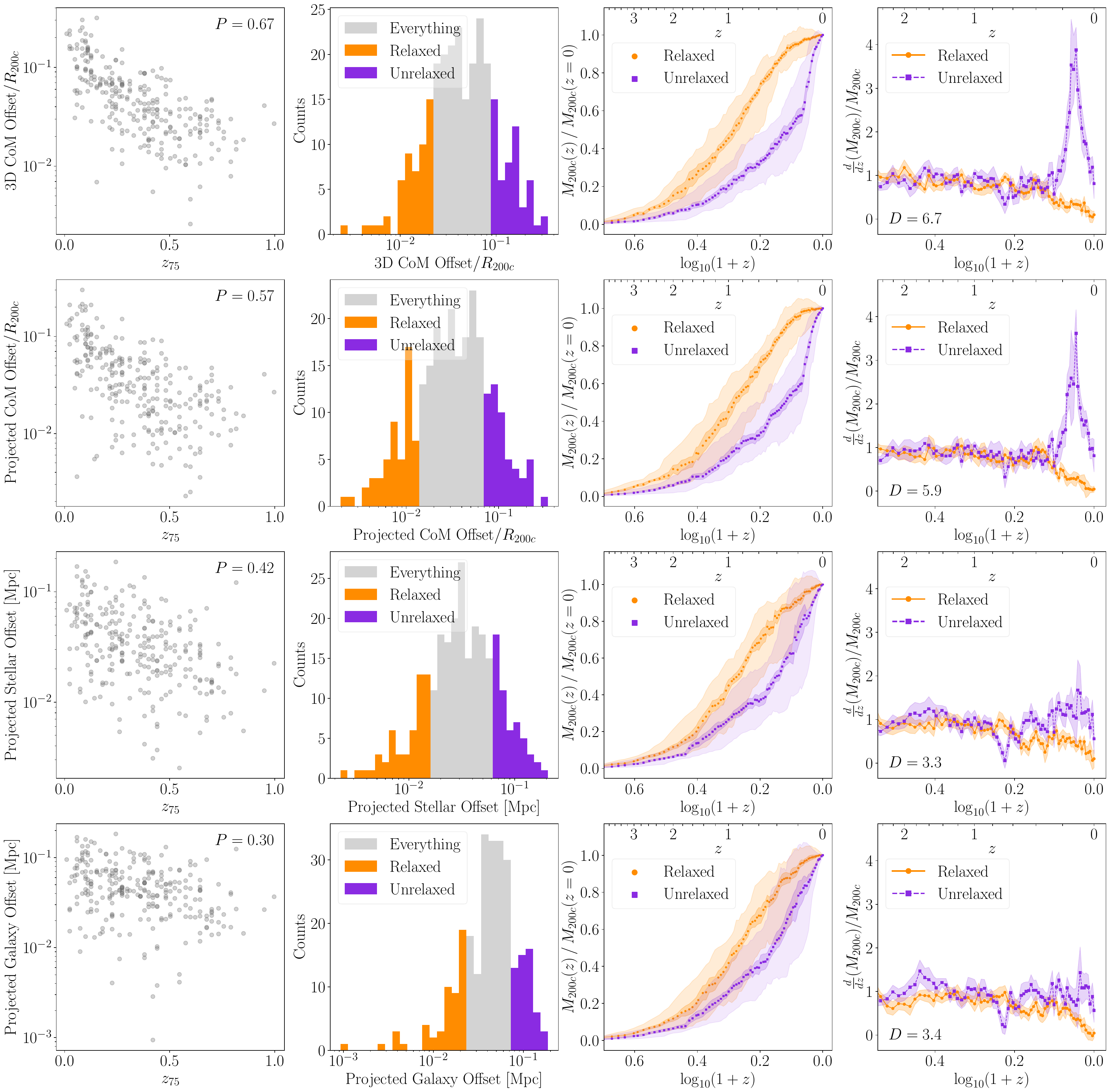}
    \caption{Parameter correlations and split sample properties, as in Figure~\ref{fig:offsets-panels}, but for halos with masses $1\times10^{14}\leq M_{200c}/M_\odot\leq5\times10^{14}$.}
    \label{fig:highmass-offsets-panels}
\end{figure*}

In Figure~\ref{fig:highmass-offsets-panels}, we present the parameter distributions and mass accretion histories of relaxed and unrelaxed halo samples selected on centre of mass offsets. These results are analogous to those in Section~\ref{subsec:structure-offsets}. In the leftmost column, we see that the 3D centre of mass offset is most strongly correlated with $z_{75}$, while the projected, stellar and galaxy offsets have progressively more scatter with respect to $z_{75}$. In the third and rightmost columns, we see that the differences between the median MAHs and MARs of the relaxed and unrelaxed groups shrink with each row as the measurements transition from most intrinsic to most easily observable. All of these trends are consistent with what is presented for group-mass halos in Section~\ref{subsec:structure-offsets}.

Using the maximum normalised difference $D$ of the MARs, we can compare the results for cluster-mass halos in Figure~\ref{fig:highmass-offsets-panels} to those for group-mass halos in Figure~\ref{fig:offsets-panels}. For each of the four offset measurements, we see that $D$ is greater for group-mass halos than it is for cluster-mass halos. However, this is due to the greater uncertainty on the median MARs of cluster-mass halos, resulting from the smaller size of the more massive sample. The vertical axis scales in the rightmost panels of Figures~\ref{fig:offsets-panels} and \ref{fig:highmass-offsets-panels} indicate that the difference between the MARs of the relaxed and unrelaxed subsamples is more pronounced for more massive halos.

Though the figures are omitted for brevity, we achieved similar results for all of the other structural measurements tested in this work. The differences between the subsample MARs are generally more pronounced for higher mass halos, but the results are noisier due to there being fewer systems to compare. This is consistent with work by \citet{Magnus2026}, who found that correlations between dynamical state indicators and halo mass accretion are typically stronger in more massive clusters at low redshift.


\bsp	
\label{lastpage}
\end{document}